\documentclass{article}
\usepackage{amssymb}
\usepackage{amsmath}

 \allowdisplaybreaks

\begin{document}

\title{On the uniqueness of $D=11$ interactions among a graviton, a massless gravitino and a three-form.\\
IV: Putting things together}

\author{E. M. Cioroianu\thanks{ e-mail address:
manache@central.ucv.ro}, E. Diaconu\thanks{ e-mail address:
ediaconu@central.ucv.ro}, S. C. Sararu\thanks{ e-mail
address: scsararu@central.ucv.ro}\\
Faculty of Physics, University of Craiova,\\ 13 Al. I. Cuza Street
Craiova, 200585, Romania}

\maketitle

\begin{abstract}
Under the hypotheses of smoothness of the interactions in the
coupling constant, locality, Poincar\'{e} invariance, Lorentz
covariance, and the preservation of the number of derivatives on
each field in the Lagrangian of the interacting theory (the same
number of derivatives like in the free Lagrangian), we prove that
the only consistent interactions in $D=11$ among massless gravitini,
a graviton, and a $3$-form are described by $N=1$, $D=11 $ SUGRA.


\end{abstract}

\section{Introduction}

In this part we use the results from Refs.~\cite{pI}, \cite{pII},
and \cite{pIII} and approach the fourth (and final) step of
constructing all possible interactions in $D=11$ among a graviton, a
massless Majorana spin-$3/2$ field, and a three-form gauge field. Of
course, we maintain the same working hypotheses like in the first
three parts, namely smoothness of interactions in the coupling
constant, locality, Poincar\'{e} invariance, Lorentz covariance, and
the preservation of the number of derivatives on each field in the
Lagrangian density of the interacting theory. First, we put all the
fields together and investigate if there are consistent interactions
vertices \emph{at order one} in the coupling constant involving
\emph{all} of them. The answer is negative, such that the
first-order deformation of the solution to the master equation is
completely known from the previous steps. Second, we analyze the
consistency of the first-order deformation at order two in the
coupling constant. This restricts the six constants that
parameterize the first-order deformation to satisfy a simple,
algebraic system. There are two types of solutions, but only one is
interesting from the point of view of interactions (the other allows
at most the interactions between a graviton and a $3$-form). Third,
we analyze this solution and observe that it systematically
reproduces the Lagrangian formulation of $D=11
$, $N=1$ SUGRA. Therefore, we can state that all consistent interactions in $%
D=11$ among a spin-$2$ field, a massless Majorana spin-$3/2$ field, and a
three-form that comply with our working hypotheses are uniquely described by
$D=11$, $N=1$ SUGRA.

\section{No simultaneous interactions at order one in the coupling constant}

Now, we put together all the three kinds of fields (graviton, massless
gravitini and three-form) and start from the free action
\begin{equation}
S_{0}^{\mathrm{L}}\left[ h_{\mu \nu },A_{\mu \nu \rho },\psi _{\mu }\right]
=\int d^{11}x\left( \mathcal{L}_{0}^{\mathrm{h}}+\mathcal{L}_{0}^{%
\mathrm{A}}+\mathcal{L}_{0}^{\mathrm{\psi }}\right)  \label{int0}
\end{equation}%
in $D=11$, where $\mathcal{L}_{0}^{\mathrm{h}}$, $\mathcal{L}_{0}^{\mathrm{A}%
}$, and $\mathcal{L}_{0}^{\mathrm{\psi }}$ denote the Lagrangian
densities of the Pauli-Fierz model, of an Abelian three-form, and of
a massless Rarita-Schwinger field respectively (see Section 2 from
Ref.~\cite{pI} and also Section 2 from Ref.~\cite{pII}).
Consequently, the BRST symmetry of the free
model (\ref{int0}) is written as%
\begin{equation}
s=\frac{1}{2}\left( s^{\mathrm{h,A}}+s^{\mathrm{A,\psi }}+s^{\mathrm{h,\psi }%
}\right) ,  \label{int01}
\end{equation}%
where $s^{\mathrm{h,A}}$, $s^{\mathrm{A,\psi }}$, and
$s^{\mathrm{h,\psi }}$ denote the BRST symmetries of the free models
respectively approached in Refs.~\cite{pI}, \cite{pII}, and
\cite{pIII}.
The overall BRST differential (\ref%
{int01}) further decomposes as%
\begin{equation}
s=\delta +\gamma ,  \label{int02}
\end{equation}%
where $\delta $ stands for the full Koszul-Tate differential and $\gamma $
represents the total longitudinal exterior derivative. Both operators from
the right-hand side of (\ref{int02}) can be written in a manner similar to (%
\ref{int01}), but in terms of the corresponding operators built in
Refs.~\cite{pI}--\cite{pIII}%
\begin{equation}
\delta =\frac{1}{2}\left( \delta ^{\mathrm{h,A}}+\delta ^{\mathrm{A,\psi }%
}+\delta ^{\mathrm{h,\psi }}\right) ,\qquad \gamma =\frac{1}{2}\left( \gamma
^{\mathrm{h,A}}+\gamma ^{\mathrm{A,\psi }}+\gamma ^{\mathrm{h,\psi }}\right)
.  \label{int03}
\end{equation}%
The actions of $\delta $ and $\gamma $ on the generators from the BRST
complex associated with theory (\ref{int0}) are given by%
\begin{eqnarray}
&&\delta h^{\ast \mu \nu }=2H^{\mu \nu },\quad \delta A^{\ast \mu \nu \rho }=%
\frac{1}{3!}\partial _{\lambda }F^{\mu \nu \rho \lambda },\quad \delta \psi
^{\ast \mu }=-\mathrm{i}\partial _{\alpha }\bar{\psi}_{\beta }\gamma
^{\alpha \beta \mu },  \label{int04a} \\
&&\delta \eta ^{\ast \mu }=-2\partial _{\nu }h^{\ast \mu \nu },\quad \delta
C^{\ast \mu \nu }=-3\partial _{\rho }A^{\ast \mu \nu \rho },\quad \delta \xi
^{\ast }=\partial _{\mu }\psi ^{\ast \mu },  \label{int04b} \\
&&\delta C^{\ast \mu }=-2\partial _{\nu }C^{\ast \mu \nu },\quad \delta
C^{\ast }=-\partial _{\mu }C^{\ast \mu },\quad \delta \chi ^{\Omega }=0,
\label{int04c} \\
&&\gamma \chi _{\Omega }^{\ast }=0,\quad \gamma h_{\mu \nu }=\partial _{(\mu
}\eta _{\nu )},\quad \gamma A_{\mu \nu \rho }=\partial _{\lbrack \mu }C_{\nu
\rho ]},\quad \gamma \psi _{\mu }=\partial _{\mu }\xi ,  \label{int04d} \\
&&\gamma \eta _{\mu }=0,\quad \gamma C_{\mu \nu }=\partial _{\lbrack \mu
}C_{\nu ]},\quad \gamma \xi =0,\quad \gamma C_{\mu }=\partial _{\mu }C,\quad
\gamma C=0.  \label{int04e}
\end{eqnarray}%
In formulas (\ref{int04a})--(\ref{int04e}) we denoted by $\chi ^{\Omega }$
the entire field/ghost spectrum and by $\chi _{\Omega }^{\ast }$ their
antifields.

At this point, we will rely on the previous results exposed in
Refs.~\cite{pI}--\cite{pIII} related to the first-order deformation
of the solution to the master equation in the various sectors of
theory (\ref{int0}) and to the associated local BRST cohomologies in
order to determine the expression of the first-order deformation for
the full model.

Let us denote by $S_{1}$ the first-order deformation of the solution to the
master equation for theory (\ref{int0}), which is solution to the equation
\begin{equation}
sS_{1}=0.  \label{firstorder}
\end{equation}%
The functional $S_{1}$ naturally decomposes into%
\begin{equation}
S_{1}=S_{1}^{\mathrm{h}}+S_{1}^{\mathrm{A}}+S_{1}^{\mathrm{\psi }}+S_{1}^{%
\mathrm{h-A}}+S_{1}^{\mathrm{A-\psi }}+S_{1}^{\mathrm{h-\psi }}+S_{1}^{%
\mathrm{int}}.  \label{descfirstorder}
\end{equation}%
Some of the terms from the right-hand side of (\ref{descfirstorder})
have already been constructed in Refs.~\cite{pI}--\cite{pIII}. Their
significance is as follows:

\begin{itemize}
\item $S_{1}^{\mathrm{h}}$ means the first-order deformation in the
Pauli-Fierz sector (it depends \emph{only} on the BRST generators
associated with the Pauli-Fierz model) and has been extensively
investigated in the literature. Its nonintegrated density is given
for instance in formula (47) from Ref.~\cite{pI};

\item $S_{1}^{\mathrm{A}}$ represents the first-order deformation for the $3$%
-form (it involves \emph{only} the BRST generators corresponding to
an Abelian $3$-form) and was explicitly computed in Ref.~\cite{pI},
see formula (52).

\item $S_{1}^{\mathrm{\psi }}$ signifies the component of the first-order
deformation in the Rarita-Schwinger sector (it comprises \emph{only}
the BRST generators for a massless Rarita-Schwinger vector spinor)
and was deduced in Ref.~\cite{pII}, see formula (50);

\item $S_{1}^{\mathrm{h-A}}$ denotes the first-order deformation related to
the cross-couplings between a Pauli-Fierz field and a $3$-form (it \emph{%
effectively mixes} the two sorts of BRST generators) and was built
in detail in Ref.~\cite{pI}, see formula (77);

\item $S_{1}^{\mathrm{A-\psi }}$ is the first-order deformation describing
the interactions between massless gravitini and a $3$-form (again, it \emph{%
effectively couples} the BRST generators from the vector spinor
complex with those of the $3$-form) and was approached in
Ref.~\cite{pII}, see formula (110);

\item $S_{1}^{\mathrm{h-\psi }}$ stands for the first-order deformation
expressing the cross-couplings between a spin-$2$ field and a
massless Rarita-Schwinger spinor (it \emph{effectively combines} the
Pauli-Fierz BRST generators with those from the vector spinor
sector) and was analyzed in Ref.~\cite{pIII}. Its nonintegrated
density is the sum between the components listed in formulas
(20)--(22) from Ref.~\cite{pIII};

\item Finally, $S_{1}^{\mathrm{int}}$ is the first-order deformation that
\emph{gathers simultaneously all the three types of BRST generators
} and thus describes (at least cubic) interaction vertices
containing the spin-$2$ field, the massless gravitini and the
$3$-form. It will be investigated in the sequel.
\end{itemize}

Since each of the first six components from the right-hand side of (\ref%
{descfirstorder}) satisfies an equation of the type (\ref{firstorder}), it
follows that $S_{1}^{\mathrm{int}}$ is subject to the equation%
\begin{equation}
sS_{1}^{\mathrm{int}}=0.  \label{int1}
\end{equation}%
In order to compute the general solution to this equation, let us denote by $%
a^{\mathrm{int}}$ its nonintegrated density, such that the local form of (%
\ref{int1}) is%
\begin{equation}
sa^{\mathrm{int}}=\partial ^{\mu }m_{\mu }^{\mathrm{int}},  \label{int3}
\end{equation}%
where $m_{\mu }^{\mathrm{int}}$ is a local current. Eq. (\ref{int3})
shows that $a^{\mathrm{int}}\in H^{0}\left( s|d\right) $, where $d$
is the
exterior spacetime differential in $\mathbf{M}_{11}$. The solution to (\ref%
{int3}) is unique modulo the addition of $s$-exact terms and full
divergences
\begin{equation}
a^{\mathrm{int}}\rightarrow a^{\mathrm{int}}+sc^{\mathrm{int}}+\partial
^{\mu }n_{\mu }^{\mathrm{int}}.  \label{int4}
\end{equation}%
If the general solution to Eq. (\ref{int3}) is purely trivial, $a^{%
\mathrm{int}}=sc^{\mathrm{int}}+\partial ^{\mu }n_{\mu }^{\mathrm{int}}$,
then it can be taken to vanish, $a^{\mathrm{int}}=0$.

In order to analyze Eq. (\ref{int3}), we develop $a^{\mathrm{int}}$
with respect to the antighost number%
\begin{equation}
a^{\mathrm{int}}=\sum\limits_{i=0}^{I}a_{i}^{\mathrm{int}},\qquad \mathrm{agh%
}\left( a_{i}^{\mathrm{int}}\right) =i,\qquad \mathrm{gh}\left( a_{i}^{%
\mathrm{int}}\right) =0,\qquad \varepsilon \left( a_{i}^{\mathrm{int}%
}\right) =0,  \label{int5}
\end{equation}%
and assume, without loss of generality, that decomposition (\ref{int5})
stops at some finite value of $I$. Replacing (\ref{int5}) into (\ref{int3})
and projecting it on the various values of the antighost number by means of (%
\ref{int02}), we obtain that (\ref{int3}) is equivalent to the tower of
equations
\begin{eqnarray}
\gamma a_{I}^{\mathrm{int}} &=&\partial ^{\mu }\overset{\left( I\right) }{m}%
_{\mu }^{\mathrm{int}},  \label{int6a} \\
\delta a_{I}^{\mathrm{int}}+\gamma a_{I-1}^{\mathrm{int}} &=&\partial ^{\mu }%
\overset{\left( I-1\right) }{m}_{\mu }^{\mathrm{int}},  \label{int6b} \\
\delta a_{i}^{\mathrm{int}}+\gamma a_{i-1}^{\mathrm{int}} &=&\partial ^{\mu }%
\overset{\left( i-1\right) }{m}_{\mu }^{\mathrm{int}},\qquad 1\leq i\leq I-1,
\label{int6c}
\end{eqnarray}%
where $\left( \overset{\left( i\right) }{m}_{\mu }^{\mathrm{int}}\right) _{i=%
\overline{0,I}}$ are some local currents, with $\mathrm{agh}\left( \overset{%
\left( i\right) }{m}_{\mu }^{\mathrm{int}}\right) =i$. Eq. (\ref%
{int6a}) can be replaced in strictly positive antighost numbers by
\begin{equation}
\gamma a_{I}^{\mathrm{int}}=0,\qquad I>0.  \label{int7}
\end{equation}%
Due to the second-order nilpotency of $\gamma $ ($\gamma ^{2}=0$), the
solution to (\ref{int7}) is unique up to $\gamma $-exact contributions
\begin{equation}
a_{I}^{\mathrm{int}}\rightarrow a_{I}^{\mathrm{int}}+\gamma c_{I}^{\mathrm{%
int}}.  \label{int8}
\end{equation}%
Meanwhile, if it turns out that $a_{I}^{\mathrm{int}}$ reduces to $\gamma $%
-exact terms only, $a_{I}^{\mathrm{int}}=\gamma c_{I}^{\mathrm{int}}$, then
it can be made to vanish, $a_{I}^{\mathrm{int}}=0$. In other words, the
nontriviality of the first-order deformation $a^{\mathrm{int}}$ is
translated at its highest antighost number component into the requirement
that $a_{I}^{\mathrm{int}}\in H^{I}\left( \gamma \right) $, where $%
H^{I}\left( \gamma \right) $ denotes the cohomology of the exterior
longitudinal derivative $\gamma $ in pure ghost number equal to $I$. So, in
order to solve Eq. (\ref{int3}) (equivalent with (\ref{int7}) and (\ref%
{int6b})--(\ref{int6c})), we need to compute the cohomology of $\gamma $, $%
H\left( \gamma \right) $, and, as it will be made clear below, also the
local cohomology of $\delta $, $H\left( \delta |d\right) $.

Using the results derived in Refs.~\cite{pI}--\cite{pIII} regarding
the cohomology of $\gamma $, we can state that $H\left( \gamma
\right) $ is generated on
the one hand by $\chi _{\Omega }^{\ast }$, $F_{\mu \nu \rho \lambda }$, $%
\partial _{\lbrack \mu }\psi _{\nu ]}$, and $K_{\mu \nu \alpha \beta }$,
together with their spacetime derivatives and, on the other hand, by the
undifferentiated ghost for ghost for ghost $C$, by the undifferentiated
ghost $\xi $\ as well by the ghosts $\eta _{\mu }$ and their antisymmetric
first-order derivatives $\partial _{\lbrack \mu }\eta _{\nu ]}$. So, the
most general (and nontrivial) solution to (\ref{int7}) can be written, up to
$\gamma $-exact contributions, as
\begin{equation}
a_{I}^{\mathrm{int}}=\alpha _{I}\left( \left[ F_{\mu \nu \rho \lambda }%
\right] ,\left[ \partial _{\lbrack \mu }\psi _{\nu ]}\right] ,\left[ K_{\mu
\nu \alpha \beta }\right] ,\left[ \chi _{\Omega }^{\ast }\right] \right)
\omega ^{I}\left( C,\xi ,\eta _{\mu },\partial _{\lbrack \mu }\eta _{\nu
]}\right) ,  \label{int9}
\end{equation}%
where the notation $f\left( \left[ q\right] \right) $ means that $f$ depends
on $q$ and its derivatives up to a finite order, and $\omega ^{I}$ denotes
the elements of a basis in the space of polynomials with pure ghost number $%
I $ in the corresponding ghost for ghost for ghost, Rarita-Schwinger ghost,
Pauli-Fierz ghosts and their antisymmetric first-order derivatives. The
objects $\alpha _{I}$ (obviously nontrivial in $H^{0}\left( \gamma \right) $%
) were taken to have a finite antighost number and a bounded number of
derivatives, and therefore they are polynomials in the antifields $\chi
_{\Omega }^{\ast }$, in the linearized Riemann tensor $K_{\mu \nu \alpha
\beta }$, in the antisymmetric first-order derivatives of the spin-vector, $%
\partial _{\lbrack \mu }\psi _{\nu ]}$, and in the field-strength of the
three-form $F_{\mu \nu \rho \lambda }$ as well as in their subsequent
derivatives. They are required to fulfill the property $\mathrm{agh}\left(
\alpha _{I}\right) =I$ in order to ensure that the ghost number of $a_{I}^{%
\mathrm{int}}$ is equal to zero. Due to their $\gamma $-closeness, $\gamma
\alpha _{I}=0$, and to their polynomial character, $\alpha _{I}$ will be
called invariant polynomials.

Inserting (\ref{int9}) in (\ref{int6b}), we obtain that a necessary (but not
sufficient) condition for the existence of (nontrivial) solutions $a_{I-1}$
is that the invariant polynomials $\alpha _{I}$ are (nontrivial) objects
from the local cohomology of the Koszul-Tate differential $H\left( \delta
|d\right) $ in antighost number $I>0$ and in pure ghost number zero,
\begin{equation}
\delta \alpha _{I}=\partial _{\mu }\overset{\left( I-1\right) }{j}^{\mu
},\qquad \mathrm{agh}\left( \overset{\left( I-1\right) }{j}^{\mu }\right)
=I-1,\qquad \mathrm{pgh}\left( \overset{\left( I-1\right) }{j}^{\mu }\right)
=0.  \label{int10}
\end{equation}%
We recall that the local cohomology $H\left( \delta |d\right) $ is
completely trivial in both strictly positive antighost \textit{and}
pure ghost numbers. Using the fact that the Cauchy order of the free
theory under study is equal to four, the general results from
Refs.~\cite{gen1} and \cite{gen2}, according to which the local
cohomology of the Koszul-Tate differential in pure ghost number zero
is trivial in antighost numbers strictly greater than its Cauchy
order, ensure that
\begin{equation}
H_{J}\left( \delta |d\right) =0,\qquad J>4,  \label{int11}
\end{equation}%
where $H_{J}\left( \delta |d\right) $ denotes the local cohomology of the
Koszul-Tate differential in antighost number $J$ and in pure ghost number
zero. It can be shown that any invariant polynomial that is trivial in $%
H_{J}\left( \delta |d\right) $ with $J\geq 4$ can be taken to be trivial
also in $H_{J}^{\mathrm{inv}}\left( \delta |d\right) $. ($H_{J}^{\mathrm{inv}%
}\left( \delta |d\right) $ denotes the invariant characteristic cohomology
in antighost number $J$ --- the local cohomology of the Koszul-Tate
differential in the space of invariant polynomials.) Thus:
\begin{equation}
\left( \alpha _{J}=\delta b_{J+1}+\partial _{\mu }\overset{(J)}{c}^{\mu },\
\mathrm{agh}\left( \alpha _{J}\right) =J\geq 4\right) \Rightarrow \alpha
_{J}=\delta \beta _{J+1}+\partial _{\mu }\overset{(J)}{\gamma }^{\mu },
\label{int12}
\end{equation}%
with both $\beta _{J+1}$ and $\overset{(J)}{\gamma }^{\mu }$ invariant
polynomials. Results (\ref{int11}) and (\ref{int12}) yield the conclusion
that
\begin{equation}
H_{J}^{\mathrm{inv}}\left( \delta |d\right) =0,\qquad J>4.  \label{int13}
\end{equation}%
Using the results from Refs.~\cite{pI}--\cite{pIII}, the spaces
$\left(
H_{J}\left( \delta |d\right) \right) _{J\geq 2}$ and $\left( H_{J}^{\mathrm{%
inv}}\left( \delta |d\right) \right) _{J\geq 2}$ are spanned by
\begin{eqnarray}
&&H_{4}\left( \delta |d\right) ,H_{4}^{\mathrm{inv}}\left( \delta |d\right)
:\quad \left( C^{\ast }\right) ,  \label{int14a} \\
&&H_{3}\left( \delta |d\right) ,H_{3}^{\mathrm{inv}}\left( \delta |d\right)
:\quad \left( C^{\ast \mu }\right) ,  \label{int14b} \\
&&H_{2}\left( \delta |d\right) ,H_{2}^{\mathrm{inv}}\left( \delta |d\right)
:\quad \left( C^{\ast \mu \nu },\eta ^{\ast \mu },\xi ^{\ast }\right) .
\label{int14c}
\end{eqnarray}%
In contrast to the groups $\left( H_{J}\left( \delta |d\right) \right)
_{J\geq 2}$ and $\left( H_{J}^{\mathrm{inv}}\left( \delta |d\right) \right)
_{J\geq 2}$, which are finite-dimensional, the cohomology $H_{1}\left(
\delta |d\right) $ in pure ghost number zero, known to be related to global
symmetries and ordinary conservation laws, is infinite-dimensional since the
theory is free. Fortunately, it will not be needed in the sequel.

The previous results on $H\left( \delta |d\right) $ and $H^{\mathrm{inv}%
}\left( \delta |d\right) $ in strictly positive antighost numbers are
important because they control the obstructions to removing the antifields
from the first-order deformation. Based on formulas (\ref{int11})--(\ref%
{int12}), one can successively eliminate all the pieces of antighost number
strictly greater than four from the nonintegrated density of the first-order
deformation by adding only trivial terms. Consequently, one can take
(without loss of nontrivial objects) $I\leq 4$ into the decomposition (\ref%
{int5}). In addition, the last representative reads as in (\ref{int9}),
where the invariant polynomial is necessarily a nontrivial object from $%
\left( H_{J}^{\mathrm{inv}}\left( \delta |d\right) \right) _{2\leq J\leq 4}$
or from $H_{1}\left( \delta |d\right) $ for $J=1$.

The previous discussion enforces that we can take $I=4$ in (\ref{int5}) and
work with
\begin{equation}
a^{\mathrm{int}}=a_{0}^{\mathrm{int}}+a_{1}^{\mathrm{int}}+a_{2}^{\mathrm{int%
}}+a_{3}^{\mathrm{int}}+a_{4}^{\mathrm{int}},  \label{int15}
\end{equation}%
where the components from the right-hand side of (\ref{int15}) satisfy
Eqs. (\ref{int7}) and (\ref{int6b})--(\ref{int6c}) for $I=4$. Due to (%
\ref{int9}) and (\ref{int14a}), we can write the nontrivial solution to (\ref%
{int7}) for $I=4$ in the form%
\begin{equation}
a_{4}^{\mathrm{int}}=C^{\ast }\omega ^{4}\left( C,\xi ,\eta _{\mu },\partial
_{\lbrack \mu }\eta _{\nu ]}\right) .  \label{int16}
\end{equation}%
Since $a_{4}^{\mathrm{int}}$ already depends on $C^{\ast }$, which is a BRST
generator from the $3$-form sector, we have to select from $\omega ^{4}$
only those elements of pure ghost number $4$ that depend simultaneously on
the ghosts from the Rarita-Schwinger and Pauli-Fierz sectors, namely involve
both $\xi $ and $\eta _{\mu }$ or $\partial _{\lbrack \mu }\eta _{\nu ]}$.
These are precisely%
\begin{eqnarray}
&&\left\{ \bar{\xi}\gamma _{\alpha }\xi \eta _{\nu }\eta _{\rho },\ \bar{\xi}%
\gamma _{\alpha \beta }\xi \eta _{\nu }\eta _{\rho },\ \bar{\xi}\gamma
_{\alpha \beta \gamma \delta \varepsilon }\xi \eta _{\nu }\eta _{\rho },\
\bar{\xi}\gamma _{\alpha }\xi \eta _{\mu }\partial _{\lbrack \nu }\eta
_{\rho ]},\right.  \notag \\
&&\bar{\xi}\gamma _{\alpha \beta }\xi \eta _{\mu }\partial _{\lbrack \nu
}\eta _{\rho ]},\ \bar{\xi}\gamma _{\alpha \beta \gamma \delta \varepsilon
}\xi \eta _{\mu }\partial _{\lbrack \nu }\eta _{\rho ]},\ \bar{\xi}\gamma
_{\alpha }\xi \left( \partial _{\lbrack \mu }\eta _{\nu ]}\right) \partial
_{\lbrack \rho }\eta _{\lambda ]},  \notag \\
&&\left. \bar{\xi}\gamma _{\alpha \beta }\xi \left( \partial _{\lbrack \mu
}\eta _{\nu ]}\right) \partial _{\lbrack \rho }\eta _{\lambda ]},\ \bar{\xi}%
\gamma _{\alpha \beta \gamma \delta \varepsilon }\xi \left( \partial
_{\lbrack \mu }\eta _{\nu ]}\right) \partial _{\lbrack \rho }\eta _{\lambda
]}\right\} ,  \label{int17}
\end{eqnarray}%
such that (\ref{int16}) becomes%
\begin{eqnarray}
a_{4}^{\mathrm{int}} &=&C^{\ast }\left[ \frac{q_{1}}{2}\bar{\xi}\gamma ^{\mu
\nu }\xi \eta _{\mu }\eta _{\nu }+q_{2}\sigma ^{\mu \rho }\bar{\xi}\gamma
^{\nu }\xi \eta _{\mu }\partial _{\lbrack \nu }\eta _{\rho ]}\right.  \notag
\\
&&\left. +\frac{q_{3}}{2}\sigma ^{\rho \lambda }\bar{\xi}\gamma ^{\mu \nu
}\xi \left( \partial _{\lbrack \mu }\eta _{\rho ]}\right) \partial _{\lbrack
\nu }\eta _{\lambda ]}\right] ,  \label{int18}
\end{eqnarray}%
with $q_{1}$, $q_{2}$, and $q_{3}$ some real, arbitrary constants. By
applying the operator $\delta $ on (\ref{int18}) and further using
definitions (\ref{int04a})--(\ref{int04e}), we obtain that%
\begin{eqnarray}
\delta a_{4}^{\mathrm{int}} &=&\partial _{\mu }\left\{ -C^{\ast \mu }\left[
\frac{q_{1}}{2}\bar{\xi}\gamma ^{\alpha \beta }\xi \eta _{\alpha }\eta
_{\beta }+q_{2}\sigma ^{\alpha \beta }\bar{\xi}\gamma ^{\nu }\xi \eta
_{\alpha }\partial _{\lbrack \nu }\eta _{\beta ]}\right. \right.  \notag \\
&&\left. \left. +\frac{q_{3}}{2}\sigma ^{\rho \lambda }\bar{\xi}\gamma
^{\alpha \beta }\xi \left( \partial _{\lbrack \alpha }\eta _{\rho ]}\right)
\partial _{\lbrack \beta }\eta _{\lambda ]}\right] \right\}  \notag \\
&&+\gamma \left\{ C^{\ast \mu }\left[ \frac{q_{1}}{2}\bar{\xi}\gamma
^{\alpha \beta }\xi \eta _{\alpha }h_{\mu \beta }-\frac{q_{2}}{2}\sigma
^{\alpha \beta }\bar{\xi}\gamma ^{\nu }\xi \left( h_{\mu \alpha }\partial
_{\lbrack \nu }\eta _{\beta ]}\right. \right. \right.  \notag \\
&&\left. -2\eta _{\alpha }\partial _{\lbrack \nu }h_{\beta ]\mu }\right)
+q_{3}\sigma ^{\rho \lambda }\bar{\xi}\gamma ^{\alpha \beta }\xi \left(
\partial _{\lbrack \alpha }\eta _{\rho ]}\right) \partial _{\lbrack \beta
}h_{\lambda ]\mu }  \notag \\
&&+q_{1}\bar{\xi}\gamma ^{\alpha \beta }\psi _{\mu }\eta _{\alpha }\eta
_{\beta }+2q_{2}\sigma ^{\alpha \beta }\bar{\xi}\gamma ^{\nu }\psi _{\mu
}\eta _{\alpha }\partial _{\lbrack \nu }\eta _{\beta ]}  \notag \\
&&\left. \left. +q_{3}\sigma ^{\rho \lambda }\bar{\xi}\gamma ^{\alpha \beta
}\psi _{\mu }\left( \partial _{\lbrack \alpha }\eta _{\rho ]}\right)
\partial _{\lbrack \beta }\eta _{\lambda ]}\right] \right\}  \notag \\
&&+\frac{1}{2}C^{\ast \mu }\left( q_{1}\bar{\xi}\gamma ^{\alpha \beta }\xi
\eta _{\alpha }-q_{2}\sigma ^{\alpha \beta }\bar{\xi}\gamma ^{\nu }\xi
\partial _{\lbrack \nu }\eta _{\alpha ]}\right) \partial _{\lbrack \mu }\eta
_{\beta ]}.  \label{int18a}
\end{eqnarray}%
We observe that (\ref{int18a}) cannot agree with (\ref{int6b}) for $I=4$
unless we set%
\begin{equation}
q_{1}=q_{2}=0,  \label{int18b}
\end{equation}%
which then replaced in (\ref{int18}) and (\ref{int18a}) produces%
\begin{equation}
a_{4}^{\mathrm{int}}=\frac{q_{3}}{2}\sigma ^{\rho \lambda }C^{\ast }\bar{\xi}%
\gamma ^{\alpha \beta }\xi \left( \partial _{\lbrack \alpha }\eta _{\rho
]}\right) \partial _{\lbrack \beta }\eta _{\lambda ]},  \label{int19}
\end{equation}%
\begin{equation}
a_{3}^{\mathrm{int}}=-q_{3}\sigma ^{\rho \lambda }C^{\ast \mu }\left[ \bar{%
\xi}\gamma ^{\alpha \beta }\xi \left( \partial _{\lbrack \alpha }\eta _{\rho
]}\right) \partial _{\lbrack \beta }h_{\lambda ]\mu }+\bar{\xi}\gamma
^{\alpha \beta }\psi _{\mu }\left( \partial _{\lbrack \alpha }\eta _{\rho
]}\right) \partial _{\lbrack \beta }\eta _{\lambda ]}\right] .
\label{int19a}
\end{equation}%
Acting now with $\delta $ on (\ref{int19a}) and recalling definitions (\ref%
{int04a})--(\ref{int04e}), we have that%
\begin{eqnarray}
\delta a_{3}^{\mathrm{int}} &=&\partial _{\nu }\left\{ -2q_{3}\sigma ^{\rho
\lambda }C^{\ast \mu \nu }\bar{\xi}\gamma ^{\alpha \beta }\left[ \xi \left(
\partial _{\lbrack \alpha }\eta _{\rho ]}\right) \partial _{\lbrack \beta
}h_{\lambda ]\mu }+\psi _{\mu }\left( \partial _{\lbrack \alpha }\eta _{\rho
]}\right) \partial _{\lbrack \beta }\eta _{\lambda ]}\right] \right\}  \notag
\\
&&+\gamma \left\{ q_{3}\sigma ^{\rho \lambda }C^{\ast \mu \nu }\bar{\xi}%
\gamma ^{\alpha \beta }\left[ \xi \left( \partial _{\lbrack \alpha }h_{\rho
]\nu }\right) \partial _{\lbrack \beta }h_{\lambda ]\mu }+2\psi _{\mu
}\left( \partial _{\lbrack \alpha }\eta _{\rho ]}\right) \partial _{\lbrack
\beta }h_{\lambda ]\nu }\right] \right.  \notag \\
&&\left. -q_{3}\sigma ^{\rho \lambda }C^{\ast \mu \nu }\bar{\psi}_{\mu
}\gamma ^{\alpha \beta }\psi _{\nu }\left( \partial _{\lbrack \alpha }\eta
_{\rho ]}\right) \partial _{\lbrack \beta }\eta _{\lambda ]}\right\}  \notag
\\
&&+q_{3}\sigma ^{\rho \lambda }C^{\ast \mu \nu }\bar{\xi}\gamma ^{\alpha
\beta }\left[ 2\xi \left( \partial _{\lbrack \alpha }\eta _{\rho ]}\right)
K_{\beta \lambda \mu \nu }-\left( \partial _{\lbrack \mu }\psi _{\nu
]}\right) \left( \partial _{\lbrack \alpha }\eta _{\rho ]}\right) \partial
_{\lbrack \beta }\eta _{\lambda ]}\right] ,  \label{int20}
\end{eqnarray}%
such that (\ref{int20}) is compatible with (\ref{int6b}) for $I=3$ if%
\begin{equation}
q_{3}=0.  \label{int21}
\end{equation}%
If we insert (\ref{int21}) into (\ref{int19}), then we conclude that $a_{4}^{%
\mathrm{int}}=0$, so we can take $I=3$ in (\ref{int5}).

Consequently, decomposition (\ref{int5}) reduces to%
\begin{equation}
a^{\mathrm{int}}=a_{0}^{\mathrm{int}}+a_{1}^{\mathrm{int}}+a_{2}^{\mathrm{int%
}}+a_{3}^{\mathrm{int}},  \label{int22}
\end{equation}%
where the components from the right-hand side of (\ref{int22})
satisfy Eqs. (\ref{int7}) and (\ref{int6b})--(\ref{int6c}) for
$I=3$. Taking
into account formula (\ref{int9}) and relation (\ref{int14b}), we find that%
\begin{equation}
a_{3}^{\mathrm{int}}=C^{\ast \mu }\omega _{\mu }^{3}\left( C,\xi ,\eta _{\mu
},\partial _{\lbrack \mu }\eta _{\nu ]}\right) ,  \label{int22a}
\end{equation}%
where we have to elect again among the elements $\omega ^{3}$ only those
involving simultaneously $\xi $ and $\eta _{\mu }$ or $\partial _{\lbrack
\mu }\eta _{\nu ]}$, namely%
\begin{eqnarray}
&&\left\{ \bar{\xi}\gamma _{\alpha }\xi \eta _{\mu },\ \bar{\xi}\gamma
_{\alpha \beta }\xi \eta _{\mu },\ \bar{\xi}\gamma _{\alpha \beta \gamma
\delta \varepsilon }\xi \eta _{\mu },\ \bar{\xi}\gamma _{\alpha }\xi
\partial _{\lbrack \mu }\eta _{\nu ]},\right.  \notag \\
&&\left. \bar{\xi}\gamma _{\alpha \beta }\xi \partial _{\lbrack \mu }\eta
_{\nu ]},\ \bar{\xi}\gamma _{\alpha \beta \gamma \delta \varepsilon }\xi
\partial _{\lbrack \mu }\eta _{\nu ]}\right\} .  \label{int22b}
\end{eqnarray}%
Only the second and fourth elements from (\ref{int22b}) allow the formation
of $11$-dimensional vector-like combinations, so the general form of (\ref%
{int22a}) reads as%
\begin{equation}
a_{3}^{\mathrm{int}}=\frac{1}{2}\sigma ^{\alpha \beta }C^{\ast \mu }\left(
q_{4}\bar{\xi}\gamma _{\mu \alpha }\xi \eta _{\beta }+q_{5}\bar{\xi}\gamma
_{\alpha }\xi \partial _{\lbrack \mu }\eta _{\beta ]}\right) ,  \label{int23}
\end{equation}%
with $q_{4}$ and $q_{5}$ real numbers. Next, we act like in the case $I=4$,
namely apply $\delta $ on (\ref{int23}) and then manipulate the resulting
expression with the help of definitions (\ref{int04a})--(\ref{int04e}),
which further yields%
\begin{eqnarray}
\delta a_{3}^{\mathrm{int}} &=&\partial _{\nu }\left[ \sigma ^{\alpha \beta
}C^{\ast \mu \nu }\left( q_{4}\bar{\xi}\gamma _{\mu \alpha }\xi \eta _{\beta
}+q_{5}\bar{\xi}\gamma _{\alpha }\xi \partial _{\lbrack \mu }\eta _{\beta
]}\right) \right]  \notag \\
&&+\gamma \left\{ \sigma ^{\alpha \beta }C^{\ast \mu \nu }\left[ q_{4}\bar{%
\xi}\gamma _{\mu \alpha }\left( 2\psi _{\nu }\eta _{\beta }-\xi h_{\beta \nu
}\right) \right. \right.  \notag \\
&&\left. \left. +q_{5}\bar{\xi}\gamma _{\alpha }\left( 2\psi _{\nu }\partial
_{\lbrack \mu }\eta _{\beta ]}-\xi \partial _{\lbrack \mu }h_{\beta ]\nu
}\right) \right] \right\}  \notag \\
&&-\frac{q_{4}}{2}\sigma ^{\alpha \beta }C^{\ast \mu \nu }\bar{\xi}\gamma
_{\mu \alpha }\xi \partial _{\lbrack \nu }\eta _{\beta ]}.  \label{int24}
\end{eqnarray}%
Formula (\ref{int24}) does not concur with (\ref{int6b}) for $I=3$ unless
\begin{equation}
q_{4}=0,  \label{int25}
\end{equation}%
which substituted in (\ref{int23}) and (\ref{int24}) provides%
\begin{equation}
a_{3}^{\mathrm{int}}=\frac{q_{5}}{2}C^{\ast \mu }\bar{\xi}\gamma ^{\alpha
}\xi \partial _{\lbrack \mu }\eta _{\alpha ]},  \label{int26}
\end{equation}%
\begin{equation}
a_{2}^{\mathrm{int}}=-q_{5}C^{\ast \mu \nu }\bar{\xi}\gamma ^{\alpha }\left(
2\psi _{\nu }\partial _{\lbrack \mu }\eta _{\alpha ]}-\xi \partial _{\lbrack
\mu }h_{\alpha ]\nu }\right) .  \label{int27}
\end{equation}%
Acting with $\delta $ on (\ref{int27}), we can write that%
\begin{eqnarray}
\delta a_{2}^{\mathrm{int}} &=&\partial _{\rho }\left[ 3q_{5}A^{\ast \mu \nu
\rho }\bar{\xi}\gamma ^{\alpha }\left( 2\psi _{\nu }\partial _{\lbrack \mu
}\eta _{\alpha ]}-\xi \partial _{\lbrack \mu }h_{\alpha ]\nu }\right) \right]
\notag \\
&&+\gamma \left[ -3q_{5}A^{\ast \mu \nu \rho }\left( \bar{\psi}_{\nu }\gamma
^{\alpha }\psi _{\rho }\partial _{\lbrack \mu }\eta _{\alpha ]}+\bar{\xi}%
\gamma ^{\alpha }\psi _{\nu }\partial _{\lbrack \mu }h_{\rho ]\alpha
}\right) \right]  \notag \\
&&+3q_{5}A^{\ast \mu \nu \rho }\bar{\xi}\gamma ^{\alpha }\left( \partial
_{\lbrack \nu }\psi _{\rho ]}\right) \partial _{\lbrack \mu }\eta _{\alpha
]},  \label{int28}
\end{eqnarray}%
so Eq. (\ref{int6c}) for $I=2$ cannot hold except for the case where
\begin{equation}
q_{5}=0.  \label{int29}
\end{equation}%
Introducing (\ref{int25}) and (\ref{int29}) in (\ref{int23}), we deduce that
we can take $a_{3}^{\mathrm{int}}=0$.

The following possibility is to stop at antighost number $2$, in which
situation
\begin{equation}
a^{\mathrm{int}}=a_{0}^{\mathrm{int}}+a_{1}^{\mathrm{int}}+a_{2}^{\mathrm{int%
}},  \label{int30}
\end{equation}%
where the components of $a^{\mathrm{int}}$ are subject to Eqs. (\ref%
{int7}) and (\ref{int6b})--(\ref{int6c}) for $I=2$. Due to result (\ref%
{int14c}) and formula (\ref{int9}), the nontrivial solution to (\ref{int7})
for $I=2$ takes the form
\begin{eqnarray}
a_{2}^{\mathrm{int}} &=&\xi ^{\ast }\hat{\omega}^{2}\left( C,\xi ,\eta _{\mu
},\partial _{\lbrack \mu }\eta _{\nu ]}\right) +\eta ^{\ast \mu }\omega
_{\mu }^{2}\left( C,\xi ,\eta _{\mu },\partial _{\lbrack \mu }\eta _{\nu
]}\right)  \notag \\
&&+C^{\ast \mu \nu }\omega _{\mu \nu }^{2}\left( C,\xi ,\eta _{\mu
},\partial _{\lbrack \mu }\eta _{\nu ]}\right) .  \label{int31}
\end{eqnarray}%
The elements $\hat{\omega}^{2}$, $\omega _{\mu }^{2}$, and $\omega _{\mu \nu
}^{2}$ have the pure ghost number equal to $2$ and must at least contain
ghosts belonging to the sectors respectively complementary to that including
the antifield coupled to them. In other words, $\hat{\omega}^{2}$ compulsory
contains both $C$ and $\eta _{\mu }$ or $\partial _{\lbrack \mu }\eta _{\nu
]}$, $\omega _{\mu }^{2}$ must depend on both $C$ and $\xi $, and $\omega
_{\mu \nu }^{2}$ are restricted to involve both $\xi $ and $\eta _{\mu }$ or
$\partial _{\lbrack \mu }\eta _{\nu ]}$. Since the pure ghost number of $C$
is already $3$, it follows that $\hat{\omega}^{2}$ and $\omega _{\mu }^{2}$
must be discarded by putting%
\begin{equation}
\hat{\omega}^{2}\left( C,\xi ,\eta _{\mu },\partial _{\lbrack \mu }\eta
_{\nu ]}\right) =0,  \label{int32a}
\end{equation}%
\begin{equation}
\omega _{\mu }^{2}\left( C,\xi ,\eta _{\mu },\partial _{\lbrack \mu }\eta
_{\nu ]}\right) =0.  \label{int32b}
\end{equation}%
Regarding $\omega _{\mu \nu }^{2}$, we observe that the ghost $\xi $ is a
spinor, so it can be mixed with $\eta _{\mu }$ or $\partial _{\lbrack \mu
}\eta _{\nu ]}$ into an antisymmetric tensor through an at least cubic
combination, simultaneously involving $\bar{\xi}$, $\xi $, and $\eta _{\mu }$
or $\partial _{\lbrack \mu }\eta _{\nu ]}$, which therefore displays a pure
ghost number greater or equal to $3$. For this reason we must also give up $%
\omega _{\mu \nu }^{2}$%
\begin{equation}
\omega _{\mu \nu }^{2}\left( C,\xi ,\eta _{\mu },\partial _{\lbrack \mu
}\eta _{\nu ]}\right) =0.  \label{int32c}
\end{equation}%
The results expressed by (\ref{int32a})--(\ref{int32c}) ensure, via (\ref%
{int31}), that $a_{2}^{\mathrm{int}}=0$, so we have to consider the case $%
I=1 $ in (\ref{int5}).

Consequently, we have that%
\begin{equation}
a^{\mathrm{int}}=a_{0}^{\mathrm{int}}+a_{1}^{\mathrm{int}},  \label{int33}
\end{equation}%
where $a_{0}^{\mathrm{int}}$ and $a_{1}^{\mathrm{int}}$ fulfill the equations%
\begin{eqnarray}
\gamma a_{1}^{\mathrm{int}} &=&0,  \label{int34a} \\
\delta a_{1}^{\mathrm{int}}+\gamma a_{0}^{\mathrm{int}} &=&\partial ^{\mu }%
\overset{(0)}{m}_{\mu }^{\mathrm{int}}.  \label{int34b}
\end{eqnarray}%
According to (\ref{int9}), the general, nontrivial solution to (\ref{int34a}%
) is given by%
\begin{eqnarray}
a_{1}^{\mathrm{int}} &=&\psi ^{\ast \mu }\left( M_{\mu }\xi +M_{\mu
}^{\alpha }\eta _{\alpha }+M_{\mu }^{\alpha \beta }\partial _{\lbrack \alpha
}\eta _{\beta ]}\right)   \notag \\
&&+h^{\ast \mu \nu }\left( N_{\mu \nu }\xi +N_{\mu \nu }^{\alpha }\eta
_{\alpha }+N_{\mu \nu }^{\alpha \beta }\partial _{\lbrack \alpha }\eta
_{\beta ]}\right)   \notag \\
&&+A^{\ast \mu \nu \rho }\left( P_{\mu \nu \rho }\xi +P_{\mu \nu \rho
}^{\alpha }\eta _{\alpha }+P_{\mu \nu \rho }^{\alpha \beta }\partial
_{\lbrack \alpha }\eta _{\beta ]}\right) ,  \label{int35}
\end{eqnarray}%
where the objects generically denoted by $M$, $N$ or $P$ are gauge invariant
quantities. In order to provide interactions among all the three kinds of
fields, each of them is required to depend at least on those gauge invariant
combinations constructed out of the fields from the sector(s) that are
complementary to the respectively coupled antifields/ghosts. Regarding their
tensorial properties, the elements $M_{\mu }^{\alpha \beta }$, $N_{\mu \nu
}^{\alpha \beta }$, and $P_{\mu \nu \rho }^{\alpha \beta }$ are
antisymmetric in their upper indices $\alpha $ and $\beta $, all the
quantities of the type $N$ are symmetric in their lower indices $\mu $ and $%
\nu $, and all the quantities denoted by $P$ are completely antisymmetric in
their lower indices $\mu $, $\nu $, and $\rho $. Moreover, each $M_{\mu }$
is a $2^{5}\times 2^{5}$ matrix with bosonic, gauge invariant functions as
elements. Furthermore, each of $M_{\mu }^{\alpha }$, $M_{\mu }^{\alpha \beta
}$, $N_{\mu \nu }$, or $P_{\mu \nu \rho }$ is a fermionic, gauge invariant
spinor tensor. Since the only fermionic, spinor fields are the gravitini and
the gauge invariant quantities built out of them are their antisymmetric
first-order derivatives, $\partial _{\lbrack \gamma }\psi _{\delta ]}$, it
follows that $M_{\mu }^{\alpha }$, $M_{\mu }^{\alpha \beta }$, $N_{\mu \nu }$%
, and $P_{\mu \nu \rho }$ can be further represented in terms of some $%
2^{5}\times 2^{5}$ matrices with bosonic, gauge invariant elements, of the
type:%
\begin{eqnarray}
M_{\mu }^{\alpha } &=&\bar{M}_{\mu }^{\alpha \mid \gamma \delta }\partial
_{\lbrack \gamma }\psi _{\delta ]},\qquad M_{\mu }^{\alpha \beta }=\bar{M}%
_{\mu }^{\alpha \beta \mid \gamma \delta }\partial _{\lbrack \gamma }\psi
_{\delta ]},  \label{int36a} \\
N_{\mu \nu } &=&\partial _{\lbrack \gamma }\bar{\psi}_{\delta ]}\bar{N}_{\mu
\nu }^{\gamma \delta },\qquad P_{\mu \nu \rho }=\partial _{\lbrack \gamma }%
\bar{\psi}_{\delta ]}\bar{P}_{\mu \nu \rho }^{\gamma \delta },
\label{int36b}
\end{eqnarray}%
where $\bar{M}_{\mu }^{\alpha \mid \gamma \delta }$, $\bar{M}_{\mu }^{\alpha
\beta \mid \gamma \delta }$, $\bar{N}_{\mu \nu }^{\gamma \delta }$, or $\bar{%
P}_{\mu \nu \rho }^{\gamma \delta }$ may contain in principle additional
spacetime derivatives. At this point we ask that the corresponding $a_{0}^{%
\mathrm{int}}$ (as solution to (\ref{int34b})) leads to interacting field
equations preserving the derivative order of the free ones (derivative order
assumption). This further requires that the maximum derivative order of $%
a_{0}^{\mathrm{int}}$ is equal to two, with the precaution that each
interacting field equation contains at most one spacetime derivative acting
on the gravitini. In the sequel we will argue that each of the terms from
the right-hand side of (\ref{int35}), if consistent, would produce in the
interacting Lagrangian terms forbidden by the derivative order assumption.

Related to the first term, $\psi ^{\ast \mu }M_{\mu }\xi $, since both $\psi
^{\ast \mu }$ and $\xi $ belong to the Rarita-Schwinger sector, it follows
that each element of the matrices $M_{\mu }$ is constrained to be at least
linear in both the linearized Riemann tensor $K_{\mu \nu \rho \lambda }$ and
the field strength $F_{\mu \nu \rho \lambda }$ of the $3$-form, so it
contains at least three spacetime derivatives. If consistent, each of these
terms would lead to an interacting Lagrangian density with minimum three
derivatives, which is unacceptable, so we must set%
\begin{equation}
M_{\mu }=0.  \label{int37a}
\end{equation}%
The second element, $\psi ^{\ast \mu }M_{\mu }^{\alpha }\eta _{\alpha }$,
already contain generators from the Rarita-Schwinger and Pauli-Fierz
sectors, so $M_{\mu }^{\alpha }$ are bound to be at least linear in $F_{\mu
\nu \rho \lambda }$. Due to the former relation from (\ref{int36a}), we
conclude that $\psi ^{\ast \mu }M_{\mu }^{\alpha }\eta _{\alpha }$ has at
least two spacetime derivatives, among which one already acts on the
gravitini, so it would provide field equations with at least two derivatives
acting on the gravitini. We have to forbid this by setting%
\begin{equation}
M_{\mu }^{\alpha }=0.  \label{int37b}
\end{equation}%
Using exactly the same arguments we eliminate the third piece, $\psi ^{\ast
\mu }M_{\mu }^{\alpha \beta }\partial _{\lbrack \alpha }\eta _{\beta ]}$, by
putting%
\begin{equation}
M_{\mu }^{\alpha \beta }=0.  \label{int37c}
\end{equation}%
A simple analysis of the fourth component, $h^{\ast \mu \nu }N_{\mu \nu }\xi
$, shows that $N_{\mu \nu }$ is compelled to be at least linear in $F_{\mu
\nu \rho \lambda }$, and so, according to the first formula in (\ref{int36b}%
), it contains at least two derivatives. Since $\delta h^{\ast \mu \nu }$
also contains two derivatives, this component would generate an interacting
Lagrangian density with at least three derivatives. The derivative order
assumption is again broken, so we must take%
\begin{equation}
N_{\mu \nu }=0.  \label{int37d}
\end{equation}%
Looking at the fifth constituent, $h^{\ast \mu \nu }N_{\mu \nu }^{\alpha
}\eta _{\alpha }$, since both $h^{\ast \mu \nu }$ and $\eta _{\alpha }$
pertain to the Pauli-Fierz sector, it results that the bosonic, $\gamma $%
-invariant tensor $N_{\mu \nu }^{\alpha }$ is simultaneously at least
quadratic in the antisymmetric first-order derivatives of the
Rarita-Schwinger spinors and linear in $F_{\mu \nu \rho \lambda }$, which
amounts to at least three derivatives. Thus, if consistent, this term would
give rise to a Lagrangian density with at least four derivatives. The same
reason can be used to eliminate $h^{\ast \mu \nu }N_{\mu \nu }^{\alpha \beta
}\partial _{\lbrack \alpha }\eta _{\beta ]}$ from $a_{1}^{\mathrm{int}}$,
and hence we can write
\begin{equation}
N_{\mu \nu }^{\alpha }=0,\qquad N_{\mu \nu }^{\alpha \beta }=0.
\label{int37e}
\end{equation}%
The seventh term, $A^{\ast \mu \nu \rho }P_{\mu \nu \rho }\xi $, contains
the fermionic, gauge invariant spinor tensor $P_{\mu \nu \rho }$, which is
required to involve the Pauli-Fierz field, so it effectively depends on $%
K_{\mu \nu \rho \lambda }$. Joining this observation to the second relation
from (\ref{int36b}), we get that $P_{\mu \nu \rho }$ includes at least three
derivatives, and thus the corresponding Lagrangian density (if any) would
furnish interaction vertices with at least four derivatives. This is again
in contradiction with the derivative order assumption, so we must discard
this term by choosing
\begin{equation}
P_{\mu \nu \rho }=0.  \label{int37f}
\end{equation}%
Finally, the last two pieces from the right-hand side of (\ref{int35}), $%
A^{\ast \mu \nu \rho }P_{\mu \nu \rho }^{\alpha }\eta _{\alpha }$ and $%
A^{\ast \mu \nu \rho }P_{\mu \nu \rho }^{\alpha \beta }\partial _{\lbrack
\alpha }\eta _{\beta ]}$, involve the bosonic, gauge invariant tensors $%
P_{\mu \nu \rho }^{\alpha }$ and $P_{\mu \nu \rho }^{\alpha \beta }$, which
are required to depend on the Rarita-Schwinger spinors, and therefore they
are at least quadratic in the antisymmetric first-order derivatives of
gravitini. If consistent, these objects would imply interaction vertices
with at least three and respectively four derivatives, and therefore must be
canceled through%
\begin{equation}
P_{\mu \nu \rho }^{\alpha }=0,\qquad P_{\mu \nu \rho }^{\alpha \beta }=0.
\label{int37g}
\end{equation}%
Inserting the previous results, (\ref{int37a})--(\ref{int37g}), into (\ref%
{int35}), we obtain $a_{1}^{\mathrm{int}}=0$, such that the first-order
deformation of the solution to the master equation can only reduce to its
antighost number zero component (we can only have $I=0$ in (\ref{int5})).

This final possibility is described by%
\begin{equation}
a^{\mathrm{int}}=a_{0}^{\mathrm{int}},  \label{int38}
\end{equation}%
where $a_{0}^{\mathrm{int}}$ is subject to the equation\footnote{%
One can no longer replace Eq. (\ref{int39}) with the homogeneous
one, like in the previous cases, since now we reached the bottom
value of the
antighost number, namely zero.}%
\begin{equation}
\gamma a_{0}^{\mathrm{int}}=\partial ^{\mu }\overset{(0)}{m}_{\mu }^{\mathrm{%
int}}.  \label{int39}
\end{equation}%
In order to analyze properly the solution to (\ref{int39}), we split its
solution as%
\begin{equation}
a_{0}^{\mathrm{int}}=\bar{a}_{0}^{\mathrm{int}}+\tilde{a}_{0}^{\mathrm{int}},
\label{int40}
\end{equation}%
where
\begin{eqnarray}
\gamma \bar{a}_{0}^{\mathrm{int}} &=&0,  \label{int41a} \\
\gamma \tilde{a}_{0}^{\mathrm{int}} &=&\partial ^{\mu }\overset{(0)}{m}_{\mu
}^{\mathrm{int}},  \label{int41b}
\end{eqnarray}%
with $\overset{(0)}{m}_{\mu }^{\mathrm{int}}\neq 0$.

Due to (\ref{int41a}), $\bar{a}_{0}^{\mathrm{int}}$ is a bosonic, gauge
invariant object, which is required to depend on all three kinds of fields.
Consequently, $\bar{a}_{0}^{\mathrm{int}}$ is at least quadratic in the
antisymmetric first-order derivatives of the spinors, $\partial _{\lbrack
\mu }\psi _{\nu ]}$, and at least linear in both $K_{\mu \nu \rho \lambda }$
and $F_{\mu \nu \rho \lambda }$, so it contains at least five spacetime
derivatives, which disagrees with the derivative order assumption. In
conclusion, we eliminate it from the interacting Lagrangian density by
putting%
\begin{equation}
\bar{a}_{0}^{\mathrm{int}}=0.  \label{int42}
\end{equation}

Now, we approach Eq. (\ref{int41b}) in a standard manner. Namely, we
decompose $\tilde{a}_{0}^{\mathrm{int}}$ with respect to the total
number of
derivatives into%
\begin{equation}
\tilde{a}_{0}^{\mathrm{int}}=\omega _{0}+\omega _{1}+\omega _{2},
\label{int43}
\end{equation}%
where $\left( \omega _{k}\right) _{k=\overline{0,2}}$ comprises $k$
derivatives. By projecting (\ref{int41b}) on the different possible values
of the number of derivatives, we find that it becomes equivalent to three
equations, one for each component%
\begin{equation}
\gamma \omega _{k}=\partial _{\mu }l_{k}^{\mu },\qquad k=\overline{0,2},
\label{int44}
\end{equation}%
where%
\begin{equation}
\overset{(0)}{m}^{\mathrm{int}\mu }=l_{0}^{\mu }+l_{1}^{\mu }+l_{2}^{\mu }.
\label{int45}
\end{equation}%
In the sequel we solve (\ref{int44}) for each value of $k$.

We start with (\ref{int44}) for $k=0$ and recall definitions (\ref{int04d}),
which produce%
\begin{eqnarray}
\gamma \omega _{0} &=&\frac{\partial ^{\mathrm{R}}\omega _{0}}{\partial \psi
_{\mu }}\partial _{\mu }\xi +\frac{\partial \omega _{0}}{\partial h_{\mu \nu
}}\partial _{(\mu }\eta _{\nu )}+\frac{\partial \omega _{0}}{\partial A_{\mu
\nu \rho }}\partial _{\lbrack \mu }C_{\nu \rho ]}  \notag \\
&=&\partial _{\mu }\left( \frac{\partial ^{\mathrm{R}}\omega _{0}}{\partial
\psi _{\mu }}\xi +2\frac{\partial \omega _{0}}{\partial h_{\mu \nu }}\eta
_{\nu }+3\frac{\partial \omega _{0}}{\partial A_{\mu \nu \rho }}C_{\nu \rho
}\right)  \notag \\
&&-\left( \partial _{\mu }\frac{\partial ^{\mathrm{R}}\omega _{0}}{\partial
\psi _{\mu }}\right) \xi -2\left( \partial _{\mu }\frac{\partial \omega _{0}%
}{\partial h_{\mu \nu }}\right) \eta _{\nu }-3\left( \partial _{\mu }\frac{%
\partial \omega _{0}}{\partial A_{\mu \nu \rho }}\right) C_{\nu \rho }.
\label{int46}
\end{eqnarray}%
From (\ref{int46}) we observe that (\ref{int44}) for $k=0$ cannot hold unless%
\begin{equation}
\partial _{\mu }\frac{\partial ^{\mathrm{R}}\omega _{0}}{\partial \psi _{\mu
}}=0,\qquad \partial _{\mu }\frac{\partial \omega _{0}}{\partial h_{\mu \nu }%
}=0,\qquad \partial _{\mu }\frac{\partial \omega _{0}}{\partial A_{\mu \nu
\rho }}=0.  \label{int47}
\end{equation}%
But $\omega _{0}$ has no derivatives acting on the fields, such that the
only solution to (\ref{int47}) is purely trivial%
\begin{equation}
\omega _{0}=0.  \label{int48}
\end{equation}

For $k=1$ we have that%
\begin{eqnarray}
\gamma \omega _{1} &=&\frac{\partial ^{\mathrm{R}}\omega _{1}}{\partial \psi
_{\mu }}\partial _{\mu }\xi +\frac{\partial \omega _{1}}{\partial h_{\mu \nu
}}\partial _{(\mu }\eta _{\nu )}+\frac{\partial \omega _{1}}{\partial A_{\mu
\nu \rho }}\partial _{\lbrack \mu }C_{\nu \rho ]}  \notag \\
&&+\frac{\partial ^{\mathrm{R}}\omega _{1}}{\partial \left( \partial
_{\alpha }\psi _{\mu }\right) }\partial _{\alpha }\partial _{\mu }\xi +\frac{%
\partial \omega _{1}}{\partial \left( \partial _{\alpha }h_{\mu \nu }\right)
}\partial _{\alpha }\partial _{(\mu }\eta _{\nu )}+\frac{\partial \omega _{1}%
}{\partial \left( \partial _{\alpha }A_{\mu \nu \rho }\right) }\partial
_{\alpha }\partial _{\lbrack \mu }C_{\nu \rho ]}  \notag \\
&=&\partial _{\mu }\left[ \frac{\partial ^{\mathrm{R}}\omega _{1}}{\partial
\psi _{\mu }}\xi +2\frac{\partial \omega _{1}}{\partial h_{\mu \nu }}\eta
_{\nu }+3\frac{\partial \omega _{1}}{\partial A_{\mu \nu \rho }}C_{\nu \rho
}+\frac{\partial ^{\mathrm{R}}\omega _{1}}{\partial \left( \partial _{\mu
}\psi _{\alpha }\right) }\partial _{\alpha }\xi \right.  \notag \\
&&+\frac{\partial \omega _{1}}{\partial \left( \partial _{\mu }h_{\alpha
\beta }\right) }\partial _{(\alpha }\eta _{\beta )}+\frac{\partial \omega
_{1}}{\partial \left( \partial _{\mu }A_{\alpha \beta \gamma }\right) }%
\partial _{\lbrack \alpha }C_{\beta \gamma ]}-\left( \partial _{\alpha }%
\frac{\partial ^{\mathrm{R}}\omega _{1}}{\partial \left( \partial _{\alpha
}\psi _{\mu }\right) }\right) \xi  \notag \\
&&\left. -2\left( \partial _{\alpha }\frac{\partial \omega _{1}}{\partial
\left( \partial _{\alpha }h_{\mu \nu }\right) }\right) \eta _{\nu }-3\left(
\partial _{\alpha }\frac{\partial \omega _{1}}{\partial \left( \partial
_{\alpha }A_{\mu \nu \rho }\right) }\right) C_{\nu \rho }\right]  \notag \\
&&-\left( \partial _{\mu }\frac{\delta ^{\mathrm{R}}\omega _{1}}{\delta \psi
_{\mu }}\right) \xi -2\left( \partial _{\mu }\frac{\delta \omega _{1}}{%
\delta h_{\mu \nu }}\right) \eta _{\nu }-3\left( \partial _{\mu }\frac{%
\delta \omega _{1}}{\delta A_{\mu \nu \rho }}\right) C_{\nu \rho },
\label{int49}
\end{eqnarray}%
so (\ref{int49}) complies with (\ref{int44}) for $k=1$ if%
\begin{equation}
\partial _{\mu }\frac{\delta ^{\mathrm{R}}\omega _{1}}{\delta \psi _{\mu }}%
=0,\qquad \partial _{\mu }\frac{\delta \omega _{1}}{\delta h_{\mu \nu }}%
=0,\qquad \partial _{\mu }\frac{\delta \omega _{1}}{\delta A_{\mu \nu \rho }}%
=0.  \label{int50}
\end{equation}%
The general solutions to (\ref{int50}) are expressed by%
\begin{equation}
\frac{\delta ^{\mathrm{R}}\omega _{1}}{\delta \psi _{\mu }}=\partial _{\nu
}L^{\mu \nu },\qquad \frac{\delta \omega _{1}}{\delta h_{\mu \nu }}=\partial
_{\rho }L^{\rho \mu \nu },\qquad \frac{\delta \omega _{1}}{\delta A_{\mu \nu
\rho }}=\partial _{\lambda }L^{\mu \nu \rho \lambda },  \label{int60}
\end{equation}%
where all the quantities generically denoted by $L$ depend only on the
undifferentiated fields (have no spacetime derivatives). In addition, $%
L^{\mu \nu }$ is a fermionic, spinor tensor, antisymmetric in its Lorentz
indices, $L^{\mu \nu \rho \lambda }$ is a bosonic, completely antisymmetric
tensor, and $L^{\rho \mu \nu }$ is also a bosonic tensor, antisymmetric in
its first two indices%
\begin{equation}
L^{\rho \mu \nu }=-L^{\mu \rho \nu }.  \label{int61a}
\end{equation}%
As $\delta \omega _{1}/\delta h_{\mu \nu }$ is symmetric and $L^{\rho \mu
\nu }$ is derivative-free, it follows that this tensor must be symmetric in
its last two indices%
\begin{equation}
L^{\rho \mu \nu }=L^{\rho \nu \mu }.  \label{int61b}
\end{equation}%
Using repeatedly properties (\ref{int61a}) and (\ref{int61b}), it is easy to
obtain $L^{\rho \mu \nu }=0$, and hence%
\begin{equation}
\frac{\delta \omega _{1}}{\delta h_{\mu \nu }}=0.  \label{int61d}
\end{equation}%
This means that $\omega _{1}$ may depend on the Pauli-Fierz field only
through trivial combinations (full divergences), which bring no
contributions to the interacting Lagrangian density, and therefore $\omega
_{1}$ cannot assemble all the three sectors in a nontrivial way and can be
taken to vanish%
\begin{equation}
\omega _{1}=0.  \label{int62}
\end{equation}

Finally, we solve Eq. (\ref{int44}) for $k=2$. If we make the notations%
\begin{equation}
\digamma ^{\mu }=\frac{\delta ^{\mathrm{R}}\omega _{2}}{\delta \psi _{\mu }}%
,\qquad D^{\mu \nu }=\frac{\delta \omega _{2}}{\delta h_{\mu \nu }},\qquad
D^{\mu \nu \rho }=\frac{\delta \omega _{2}}{\delta A_{\mu \nu \rho }},
\label{int63}
\end{equation}%
then we can write
\begin{equation}
\gamma \omega _{2}=-\left( \partial _{\mu }\digamma ^{\mu }\right) \xi
-2\left( \partial _{\mu }D^{\mu \nu }\right) \eta _{\nu }-3\left( \partial
_{\mu }D^{\mu \nu \rho }\right) C_{\nu \rho }+\partial _{\mu }u^{\mu },
\label{int64}
\end{equation}%
with $u^{\mu }$ a local current. From (\ref{int64}) we infer that $\omega
_{2}$ cannot be solution to (\ref{int44}) unless%
\begin{equation}
\partial _{\mu }\digamma ^{\mu }=0,\qquad \partial _{\mu }D^{\mu \nu
}=0,\qquad \partial _{\mu }D^{\mu \nu \rho }=0.  \label{int65}
\end{equation}%
The solutions to the last equations are known and take the general form%
\begin{equation}
\digamma ^{\mu }=\partial _{\nu }\digamma ^{\mu \nu },\qquad D^{\mu \nu
}=\partial _{\alpha }\partial _{\beta }U^{\mu \alpha \nu \beta },\qquad
D^{\mu \nu \rho }=\partial _{\lambda }D^{\mu \nu \rho \lambda },
\label{int66}
\end{equation}%
where $\digamma ^{\mu \nu }$ and $D^{\mu \nu \rho \lambda }$ are completely
antisymmetric in their Lorentz indices and $U^{\mu \alpha \nu \beta }$
possesses the mixed symmetry of the Riemann tensor. In addition, $\digamma
^{\mu \nu }$ and $D^{\mu \nu \rho \lambda }$ contain precisely one spacetime
derivative of the fields and $U^{\mu \alpha \nu \beta }$ depends only on the
undifferentiated fields. At this stage it is useful to introduce a
derivation in the algebra of the fields and of their derivatives that counts
the powers of the fields and their derivatives, defined by%
\begin{eqnarray}
N &=&\sum\limits_{k\geq 0}\left[ \frac{\partial ^{\mathrm{R}}}{\partial
\left( \partial _{\mu _{1}\cdots \mu _{k}}\psi _{\mu }\right) }\left(
\partial _{\mu _{1}\cdots \mu _{k}}\psi _{\mu }\right) +\left( \partial
_{\mu _{1}\cdots \mu _{k}}h_{\mu \nu }\right) \frac{\partial }{\partial
\left( \partial _{\mu _{1}\cdots \mu _{k}}h_{\mu \nu }\right) }\right.
\notag \\
&&\left. +\left( \partial _{\mu _{1}\cdots \mu _{k}}A_{\mu \nu \rho }\right)
\frac{\partial }{\partial \left( \partial _{\mu _{1}\cdots \mu _{k}}A_{\mu
\nu \rho }\right) }\right] .  \label{int67}
\end{eqnarray}%
Then, it is easy to see that for every nonintegrated density $\chi $, we
have that
\begin{equation}
N\chi =\frac{\delta ^{\mathrm{R}}\chi }{\delta \psi _{\mu }}\psi _{\mu }+%
\frac{\delta \chi }{\delta h_{\mu \nu }}h_{\mu \nu }+\frac{\delta \chi }{%
\delta A_{\mu \nu \rho }}A_{\mu \nu \rho }+\partial _{\mu }s^{\mu }.
\label{int68}
\end{equation}%
If $\chi ^{\left( l\right) }$ is a homogeneous polynomial of order $l>0$ in
the fields and their derivatives, then
\begin{equation}
N\chi ^{\left( l\right) }=l\chi ^{\left( l\right) }.  \label{int69}
\end{equation}%
Using (\ref{int63}), (\ref{int66}), and (\ref{int68}), we deduce that%
\begin{equation}
N\omega _{2}=\frac{1}{2}\digamma ^{\mu \nu }\partial _{\lbrack \mu }\psi
_{\nu ]}-\frac{1}{2}K_{\mu \alpha \nu \beta }U^{\mu \alpha \nu \beta }+\frac{%
1}{4}F_{\mu \nu \rho \lambda }D^{\mu \nu \rho \lambda }+\partial _{\mu
}v^{\mu }.  \label{int70}
\end{equation}%
We expand $\omega _{2}$ as%
\begin{equation}
\omega _{2}=\sum\limits_{l>0}\omega _{2}^{\left( l\right) },  \label{ww77}
\end{equation}%
where $N\omega _{2}^{\left( l\right) }=l\omega _{2}^{\left( l\right) }$,
such that
\begin{equation}
N\omega _{2}=\sum\limits_{l>0}l\omega _{2}^{\left( l\right) }.  \label{ww78}
\end{equation}%
Comparing (\ref{int70}) with (\ref{ww78}), we reach the conclusion that the
decomposition (\ref{ww77}) induces a similar decomposition with respect to $%
\digamma ^{\mu \nu }$, $U^{\mu \alpha \nu \beta }$, and $D^{\mu \nu \rho
\lambda }$, i.e.
\begin{equation}
\digamma ^{\mu \nu }=\sum\limits_{l>0}\digamma _{\left( l-1\right) }^{\mu
\nu },\qquad U^{\mu \alpha \nu \beta }=\sum\limits_{l>0}U_{\left( l-1\right)
}^{\mu \alpha \nu \beta },\qquad D^{\mu \nu \rho \lambda
}=\sum\limits_{l>0}D_{\left( l-1\right) }^{\mu \nu \rho \lambda }.
\label{ww79}
\end{equation}%
Substituting (\ref{ww79}) in (\ref{int70}) and comparing the resulting
expression with (\ref{ww78}), we obtain that
\begin{equation}
\omega _{2}^{\left( l\right) }=\frac{1}{2l}\left( \digamma _{\left(
l-1\right) }^{\mu \nu }\partial _{\lbrack \mu }\psi _{\nu ]}-K_{\mu \alpha
\nu \beta }U_{\left( l-1\right) }^{\mu \alpha \nu \beta }+\frac{1}{2}F_{\mu
\nu \rho \lambda }D_{\left( l-1\right) }^{\mu \nu \rho \lambda }\right)
+\partial _{\mu }\bar{v}_{(l)}^{\mu }.  \label{prform}
\end{equation}%
Introducing (\ref{prform}) in (\ref{ww77}), we arrive at
\begin{equation}
\omega _{2}=\frac{1}{2}\check{\digamma}^{\mu \nu }\partial _{\lbrack \mu
}\psi _{\nu ]}-\frac{1}{2}K_{\mu \alpha \nu \beta }\check{U}^{\mu \alpha \nu
\beta }+\frac{1}{4}F_{\mu \nu \rho \lambda }\check{D}^{\mu \nu \rho \lambda
}+\partial _{\mu }\bar{v}^{\mu },  \label{ww81}
\end{equation}%
where
\begin{equation}
\check{\digamma}^{\mu \nu }=\sum\limits_{l>0}\frac{1}{l}\digamma _{\left(
l-1\right) }^{\mu \nu },\qquad \check{U}^{\mu \alpha \nu \beta
}=\sum\limits_{l>0}\frac{1}{l}U_{\left( l-1\right) }^{\mu \alpha \nu \beta
},\qquad \check{D}^{\mu \nu \rho \lambda }=\sum\limits_{l>0}\frac{1}{l}%
D_{\left( l-1\right) }^{\mu \nu \rho \lambda }.  \label{ww82}
\end{equation}%
We will show that the second term from the right-hand side of (\ref{ww81})
does not comply with the derivative order assumption. Indeed, the tensor $%
\check{U}^{\mu \alpha \nu \beta }$ effectively depends on both $A_{\mu \nu
\rho }$ and $\psi _{\mu }$ (and possibly on their derivatives) in order to
describe simultaneous interactions among all the fields. Due to the presence
of the linearized Riemann tensor $K_{\mu \alpha \nu \beta }$, the term from $%
\omega _{2}$ containing $\check{U}^{\mu \alpha \nu \beta }$ will contribute
to the field equations for the spin-$2$ field with quantities involving at
least two spacetime derivatives acting on $\psi _{\mu }$, which breaks the
derivative order assumption. Consequently, we must set%
\begin{equation}
\check{U}^{\mu \alpha \nu \beta }=0.  \label{ww83}
\end{equation}%
The last result implies (via the second formula in (\ref{ww82})) $U_{\left(
l-1\right) }^{\mu \alpha \nu \beta }=0$ for all $l>0$, which further yields
(due to the second relation from (\ref{ww79})) that $U^{\mu \alpha \nu \beta
}=0$, such that the second result from (\ref{int66}) finally leads to $%
D^{\mu \nu }=0$. Recalling the second notation from (\ref{int63}) this is
the same with
\begin{equation*}
\frac{\delta \omega _{2}}{\delta h_{\mu \nu }}=0,
\end{equation*}%
meaning that $\omega _{2}$ is enabled to depend on the spin-$2$ field only
through a (trivial) full divergence, which brings no contribution to the
Lagrangian action of the interacting model. We conclude that there is no
nontrivial $\omega _{2}$ that mixes all the three field sectors, so we can
take%
\begin{equation}
\omega _{2}=0  \label{ww84}
\end{equation}%
without loss of generality.

Inserting (\ref{int48}), (\ref{int62}), and (\ref{ww84}) in (\ref{int43}) we
find that
\begin{equation}
\tilde{a}_{0}^{\mathrm{int}}=0,  \label{int71}
\end{equation}%
such that results (\ref{int42}) and (\ref{int71}) substituted in (\ref{int40}%
) provide
\begin{equation*}
a_{0}^{\mathrm{int}}=0
\end{equation*}%
and hence the first-order deformation of the solution to the master equation
that mixes all the fields cannot reduce nontrivially to its component of
antighost number zero.

Since we have exhausted all the possibilities of constructing a nontrivial $%
a^{\mathrm{int}}$ as in (\ref{int5}) ($I=4,3,2,1,0$), we conclude that the
general solution to (\ref{int1}) that complies with all our working
hypotheses is%
\begin{equation}
S_{1}^{\mathrm{int}}=0.  \label{intS1}
\end{equation}%
In conclusion, the full expression of the first-order deformation of the
solution to the master equation associated with the free theory described by
(\ref{int0}) decomposes as
\begin{equation}
S_{1}=S_{1}^{\mathrm{h}}+S_{1}^{\mathrm{A}}+S_{1}^{\mathrm{\psi }}+S_{1}^{%
\mathrm{h-A}}+S_{1}^{\mathrm{A-\psi }}+S_{1}^{\mathrm{h-\psi }},
\label{descfinfirstorder}
\end{equation}%
where all the terms from the right-hand side of
(\ref{descfinfirstorder}) have been reported in Refs.~%
\cite{pI}--\cite{pIII}.

\section{Second-order deformation}

The scope of this section is to investigate the consistency of the
first-order deformation and hence to determine the expression of the
second-order deformation of the solution to the master equation. In view of
this, we start from the equation%
\begin{equation}
\left( S_{1},S_{1}\right) +2sS_{2}=0,  \label{ecdef2}
\end{equation}%
where $S_{1}$ reads as in (\ref{descfinfirstorder}). By direct computation
we find that the antibracket $\left( S_{1},S_{1}\right) $ naturally
decomposes into%
\begin{eqnarray}
\left( S_{1},S_{1}\right)  &=&\left( S_{1},S_{1}\right) ^{\mathrm{h}}+\left(
S_{1},S_{1}\right) ^{\mathrm{A}}+\left( S_{1},S_{1}\right) ^{\mathrm{\psi }%
}+\left( S_{1},S_{1}\right) ^{\mathrm{h-A}}  \notag \\
&&+\left( S_{1},S_{1}\right) ^{\mathrm{A-\psi }}+\left( S_{1},S_{1}\right) ^{%
\mathrm{h-\psi }}+\left( S_{1},S_{1}\right) ^{\mathrm{int}},
\label{descantip}
\end{eqnarray}%
where $\left( S_{1},S_{1}\right) ^{\mathrm{sector(s)}}$ is the projection of
$\left( S_{1},S_{1}\right) $ on the respectively mentioned \textrm{sectors(s)%
}. Clearly, (\ref{descantip}) induces a similar decomposition with respect
to the second-order deformation%
\begin{equation}
S_{2}=S_{2}^{\mathrm{h}}+S_{2}^{\mathrm{A}}+S_{2}^{\mathrm{\psi }}+S_{2}^{%
\mathrm{h-A}}+S_{2}^{\mathrm{A-\psi }}+S_{2}^{\mathrm{h-\psi }}+S_{2}^{%
\mathrm{int}}.  \label{rf1}
\end{equation}%
The projection of (\ref{ecdef2}) on the various sectors makes (\ref{ecdef2})
equivalent to the tower of equations
\begin{eqnarray}
\left( S_{1},S_{1}\right) ^{\mathrm{h}}+2sS_{2}^{\mathrm{h}} &=&0,
\label{ecdf2p1} \\
\left( S_{1},S_{1}\right) ^{\mathrm{A}}+2sS_{2}^{\mathrm{A}} &=&0,
\label{ecdf2p2} \\
\left( S_{1},S_{1}\right) ^{\mathrm{\psi }}+2sS_{2}^{\mathrm{\psi }} &=&0,
\label{ecdf2p3} \\
\left( S_{1},S_{1}\right) ^{\mathrm{h-A}}+2sS_{2}^{\mathrm{h-A}} &=&0,
\label{ecdf2p4} \\
\left( S_{1},S_{1}\right) ^{\mathrm{A-\psi }}+2sS_{2}^{\mathrm{A-\psi }}
&=&0,  \label{ecdf2p5} \\
\left( S_{1},S_{1}\right) ^{\mathrm{h-\psi }}+2sS_{2}^{\mathrm{h-\psi }}
&=&0,  \label{ecdf2p6} \\
\left( S_{1},S_{1}\right) ^{\mathrm{int}}+2sS_{2}^{\mathrm{int}} &=&0.
\label{ecdf2p7}
\end{eqnarray}%
If we denote by $\Delta ^{\mathrm{sector(s)}}$ and $b^{\mathrm{sector(s)}}$
the nonintegrated densities of the functionals $\left( S_{1},S_{1}\right) ^{%
\mathrm{sector(s)}}$ and $S_{2}^{\mathrm{sector(s)}}$ respectively, then
Eqs. (\ref{ecdf2p1})--(\ref{ecdf2p7}) take the local form%
\begin{eqnarray}
\Delta ^{\mathrm{h}} &=&-2sb^{\mathrm{h}}+\partial ^{\mu }n_{\mu }^{\mathrm{h%
}},  \label{ecdf2p1l} \\
\Delta ^{\mathrm{A}} &=&-2sb^{\mathrm{A}}+\partial ^{\mu }n_{\mu }^{\mathrm{A%
}}  \label{ecdf2p2l} \\
\Delta ^{\mathrm{\psi }} &=&-2sb^{\mathrm{\psi }}+\partial ^{\mu }n_{\mu }^{%
\mathrm{\psi }},  \label{ecdf2p3l} \\
\Delta ^{\mathrm{h-A}} &=&-2sb^{\mathrm{h-A}}+\partial ^{\mu }n_{\mu }^{%
\mathrm{h-A}},  \label{ecdf2p4l} \\
\Delta ^{\mathrm{A-\psi }} &=&-2sb^{\mathrm{A-\psi }}+\partial ^{\mu }n_{\mu
}^{\mathrm{A-\psi }},  \label{ecdf2p5l} \\
\Delta ^{\mathrm{h-\psi }} &=&-2sb^{\mathrm{h-\psi }}+\partial ^{\mu }n_{\mu
}^{\mathrm{h-\psi }},  \label{ecdf2p6l} \\
\Delta ^{\mathrm{int}} &=&-2sb^{\mathrm{int}}+\partial ^{\mu }n_{\mu }^{%
\mathrm{int}},  \label{ecdf2p7l}
\end{eqnarray}%
with
\begin{equation}
\mathrm{gh}\left( \Delta ^{\mathrm{sector(s)}}\right) =1,\qquad \mathrm{gh}%
\left( b^{\mathrm{sector(s)}}\right) =0,\qquad \mathrm{gh}\left( n_{\mu }^{%
\mathrm{sector(s)}}\right) =1,  \label{rf5}
\end{equation}%
for some local currents $n_{\mu }^{\mathrm{sector(s)}}$. Recalling
decomposition (\ref{descfinfirstorder}) of the first-order deformation as
well as the concrete expressions of its components, we find that
\begin{equation*}
\left( S_{1},S_{1}\right) ^{\mathrm{h}}=\left( S_{1}^{\mathrm{h}},S_{1}^{%
\mathrm{h}}\right) .
\end{equation*}%
By direct computation we deduce
\begin{eqnarray}
&&S_{2}^{\mathrm{h}}=\int d^{11}x\left\{ \mathcal{L}_{2}^{\mathrm{EH%
}}-\frac{\Lambda }{2}\left( h^{2}-2h_{\mu \nu }h^{\mu \nu }\right) -\frac{1}{%
4}h^{\ast \mu \nu }\left[ h_{\mu }^{\lambda }\partial _{\nu }\left( h_{\rho
\lambda }\eta ^{\lambda }\right) \right. \right.   \notag \\
&&\left. +\frac{1}{2}h_{\rho \lambda }\left( \partial ^{\lambda }h_{\mu \nu
}\right) \eta ^{\rho }+\frac{3}{2}\left( \partial _{(\mu }h_{\nu )\lambda
}-\partial _{\lambda }h_{\mu \nu }\right) h_{\rho }^{\lambda }\eta ^{\rho }%
\right]   \notag \\
&&\left. +\frac{1}{8}\eta ^{\ast \mu }\eta ^{\nu }\left( h_{\mu }^{\rho
}\partial _{\lbrack \nu }\eta _{\rho ]}-h_{\nu }^{\rho }\partial _{\lbrack
\mu }\eta _{\rho ]}-\eta ^{\rho }\partial _{\lbrack \nu }h_{\rho ]\mu
}\right) \right\} ,  \label{sopf}
\end{eqnarray}%
where $\mathcal{L}_{2}^{\mathrm{EH}}$ is the quartic vertex of the
Einstein-Hilbert Lagrangian. Meanwhile, it results that%
\begin{equation*}
\left( S_{1},S_{1}\right) ^{\mathrm{A}}=0,
\end{equation*}%
so Eq. (\ref{ecdf2p2}) reduces to%
\begin{equation}
sS_{2}^{\mathrm{A}}=0  \label{ecdf2p2a}
\end{equation}%
and has been solved in Ref.~\cite{pI}. Namely, we have argued that
the solution to (\ref{ecdf2p2a}) can be taken as trivial modulo a
redefinition of the
constant $q$ that parameterizes $S_{1}^{\mathrm{A}}$%
\begin{equation}
S_{2}^{\mathrm{A}}=0.  \label{so3f}
\end{equation}%
Eq. (\ref{ecdf2p3}) has been tackled in Section 6 from
Ref.~\cite{pIII},
where we proved that the parameters $\bar{k}$, $\tilde{k}$, $m$, and $%
\Lambda $ are restricted to satisfy the relations%
\begin{equation}
\tilde{k}^{2}+\frac{\bar{k}^{2}}{32}=0,\qquad 180m^{2}-\bar{k}\Lambda =0,
\label{ecc1}
\end{equation}%
which then grant the nonintegrated density of the second-order deformation
in the Rarita-Schwinger sector to be expressed as the sum between the pieces
listed in formulas (71), (72), and (74) from  Ref.~\cite{pIII}. Eq. (\ref%
{ecdf2p4}) has been worked out in detail in Ref.~\cite{pI}, where it
was shown that the constant $k$ (parameterizing the cross-couplings
between the spin-$2 $ field and the $3$-form) is subject to the
relation
\begin{equation}
k\left( k+1\right) =0.  \label{ecc2}
\end{equation}%
Taking the nontrivial solution of (\ref{ecc2}) ($k=-1$), it follows
that the second-order deformation in the mixed sector
graviton-$3$-form is described by formula (117) din Ref.~\cite{pI}.

Let us investigate now Eq. (\ref{ecdf2p5}). It is easy to see that%
\begin{eqnarray}
\left( S_{1},S_{1}\right) ^{\mathrm{A-\psi }} &=&\left( S_{1}^{\mathrm{%
A-\psi }},S_{1}^{\mathrm{A-\psi }}\right) +2\left( S_{1}^{\mathrm{A}},S_{1}^{%
\mathrm{A-\psi }}\right)   \notag \\
&&+2\left( S_{1}^{\mathrm{\psi }},S_{1}^{\mathrm{A-\psi }}\right) +2\left(
S_{1}^{\mathrm{h-A}},S_{1}^{\mathrm{h-\psi }}\right) .  \label{rf3}
\end{eqnarray}%
Recalling that $\Delta ^{\mathrm{A-\psi }}$ denotes the nonintegrated
density of $\left( S_{1},S_{1}\right) ^{\mathrm{A-\psi }}$ and performing
the necessary computations in the right-hand side of (\ref{rf3}), we get
that $\Delta ^{\mathrm{A-\psi }}$ decomposes into
\begin{equation}
\Delta ^{\mathrm{A-\psi }}=\sum_{I=0}^{4}\Delta _{I}^{\mathrm{A-\psi }%
},\qquad \mathrm{agh}\left( \Delta _{I}^{\mathrm{A-\psi }}\right) =I,\qquad
I=\overline{0,4},  \label{rf6}
\end{equation}%
with
\begin{equation}
\Delta _{4}^{\mathrm{A-\psi }}=\gamma \left( \frac{\mathrm{i}k\bar{k}}{8}%
C^{\ast }C^{\mu }\bar{\xi}\gamma _{\mu }\xi \right) +\partial _{\mu }\tau
_{4}^{\mathrm{A-\psi ~}\mu },  \label{rf7}
\end{equation}%
\begin{equation}
\Delta _{3}^{\mathrm{A-\psi }}=\delta \left( \frac{\mathrm{i}k\bar{k}}{8}%
C^{\ast }C^{\mu }\bar{\xi}\gamma _{\mu }\xi \right) +\gamma \left[ -\frac{%
\mathrm{i}k\bar{k}}{8}C^{\ast \mu }\left( C_{\mu \nu }\bar{\xi}\gamma ^{\nu
}\xi +C^{\nu }\xi \gamma _{(\mu }\psi _{\nu )}\right) \right] +\partial
_{\mu }\tau _{3}^{\mathrm{A-\psi ~}\mu },  \label{rf8}
\end{equation}%
\begin{eqnarray}
\Delta _{2}^{\mathrm{A-\psi }} &=&\delta \left[ -\frac{\mathrm{i}k\bar{k}}{8}%
C^{\ast \mu }\left( C_{\mu \nu }\bar{\xi}\gamma ^{\nu }\xi +C^{\nu }\xi
\gamma _{(\mu }\psi _{\nu )}\right) \right] +\gamma \left[ \frac{\mathrm{i}k%
\bar{k}}{8}C^{\ast \mu \nu }\left( A_{\mu \nu \rho }\bar{\xi}\gamma ^{\rho
}\xi \right. \right.   \notag \\
&&\left. \left. -2C_{\mu }^{\quad \rho }\bar{\xi}\gamma _{(\nu }\psi _{\rho
)}\right) \right] +\partial _{\mu }\tau _{2}^{\mathrm{A-\psi ~}\mu },
\label{rf9}
\end{eqnarray}%
\begin{eqnarray}
\Delta _{1}^{\mathrm{h-A}} &=&\delta \left[ \frac{\mathrm{i}k\bar{k}}{8}%
C^{\ast \mu \nu }\left( A_{\mu \nu \rho }\bar{\xi}\gamma ^{\rho }\xi
-2C_{\mu }^{\quad \rho }\bar{\xi}\gamma _{(\nu }\psi _{\rho )}\right) \right]
\notag \\
&&+\gamma \left( -\frac{3\mathrm{i}k\bar{k}}{8}A^{\ast \mu \nu \rho }A_{\mu
\nu }^{\quad \lambda }\bar{\xi}\gamma _{(\rho }\psi _{\lambda )}\right) +
\notag \\
&&+4\mathrm{i}\left( \tilde{k}^{2}-\frac{k\bar{k}}{32}\right) A^{\ast \mu
\nu \rho }F_{\mu \nu \rho \lambda }\bar{\xi}\gamma ^{\lambda }\xi +\partial
_{\mu }\tau _{1}^{\mathrm{A-\psi ~}\mu },  \label{rf10}
\end{eqnarray}%
and
\begin{eqnarray}
\Delta _{0}^{\mathrm{h-A}} &=&\delta \left( -\frac{3\mathrm{i}k\bar{k}}{8}%
A^{\ast \mu \nu \rho }A_{\mu \nu }^{\quad \lambda }\bar{\xi}\gamma _{(\rho
}\psi _{\lambda )}\right) +  \notag \\
&&+\frac{4\mathrm{i}}{3}\left( \tilde{k}^{2}-\frac{k\bar{k}}{32}\right)
F^{\mu \nu \rho \alpha }F_{\mu \nu \rho }^{\qquad \beta }\left( \bar{\xi}%
\gamma _{\alpha }\psi _{\beta }-\frac{1}{8}\sigma _{\alpha \beta }\bar{\xi}%
\gamma ^{\lambda }\psi _{\lambda }\right)   \notag \\
&&-18\tilde{k}\left( q+\frac{\tilde{k}}{3\cdot \left( 12\right) ^{3}}\right)
\varepsilon ^{\mu _{1}\cdots \mu _{11}}\bar{\xi}\gamma _{\mu _{1}\mu
_{2}}\psi _{\mu _{3}}F_{\mu _{4}\cdots \mu _{7}}F_{\mu _{8}\cdots \mu _{11}}
\notag \\
&&+\mathrm{i}m\tilde{k}F^{\mu \nu \rho \lambda }\bar{\xi}\gamma _{\mu \nu
\rho \lambda \sigma }\psi ^{\sigma }+\partial _{\mu }\tau _{0}^{\mathrm{%
A-\psi ~}\mu }.  \label{rf11}
\end{eqnarray}%
Because $\left( S_{1},S_{1}\right) ^{\mathrm{A-\psi }}$ contains terms of
maximum antighost number equal to four, we can assume (without loss of
generality) that $b^{\mathrm{A-\psi }}$ stops at antighost number five%
\begin{eqnarray}
b^{\mathrm{A-\psi }} &=&\sum_{I=0}^{5}b_{I}^{\mathrm{A-\psi }},\qquad
\mathrm{agh}\left( b_{I}^{\mathrm{A-\psi }}\right) =I,\qquad I=\overline{0,5}%
,  \label{rf12a} \\
n^{\mathrm{A-\psi ~}\mu } &=&\sum_{I=0}^{5}n_{I}^{\mu },\qquad \mathrm{agh}%
\left( n_{I}^{\mathrm{A-\psi ~}\mu }\right) =I,\qquad I=\overline{0,5}.
\label{rf12b}
\end{eqnarray}%
By projecting Eq. (\ref{ecdf2p5l}) on the various (decreasing)
values of the antighost number, we infer the following tower of
equations
\begin{eqnarray}
\gamma b_{5}^{\mathrm{A-\psi }} &=&\partial _{\mu }\left( \frac{1}{2}n_{5}^{%
\mathrm{A-\psi ~}\mu }\right) ,  \label{rf13a} \\
\Delta _{I}^{\mathrm{A-\psi }} &=&-2\left( \delta b_{I+1}^{\mathrm{A-\psi }%
}+\gamma b_{I}^{\mathrm{A-\psi }}\right) +\partial _{\mu }n_{I}^{\mathrm{%
A-\psi ~}\mu },\qquad I=\overline{0,4}.  \label{rf13b}
\end{eqnarray}%
Eq. (\ref{rf13a}) can always be replaced with
\begin{equation}
\gamma b_{5}^{\mathrm{A-\psi }}=0.  \label{rf14}
\end{equation}%
If we compare (\ref{rf7}) with (\ref{rf13b}) for $I=4$, then we find that $%
b_{5}^{\mathrm{h-A}}$ is restricted to fulfill the equation
\begin{equation}
\delta b_{5}^{\mathrm{A-\psi }}+\gamma \tilde{b}_{4}^{\mathrm{A-\psi }%
}=\partial _{\mu }\tilde{n}_{4}^{\mathrm{A-\psi ~}\mu },  \label{rf15}
\end{equation}%
where
\begin{equation}
b_{4}^{\mathrm{A-\psi }}=-\frac{\mathrm{i}k\bar{k}}{16}C^{\ast }C^{\mu }\bar{%
\xi}\gamma _{\mu }\xi +\tilde{b}_{4}^{\mathrm{A-\psi }}.  \label{rf16}
\end{equation}%
The solution to (\ref{rf14}) reads as
\begin{equation}
b_{5}^{\mathrm{A-\psi }}=\beta _{5}^{\mathrm{A-\psi }}(\left[ F_{\mu \nu
\rho \lambda }\right] ,\left[ \partial _{\lbrack \mu }\psi _{\nu ]}\right] ,%
\left[ \chi _{\Delta }^{\ast }\right] )\omega ^{5}\left( C,\xi \right) .
\label{rf17}
\end{equation}%
Substituting the above form of $b_{5}^{\mathrm{A-\psi }}$ into (\ref{rf15}),
we infer that a necessary condition for (\ref{rf15}) to possess solutions is
that $\beta _{5}^{\mathrm{A-\psi }}$ belongs to $H_{5}\left( \delta
|d\right) $. Since for the model under consideration we know that $%
H_{5}\left( \delta |d\right) =0$ and $H_{5}^{\mathrm{inv}}\left( \delta
|d\right) =0$, it follows that we can take
\begin{equation}
b_{5}^{\mathrm{A-\psi }}=0,  \label{rf18}
\end{equation}%
such that Eq. (\ref{rf15}) reduces to $\gamma \tilde{b}_{4}^{\mathrm{%
A-\psi }}=\partial _{\mu }\tilde{n}_{4}^{\mu }$, that can always be
replaced
(as it stands in a strictly positive value of the antighost number) with $%
\gamma \tilde{b}_{4}^{\mathrm{A-\psi }}=0$. The last equation was
investigated in Ref.~\cite{pII} and was shown to possess only the
trivial solution
\begin{equation}
\tilde{b}_{4}^{\mathrm{A-\psi }}=0.  \label{rf19}
\end{equation}%
Due to (\ref{rf16}) and (\ref{rf19}), we observe that relations (\ref{rf7}%
)--(\ref{rf9}) agree with Eq. (\ref{rf13b}) for $I=4$, $I=3$ and
$I=2$
respectively. On the contrary, $\Delta _{1}^{\mathrm{A-\psi }}$ given in (%
\ref{rf10}) cannot be written like in (\ref{rf13b}) for $I=1$ unless
\begin{equation}
\chi ^{\mathrm{A-\psi }}=4\mathrm{i}\left( \tilde{k}^{2}-\frac{k\bar{k}}{32}%
\right) A^{\ast \mu \nu \rho }F_{\mu \nu \rho \lambda }\bar{\xi}\gamma
^{\lambda }\xi ,  \label{rf20}
\end{equation}%
can be expressed like
\begin{equation}
\chi ^{\mathrm{A-\psi }}=\delta \varphi ^{\mathrm{A-\psi }}+\gamma \omega ^{%
\mathrm{A-\psi }}+\partial _{\mu }l^{\mathrm{A-\psi ~}\mu }.  \label{rf21}
\end{equation}%
Assume that (\ref{rf21}) holds. Then, by acting with $\delta $ on it from
the left, we infer that
\begin{equation}
\delta \chi ^{\mathrm{A-\psi }}=\gamma \left( -\delta \omega ^{\mathrm{%
A-\psi }}\right) +\partial _{\mu }\left( \delta l^{\mathrm{A-\psi ~}\mu
}\right) .  \label{rf22}
\end{equation}%
On the other hand, using the concrete expression of $\chi $, we have that
\begin{equation}
\delta \chi ^{\mathrm{A-\psi }}=4\mathrm{i}\left( \tilde{k}^{2}-\frac{k\bar{k%
}}{32}\right) \left[ \gamma \left( -T^{\mu \nu }\bar{\xi}\gamma _{(\mu }\psi
_{\nu )}\right) +\partial _{\mu }\left( T^{\mu \nu }\bar{\xi}\gamma _{\nu
}\xi \right) \right] ,  \label{rf23}
\end{equation}%
where
\begin{equation}
T^{\mu \nu }=\frac{1}{3!}F^{\rho \lambda \sigma \mu }F_{\rho \lambda \sigma
}^{\;\;\;\;\;\nu }-\frac{\sigma ^{\mu \nu }}{2\cdot 4!}F^{\rho \lambda
\sigma \varepsilon }F_{\rho \lambda \sigma \varepsilon }  \label{stressen}
\end{equation}%
is the stress-energy tensor of the Abelian three-form gauge field. The
right-hand side of (\ref{rf23}) can be written like in the right-hand side
of (\ref{rf22}) if the following conditions are simultaneously satisfied
\begin{eqnarray}
-\delta \omega ^{\mathrm{A-\psi }} &=&-4\mathrm{i}\left( \tilde{k}^{2}-\frac{%
k\bar{k}}{32}\right) T^{\mu \nu }\bar{\xi}\gamma _{(\mu }\psi _{\nu )},
\label{rf24} \\
\delta l^{\mathrm{A-\psi ~}\mu } &=&4\mathrm{i}\left( \tilde{k}^{2}-\frac{k%
\bar{k}}{32}\right) T^{\mu \nu }\bar{\xi}\gamma _{\nu }\xi .  \label{rf25}
\end{eqnarray}%
Since none of the quantities $\psi _{\mu }$, or $\xi $ are $\delta $-exact,
we deduce that the last relations hold if stress-energy tensor of the
Abelian three-form gauge field is $\delta $-exact%
\begin{equation}
T^{\mu \nu }=\delta \Omega ^{\mu \nu }.  \label{rf26}
\end{equation}%
We have shown in Section 4.3 of Ref.~\cite{pI} that relation
(\ref{rf26}) is not
valid, and thus neither are (\ref{rf24})--(\ref{rf25}). As a consequence, $%
\chi ^{\mathrm{A-\psi }}$ must vanish, which further implies
\begin{equation}
\tilde{k}^{2}-\frac{k\bar{k}}{32}=0.  \label{ecc3}
\end{equation}%
Inserting (\ref{ecc3}) in (\ref{rf11}), we deduce that%
\begin{eqnarray}
\Delta _{0}^{\mathrm{A-\psi }} &=&\delta \left( -\frac{3\mathrm{i}k\bar{k}}{8%
}A^{\ast \mu \nu \rho }A_{\mu \nu }^{\quad \lambda }\bar{\xi}\gamma _{(\rho
}\psi _{\lambda )}\right) +  \notag \\
&&-18\tilde{k}\left( q+\frac{\tilde{k}}{3\cdot \left( 12\right) ^{3}}\right)
\varepsilon ^{\mu _{1}\cdots \mu _{11}}\bar{\xi}\gamma _{\mu _{1}\mu
_{2}}\psi _{\mu _{3}}F_{\mu _{4}\cdots \mu _{7}}F_{\mu _{8}\cdots \mu _{11}}
\notag \\
&&+\mathrm{i}m\tilde{k}F^{\mu \nu \rho \lambda }\bar{\xi}\gamma _{\mu \nu
\rho \lambda \sigma }\psi ^{\sigma }+\partial _{\mu }\tau _{0}^{\mathrm{%
A-\psi ~}\mu }.  \label{rf30}
\end{eqnarray}%
We remark that (\ref{rf30}) satisfies Eq. (\ref{rf13b}) for $I=0$ if
the quantity%
\begin{eqnarray}
\hat{\chi}^{\mathrm{A-\psi }} &=&\mathrm{i}m\tilde{k}F^{\mu \nu \rho \lambda
}\bar{\xi}\gamma _{\mu \nu \rho \lambda \sigma }\psi ^{\sigma }  \notag \\
&&-18\tilde{k}\left( q+\frac{\tilde{k}}{3\cdot \left( 12\right) ^{3}}\right)
\varepsilon ^{\mu _{1}\cdots \mu _{11}}\bar{\xi}\gamma _{\mu _{1}\mu
_{2}}\psi _{\mu _{3}}F_{\mu _{4}\cdots \mu _{7}}F_{\mu _{8}\cdots \mu _{11}},
\label{chiapsi}
\end{eqnarray}%
can be written as%
\begin{equation}
\hat{\chi}^{\mathrm{A-\psi }}=\delta \hat{\varphi}^{\mathrm{A-\psi }}+\gamma
\hat{\omega}^{\mathrm{A-\psi }}+\partial ^{\mu }\hat{l}_{\mu }^{\mathrm{%
A-\psi }}.  \label{rf31}
\end{equation}%
Let us assume that (\ref{rf31}) takes place. Then, we apply $\gamma $ on (%
\ref{rf31}) and find that
\begin{equation}
\gamma \hat{\chi}^{\mathrm{A-\psi }}=\delta \left( -\gamma \hat{\varphi}^{%
\mathrm{A-\psi }}\right) +\partial ^{\mu }\left( \gamma \hat{l}_{\mu }^{%
\mathrm{A-\psi }}\right) .  \label{rf32}
\end{equation}%
Direct computation based on (\ref{chiapsi}) provides
\begin{eqnarray}
\gamma \hat{\chi}^{\mathrm{A-\psi }} &=&\partial ^{\mu }\left[ \frac{\mathrm{%
i}m\tilde{k}}{2}F^{\nu \rho \lambda \sigma }\bar{\xi}\gamma _{\mu \nu \rho
\lambda \sigma }\xi \right.   \notag \\
&&\left. -9\tilde{k}\left( q+\frac{\tilde{k}}{3\cdot \left( 12\right) ^{3}}%
\right) \varepsilon _{\mu \mu _{1}\cdots \mu _{11}}\bar{\xi}\gamma _{\mu
_{1}\mu _{2}}\xi F_{\mu _{3}\cdots \mu _{6}}F_{\mu _{7}\cdots \mu _{10}}%
\right] .  \label{rf33}
\end{eqnarray}%
On the one hand, Eq. (\ref{rf32}) requires that the current
appearing in its right-hand side is trivial in $H\left( \gamma
\right) $. On the other hand, the current involved in the right-hand
side of (\ref{rf33}) is clearly a nontrivial element of $H\left(
\gamma \right) $. This contradiction emphasizes that relation
(\ref{rf31}) cannot be valid, and therefore we must
set $\hat{\chi}^{\mathrm{A-\psi }}=0$, which leads to the conditions%
\begin{equation}
m\tilde{k}=0,\qquad \tilde{k}\left( q+\frac{\tilde{k}}{3\cdot \left(
12\right) ^{3}}\right) =0.  \label{ecc4}
\end{equation}%
Inserting now (\ref{ecc3}) and (\ref{ecc4}) in (\ref{rf7})--(\ref{rf11}), we
are able to identify the components of the second-order deformation in the
mixed gravitini-$3$-form sector as%
\begin{equation}
b_{4}^{\mathrm{A-\psi }}=-\frac{\mathrm{i}k\bar{k}}{16}C^{\ast }C^{\mu }\bar{%
\xi}\gamma _{\mu }\xi ,  \label{rf35}
\end{equation}%
\begin{equation}
b_{3}^{\mathrm{A-\psi }}=\frac{\mathrm{i}k\bar{k}}{16}C^{\ast \mu }\left(
C_{\mu \nu }\bar{\xi}\gamma ^{\nu }\xi +C^{\nu }\xi \gamma _{(\mu }\psi
_{\nu )}\right) ,  \label{rf36}
\end{equation}%
\begin{equation}
b_{2}^{\mathrm{A-\psi }}=-\frac{\mathrm{i}k\bar{k}}{16}C^{\ast \mu \nu
}\left( A_{\mu \nu \rho }\bar{\xi}\gamma ^{\rho }\xi -2C_{\mu }^{\quad \rho }%
\bar{\xi}\gamma _{(\nu }\psi _{\rho )}\right) ,  \label{rf37}
\end{equation}%
\begin{equation}
b_{1}^{\mathrm{A-\psi }}=\frac{3\mathrm{i}k\bar{k}}{16}A^{\ast \mu \nu \rho
}A_{\mu \nu }^{\quad \lambda }\bar{\xi}\gamma _{(\rho }\psi _{\lambda )},
\label{rf38}
\end{equation}%
and
\begin{equation}
b_{0}^{\mathrm{A-\psi }}=0.  \label{rf39}
\end{equation}%
Formulas (\ref{rf35})--(\ref{rf39}) allow us to write
\begin{equation}
S_{2}^{\mathrm{A-\psi }}=\int d^{11}x\left( b_{4}^{\mathrm{A-\psi }%
}+b_{3}^{\mathrm{A-\psi }}+b_{2}^{\mathrm{A-\psi }}+b_{1}^{\mathrm{A-\psi }%
}+b_{0}^{\mathrm{A-\psi }}\right) .  \label{rf40}
\end{equation}

In the following step we approach Eq. (\ref{ecdf2p6}). From (\ref%
{descfinfirstorder}), we determine the first term from the left-hand
side of Eq. (\ref{ecdf2p6}) under the form%
\begin{equation}
\left( S_{1},S_{1}\right) ^{\mathrm{h-\psi }}=\left( S_{1}^{\mathrm{h-\psi }%
},S_{1}^{\mathrm{h-\psi }}\right) +2\left( S_{1}^{\mathrm{h}},S_{1}^{\mathrm{%
h-\psi }}\right) +2\left( S_{1}^{\mathrm{\psi }},S_{1}^{\mathrm{h-\psi }%
}\right) ,  \label{pfr2}
\end{equation}%
where $\Delta ^{\mathrm{h-\psi }}$ from the left hand-side of (\ref{ecdf2p6l}%
) (the local form of (\ref{ecdf2p6}))\ decomposes as%
\begin{equation}
\Delta ^{\mathrm{h-\psi }}=\sum_{I=0}^{2}\Delta _{I}^{\mathrm{h-\psi }%
},\qquad \mathrm{agh}\left( \Delta _{I}^{\mathrm{h-\psi }}\right) =I,\qquad
I=\overline{0,2},  \label{pfr5}
\end{equation}%
with%
\begin{eqnarray}
&&\Delta _{2}^{\mathrm{h-\psi }}=\gamma \left[ -\frac{\bar{k}}{8}\left( \xi
^{\ast }\gamma _{\mu \nu }\xi h_{\lambda }^{\mu }\partial ^{\lbrack \nu
}\eta ^{\lambda ]}+\xi ^{\ast }\gamma _{\mu \nu }\xi \eta ^{\lambda
}\partial _{\left. {}\right. }^{[\mu }h_{\lambda }^{\nu ]}-\mathrm{i}\eta
^{\ast \mu }\eta ^{\nu }\bar{\psi}_{[\mu }\gamma _{\nu ]}\xi \right) \right]
\notag \\
&&-\frac{1}{8}\bar{k}\left( \bar{k}-1\right) \left[ \xi ^{\ast }\gamma _{\mu
\nu }\xi \left( \partial ^{\lbrack \mu }\eta ^{\alpha ]}\right) \partial
^{\lbrack \nu }\eta ^{\beta ]}\sigma _{\alpha \beta }+\frac{\mathrm{i}}{2}%
\eta ^{\ast \mu }\bar{\xi}\gamma ^{\nu }\xi \partial _{\lbrack \mu }\eta
_{\nu ]}\right] +\partial _{\mu }\tau _{2}^{\mathrm{h-\psi ~}\mu },
\label{pfr6}
\end{eqnarray}%
\begin{eqnarray}
&&\Delta _{1}^{\mathrm{h-\psi }}=\delta \left[ -\frac{\bar{k}}{8}\left( \xi
^{\ast }\gamma _{\mu \nu }\xi h_{\lambda }^{\mu }\partial ^{\lbrack \nu
}\eta ^{\lambda ]}+\xi ^{\ast }\gamma _{\mu \nu }\xi \eta ^{\lambda
}\partial _{\left. {}\right. }^{[\mu }h_{\lambda }^{\nu ]}-\mathrm{i}\eta
^{\ast \mu }\eta ^{\nu }\bar{\psi}_{[\mu }\gamma _{\nu ]}\xi \right) \right.
\notag \\
&&\left. +\bar{k}^{2}\xi ^{\ast }\left( \partial _{\lbrack \mu }\bar{\psi}%
_{\nu ]}\right) \eta ^{\mu }\eta ^{\nu }+2\mathrm{i}m\bar{k}\xi ^{\ast
}\gamma _{\mu }\xi \eta ^{\mu }\right] +\gamma \left[ -\frac{\bar{k}}{8}\psi
^{\ast \mu }\gamma _{\nu \rho }\psi _{\mu }\left( h_{\lambda }^{\nu
}\partial ^{\lbrack \rho }\eta ^{\lambda ]}\right. \right.  \notag \\
&&\left. +\partial _{\left. {}\right. }^{[\nu }h_{\lambda }^{\rho ]}\eta
^{\lambda }\right) -\frac{\bar{k}}{4}\psi ^{\ast \mu }\gamma _{\nu \rho }\xi
\left( \partial _{\left. {}\right. }^{[\nu }h_{\mu }^{\lambda ]}+\frac{1}{2}%
\partial _{\mu }h^{\nu \lambda }\right) h_{\lambda }^{\rho }+\frac{\mathrm{i}%
\bar{k}}{4}h^{\ast \mu \nu }\bar{\psi}_{\mu }\left( \gamma _{\rho }\xi
h_{\nu }^{\rho }\right.  \notag \\
&&\left. -2\gamma _{\nu }\psi _{\rho }\eta ^{\rho }\right) -\frac{\bar{k}^{2}%
}{4}\psi ^{\ast \mu }\gamma _{\nu \rho }\psi ^{\lambda }\left( \partial
_{\left. {}\right. }^{[\nu }h_{\mu }^{\rho ]}\right) \eta _{\lambda }-\bar{k}%
^{2}\psi ^{\ast \mu }\left( \partial _{\lbrack \mu }\psi _{\nu ]}\right)
h_{\rho }^{\nu }\eta ^{\rho }  \notag \\
&&\left. +\frac{\bar{k}^{2}}{4}\psi ^{\ast \mu }\gamma _{\nu \rho }\psi
_{\mu }\left( \partial _{\left. {}\right. }^{[\nu }h_{\lambda }^{\rho
]}\right) \eta ^{\lambda }-\mathrm{i}m\bar{k}\psi ^{\ast \mu }\left( \gamma
_{\nu }\xi h_{\mu }^{\nu }-2\gamma _{\mu }\psi _{\nu }\eta ^{\nu }\right) %
\right]  \notag \\
&&+\bar{k}\left( \bar{k}-1\right) \left\{ \psi ^{\ast \mu }\left( \partial
_{\lbrack \mu }\psi _{\nu ]}\right) \eta _{\rho }\partial ^{\lbrack \nu
}\eta ^{\rho ]}-\frac{1}{8}\psi ^{\ast \mu }\gamma _{\nu \rho }\psi _{\mu
}\left( \partial ^{\lbrack \nu }\eta ^{\alpha ]}\right) \partial ^{\lbrack
\rho }\eta ^{\beta ]}\sigma _{\alpha \beta }\right.  \notag \\
&&-\frac{1}{2}\psi ^{\ast \mu }\gamma _{\nu \rho }\xi \left[ \eta ^{\lambda
}\partial ^{\nu }\partial _{\lbrack \mu }^{\left. {}\right. }h_{\lambda
]}^{\rho }-\frac{1}{2}\left( \partial _{\left. {}\right. }^{[\nu }h_{\mu
}^{\alpha ]}\right) \partial ^{\lbrack \rho }\eta ^{\beta ]}\sigma _{\alpha
\beta }\right]  \notag \\
&&\left. -\frac{\mathrm{i}}{8}h^{\ast \mu \nu }\left( \bar{\xi}\gamma ^{\rho
}\xi \partial _{\lbrack \mu }h_{\rho ]\nu }-2\bar{\psi}_{\mu }\gamma ^{\rho
}\xi \partial _{\lbrack \nu }\eta _{\rho ]}-4\eta ^{\rho }\bar{\xi}\gamma
_{\nu }\partial _{\lbrack \mu }\psi _{\rho ]}\right) \right\} +\partial
_{\mu }\tau _{1}^{\mathrm{h-\psi ~}\mu },  \label{pfr7}
\end{eqnarray}%
and%
\begin{eqnarray}
&&\Delta _{0}^{\mathrm{h-\psi }}=\delta \left[ -\frac{\bar{k}}{8}\psi ^{\ast
\mu }\gamma _{\nu \rho }\psi _{\mu }\left( h_{\lambda }^{\nu }\partial
^{\lbrack \rho }\eta ^{\lambda ]}+\partial _{\left. {}\right. }^{[\nu
}h_{\lambda }^{\rho ]}\eta ^{\lambda }\right) \right.  \notag \\
&&-\frac{\bar{k}}{4}\psi ^{\ast \mu }\gamma _{\nu \rho }\xi \left( \partial
_{\left. {}\right. }^{[\nu }h_{\mu }^{\lambda ]}+\frac{1}{2}\partial _{\mu
}h^{\nu \lambda }\right) h_{\lambda }^{\rho }+\frac{\mathrm{i}\bar{k}}{4}%
h^{\ast \mu \nu }\bar{\psi}_{\mu }\left( \gamma _{\rho }\xi h_{\nu }^{\rho
}-2\gamma _{\nu }\psi _{\rho }\eta ^{\rho }\right)  \notag \\
&&-\frac{\bar{k}^{2}}{4}\psi ^{\ast \mu }\gamma _{\nu \rho }\left( \psi
^{\lambda }\eta _{\lambda }\partial _{\left. {}\right. }^{[\nu }h_{\mu
}^{\rho ]}-\psi _{\mu }\eta ^{\lambda }\partial _{\left. {}\right. }^{[\nu
}h_{\lambda }^{\rho ]}\right) -\bar{k}^{2}\psi ^{\ast \mu }\partial
_{\lbrack \mu }\psi _{\nu ]}h_{\rho }^{\nu }\eta ^{\rho }  \notag \\
&&\left. -\mathrm{i}m\bar{k}\psi ^{\ast \mu }\left( \gamma _{\nu }\xi h_{\mu
}^{\nu }-\gamma _{\mu }\psi _{\nu }\eta ^{\nu }\right) \right] +\gamma
\left\{ 9m\bar{k}\left( \frac{1}{2}\bar{\psi}^{\mu }\gamma _{\mu \nu }\psi
^{\nu }h-\bar{\psi}^{\mu }\gamma _{\mu \nu }\psi ^{\rho }h_{\rho }^{\nu
}\right) \right.  \notag \\
&&+\frac{\mathrm{i}}{32}\bar{k}\left[ -2\left( \partial ^{\lbrack \mu }\bar{%
\psi}^{\nu ]}\right) \gamma _{\mu \nu \rho }\left( \psi ^{\rho }h^{\lambda
\sigma }h_{\lambda \sigma }-\psi ^{\lambda }h^{\rho \sigma }h_{\lambda
\sigma }\right) -8\bar{\psi}_{\mu }\gamma ^{\mu }\psi ^{\nu }\left( h_{\nu
}^{\rho }\partial _{\lbrack \lambda }^{\left. {}\right. }h_{\rho ]}^{\lambda
}\right. \right.  \notag \\
&&\left. -h^{\rho \lambda }\partial _{\nu }h_{\rho \lambda }+\frac{1}{2}%
h_{\nu }^{\rho }\partial _{\lambda }h_{\rho }^{\lambda }+\frac{1}{2}h^{\rho
\lambda }\partial _{\rho }h_{\lambda \nu }\right) +4\left( \partial
^{\lbrack \mu }\bar{\psi}^{\nu ]}\right) \gamma _{\nu \rho \lambda }\psi
^{\rho }h_{\mu \sigma }h^{\lambda \sigma }  \notag \\
&&-4\bar{\psi}^{\mu }\gamma ^{\nu }\psi ^{\rho }\left( h_{\mu }^{\lambda
}\partial _{\lbrack \nu }h_{\lambda ]\rho }+h_{\mu }^{\lambda }\partial
_{\lbrack \rho }h_{\lambda ]\nu }+h_{\mu }^{\lambda }\partial _{\rho }h_{\nu
\lambda }\right)  \notag \\
&&-\bar{\psi}^{\mu }\gamma _{\mu \nu \rho \lambda \sigma }\psi ^{\nu
}h_{\omega }^{\rho }\partial ^{\lbrack \lambda }h^{\sigma ]\omega }+2\bar{k}%
\left( \partial ^{\lbrack \mu }\bar{\psi}^{\nu ]}\right) \gamma _{\mu \nu
\rho }\left( \psi ^{\rho }hh-2\psi ^{\lambda }hh_{\lambda }^{\rho }+2\psi
^{\lambda }h_{\sigma }^{\rho }h_{\lambda }^{\sigma }\right.  \notag \\
&&\left. -\psi ^{\rho }h_{\lambda \sigma }h^{\lambda \sigma }\right) +4\bar{k%
}\left( \partial ^{\lbrack \lambda }\bar{\psi}^{\sigma ]}\right) \gamma
_{\mu \nu \rho }\psi ^{\mu }h_{\lambda }^{\nu }h_{\sigma }^{\rho }-8\bar{k}%
\left( \partial ^{\lbrack \mu }\bar{\psi}^{\nu ]}\right) \gamma _{\nu \rho
\lambda }\left( \psi ^{\rho }hh_{\mu }^{\lambda }-\psi ^{\rho }h_{\mu
}^{\sigma }h_{\sigma }^{\lambda }\right.  \notag \\
&&\left. +\psi ^{\sigma }h_{\mu }^{\rho }h_{\sigma }^{\lambda }\right) +8%
\bar{k}\bar{\psi}_{\mu }\gamma ^{\mu }\psi ^{\nu }\left( h^{\rho \lambda
}\partial _{\lbrack \nu }h_{\rho ]\lambda }-h_{\nu }^{\rho }\partial
_{\lbrack \lambda }^{\left. {}\right. }h_{\rho ]}^{\lambda }+h\partial
_{\lbrack \lambda }^{\left. {}\right. }h_{\nu ]}^{\lambda }\right)  \notag \\
&&\left. \left. +4\bar{k}\bar{\psi}^{\mu }\gamma ^{\nu }\psi ^{\rho }\left(
h_{\nu }^{\lambda }\partial _{\lbrack \mu }h_{\rho ]\lambda }-2h_{\mu \nu
}\partial _{\lbrack \lambda }^{\left. {}\right. }h_{\rho ]}^{\lambda
}-2h_{\mu }^{\lambda }\partial _{\lbrack \rho }h_{\lambda ]\nu }-h\partial
_{\lbrack \mu }h_{\rho ]\nu }\right) \right] \right\}  \notag \\
&&-\frac{\mathrm{i}}{16}\bar{k}\left( \bar{k}-1\right) \left\{ \bar{\psi}%
^{\mu }\gamma _{\mu \nu \rho \lambda \sigma }\left[ \psi ^{\nu }\left(
\partial _{\left. {}\right. }^{[\rho }h_{\omega }^{\lambda ]}\right)
\partial ^{\lbrack \sigma }\eta ^{\omega ]}+\frac{1}{2}\xi \left( \partial
_{\left. {}\right. }^{[\nu }h_{\omega }^{\rho ]}\right) \partial ^{\lbrack
\lambda }h^{\sigma ]\omega }\right] \right.  \notag \\
&&-4\left( \partial ^{\lbrack \mu }\bar{\psi}^{\nu ]}\right) \gamma _{\mu
\nu \rho }\left( \psi ^{\lambda }\eta ^{\sigma }\partial _{\lbrack \lambda
}h_{\sigma ]}^{\rho }-\psi ^{\rho }\eta ^{\sigma }\partial _{\lbrack \lambda
}h_{\sigma ]}^{\lambda }-\frac{1}{2}\psi ^{\lambda }h^{\rho \sigma }\partial
_{\lbrack \lambda }\eta _{\sigma ]}+\frac{1}{2}\xi h^{\rho \sigma }\partial
_{\lbrack \lambda }h_{\sigma ]}^{\lambda }\right)  \notag \\
&&+4\left( \partial ^{\lbrack \mu }\bar{\psi}^{\nu ]}\right) \gamma _{\nu
\rho \lambda }\left( \xi h^{\rho \sigma }\partial _{\lbrack \mu }^{\left.
{}\right. }h_{\sigma ]}^{\lambda }+\psi ^{\rho }h^{\lambda \sigma }\partial
_{\lbrack \mu }\eta _{\sigma ]}-2\psi ^{\rho }\eta ^{\sigma }\partial
_{\lbrack \mu }^{\left. {}\right. }h_{\sigma ]}^{\lambda }\right)  \notag \\
&&-4\bar{\psi}_{\mu }\gamma ^{\mu }\psi ^{\nu }\left[ \left( \partial
_{\lbrack \nu }\eta _{\rho ]}\right) \partial _{\left. {}\right. }^{[\lambda
}h_{\lambda }^{\rho ]}-\frac{1}{2}\left( \partial _{\lbrack \rho }\eta
_{\lambda ]}\right) \partial _{\left. {}\right. }^{[\rho }h_{\nu }^{\lambda
]}+2\eta _{\nu }\partial ^{\rho }\partial _{\lbrack \rho }h_{\lambda
]}^{\lambda }\right.  \notag \\
&&\left. +2\eta _{\lambda }\left( \partial ^{\rho }\partial _{\lbrack \nu
}^{\left. {}\right. }h_{\rho ]}^{\lambda }-\partial ^{\lambda }\partial
_{\lbrack \nu }^{\left. {}\right. }h_{\rho ]}^{\rho }\right) \right] +4\bar{%
\psi}^{\mu }\gamma ^{\nu }\psi ^{\rho }\left[ \left( \partial _{\lbrack \mu
}h_{\alpha ]\nu }\right) \partial _{\lbrack \rho }\eta _{\beta ]}\sigma
^{\alpha \beta }\right.  \notag \\
&&\left. -2\left( \partial _{\mu }\partial _{\lbrack \nu }h_{\lambda ]\rho
}\right) \eta ^{\lambda }-2\left( \partial ^{\lambda }\partial _{\lbrack \mu
}h_{\lambda ]\nu }-\partial _{\nu }\partial _{\lbrack \mu }h_{\lambda
]}^{\lambda }\right) \eta _{\rho }\right] -\bar{\psi}_{\mu }\gamma ^{\mu
}\xi \left[ \left( \partial _{\lbrack \nu }h_{\rho ]\lambda }\right)
\partial ^{\lbrack \nu }h^{\rho ]\lambda }\right.  \notag \\
&&\left. -2\left( \partial _{\lbrack \nu }h_{\rho ]}^{\nu }\right) \partial
^{\lbrack \lambda }h_{\lambda }^{\rho ]}-4h\partial ^{\nu }\partial
_{\lbrack \nu }h_{\rho ]}^{\rho }-4h_{\rho }^{\nu }\left( \partial ^{\lambda
}\partial _{\lbrack \nu }h_{\lambda ]}^{\rho }-\partial ^{\rho }\partial
_{\lbrack \nu }h_{\lambda ]}^{\lambda }\right) \right]  \notag \\
&&+4\bar{\psi}^{\mu }\gamma ^{\nu }\xi \left[ h_{\mu }^{\rho }\left(
\partial _{\nu }\partial _{\lbrack \rho }h_{\lambda ]}^{\lambda }-\partial
^{\lambda }\partial _{\lbrack \rho }h_{\lambda ]\nu }\right) +h^{\rho
\lambda }\left( \partial _{\mu }\partial _{\lbrack \nu }h_{\rho ]\lambda
}-\partial _{\lambda }\partial _{\lbrack \nu }h_{\rho ]\mu }\right) \right.
\notag \\
&&+h_{\nu }^{\rho }\left( \partial _{\mu }\partial _{\lbrack \rho
}h_{\lambda ]}^{\lambda }-\partial ^{\lambda }\partial _{\lbrack \rho
}h_{\lambda ]\mu }\right) -h\left( \partial _{\nu }\partial _{\lbrack \mu
}^{\left. {}\right. }h_{\rho ]}^{\rho }-\partial ^{\rho }\partial _{\lbrack
\mu }h_{\rho ]\nu }\right) -h_{\mu \nu }\partial ^{\rho }\partial _{\lbrack
\rho }^{\left. {}\right. }h_{\lambda ]}^{\lambda }  \notag \\
&&\left. \left. +\frac{1}{2}\left( \partial _{\lbrack \rho }h_{\lambda ]\mu
}\right) \partial _{\left. {}\right. }^{[\rho }h_{\nu }^{\lambda ]}+\left(
\partial _{\lbrack \mu }h_{\rho ]\nu }\right) \partial _{\left. {}\right.
}^{[\rho }h_{\lambda }^{\lambda ]}\right] \right\} +\partial _{\mu }\tau
_{0}^{\mathrm{h-\psi ~}\mu }.  \label{pfr8}
\end{eqnarray}%
Pursuing a reasoning similar to the previously investigated equation we
conclude that parameter $\bar{k}$ is subject to the algebraic equation%
\begin{equation}
\bar{k}\left( \bar{k}-1\right) =0.  \label{ecc5}
\end{equation}%
Introducing (\ref{ecc5}) in (\ref{pfr6})--(\ref{pfr8}) and recalling
Eq. (\ref{ecdf2p6l}), we identify the various pieces of the
nonintegrated density of the second-order deformation in the mixed
graviton-gravitini sector as%
\begin{eqnarray}
&&b_{2}^{\mathrm{h-\psi }}=\frac{\bar{k}}{16}\xi ^{\ast }\gamma _{\mu \nu
}\xi \left( h_{\lambda }^{\mu }\partial ^{\lbrack \nu }\eta ^{\lambda
]}+\eta ^{\lambda }\partial _{\left. {}\right. }^{[\mu }h_{\lambda }^{\nu
]}\right) -\frac{\mathrm{i}\bar{k}}{16}\eta ^{\ast \mu }\eta ^{\nu }\bar{\psi%
}_{[\mu }\gamma _{\nu ]}\xi  \notag \\
&&-\frac{\bar{k}^{2}}{2}\xi ^{\ast }\left( \partial _{\lbrack \mu }\bar{\psi}%
_{\nu ]}\right) \eta ^{\mu }\eta ^{\nu }-\mathrm{i}m\bar{k}\xi ^{\ast
}\gamma _{\mu }\xi \eta ^{\mu },  \label{pfr10}
\end{eqnarray}%
\begin{eqnarray}
&&b_{1}^{\mathrm{h-\psi }}=\frac{\bar{k}}{16}\psi ^{\ast \mu }\gamma _{\nu
\rho }\left[ \psi _{\mu }\left( h_{\lambda }^{\nu }\partial ^{\lbrack \rho
}\eta ^{\lambda ]}+\partial _{\left. {}\right. }^{[\nu }h_{\lambda }^{\rho
]}\eta ^{\lambda }\right) +2\xi \left( \partial _{\left. {}\right. }^{[\nu
}h_{\mu }^{\lambda ]}+\frac{1}{2}\partial _{\mu }h^{\nu \lambda }\right)
h_{\lambda }^{\rho }\right]  \notag \\
&&-\frac{\mathrm{i}\bar{k}}{8}h^{\ast \mu \nu }\bar{\psi}_{\mu }\left(
\gamma _{\rho }\xi h_{\nu }^{\rho }-2\gamma _{\nu }\psi _{\rho }\eta ^{\rho
}\right) +\frac{\bar{k}^{2}}{8}\psi ^{\ast \mu }\left[ \gamma _{\nu \rho
}\psi ^{\lambda }\left( \partial _{\left. {}\right. }^{[\nu }h_{\mu }^{\rho
]}\right) \eta _{\lambda }\right.  \notag \\
&&\left. +4\left( \partial _{\lbrack \mu }\psi _{\nu ]}\right) h_{\rho
}^{\nu }\eta ^{\rho }-\gamma _{\nu \rho }\psi _{\mu }\left( \partial
_{\left. {}\right. }^{[\nu }h_{\lambda }^{\rho ]}\right) \eta ^{\lambda }
\right] +\frac{\mathrm{i}m\bar{k}}{2}\psi ^{\ast \mu }\left( \gamma _{\nu
}\xi h_{\mu }^{\nu }-2\gamma _{\mu }\psi _{\nu }\eta ^{\nu }\right) ,
\label{pfr11}
\end{eqnarray}%
and respectively%
\begin{eqnarray}
&&b_{0}^{\mathrm{h-\psi }}=\frac{9m\bar{k}}{2}\left( \bar{\psi}^{\mu }\gamma
_{\mu \nu }\psi ^{\rho }h_{\rho }^{\nu }-\frac{1}{2}\bar{\psi}^{\mu }\gamma
_{\mu \nu }\psi ^{\nu }h\right)  \notag \\
&&+\frac{\mathrm{i}\bar{k}}{64}\bar{\psi}^{\mu }\gamma _{\mu \nu \rho
\lambda \sigma }\psi ^{\nu }h^{\rho \omega }\partial _{\left. {}\right.
}^{[\lambda }h_{\omega }^{\sigma ]}+\frac{\mathrm{i}\bar{k}}{32}\partial
^{\lbrack \mu }\bar{\psi}^{\nu ]}\gamma _{\mu \nu \rho }\left( \psi ^{\rho
}h^{\lambda \sigma }h_{\lambda \sigma }-\psi ^{\lambda }h^{\rho \sigma
}h_{\lambda \sigma }\right)  \notag \\
&&-\frac{\mathrm{i}\bar{k}}{16}\partial ^{\lbrack \mu }\bar{\psi}^{\nu
]}\gamma _{\nu \rho \lambda }\psi ^{\rho }h_{\mu \sigma }h^{\lambda \sigma }+%
\frac{\mathrm{i}\bar{k}}{8}\bar{\psi}_{\mu }\gamma ^{\mu }\psi ^{\nu }\left(
h_{\nu }^{\rho }\partial _{\lbrack \lambda }^{\left. {}\right. }h_{\rho
]}^{\lambda }-h^{\rho \lambda }\partial _{\nu }h_{\rho \lambda }\right.
\notag \\
&&\left. +\frac{1}{2}h_{\nu }^{\rho }\partial _{\lambda }h_{\rho }^{\lambda
}+\frac{1}{2}h^{\rho \lambda }\partial _{\rho }h_{\lambda \nu }\right) +%
\frac{\mathrm{i}\bar{k}}{16}\bar{\psi}^{\mu }\gamma ^{\nu }\psi ^{\rho
}\left( h_{\mu }^{\lambda }\partial _{\lbrack \nu }h_{\lambda ]\rho }+h_{\mu
}^{\lambda }\partial _{\lbrack \rho }h_{\lambda ]\nu }\right.  \notag \\
&&\left. +h_{\mu }^{\lambda }\partial _{\rho }h_{\nu \lambda }\right) -\frac{%
\mathrm{i}\bar{k}^{2}}{32}\left( \partial ^{\lbrack \mu }\bar{\psi}^{\nu
]}\right) \gamma _{\mu \nu \rho }\left( \psi ^{\rho }hh-2\psi ^{\lambda
}hh_{\lambda }^{\rho }+2\psi ^{\lambda }h_{\sigma }^{\rho }h_{\lambda
}^{\sigma }-\psi ^{\rho }h_{\lambda \sigma }h^{\lambda \sigma }\right)
\notag \\
&&+\frac{\mathrm{i}\bar{k}^{2}}{8}\left( \partial ^{\lbrack \mu }\bar{\psi}%
^{\nu ]}\right) \gamma _{\nu \rho \lambda }\left( \psi ^{\rho }hh_{\mu
}^{\lambda }-\psi ^{\rho }h_{\mu }^{\sigma }h_{\sigma }^{\lambda }+\psi
^{\sigma }h_{\mu }^{\rho }h_{\sigma }^{\lambda }\right) -\frac{\mathrm{i}%
\bar{k}^{2}}{16}\partial ^{\lbrack \lambda }\bar{\psi}^{\sigma ]}\gamma
_{\mu \nu \rho }\psi ^{\mu }h_{\lambda }^{\nu }h_{\sigma }^{\rho }  \notag \\
&&-\frac{\mathrm{i}\bar{k}^{2}}{8}\bar{\psi}_{\mu }\gamma ^{\mu }\psi ^{\nu
}\left( h^{\rho \lambda }\partial _{\lbrack \nu }h_{\rho ]\lambda }-h_{\nu
}^{\rho }\partial _{\lbrack \lambda }^{\left. {}\right. }h_{\rho ]}^{\lambda
}+h\partial _{\lbrack \lambda }^{\left. {}\right. }h_{\nu ]}^{\lambda
}\right)  \notag \\
&&-\frac{\mathrm{i}\bar{k}^{2}}{16}\bar{\psi}^{\mu }\gamma ^{\nu }\psi
^{\rho }\left( h_{\nu }^{\lambda }\partial _{\lbrack \mu }h_{\rho ]\lambda
}-2h_{\mu \nu }\partial _{\lbrack \lambda }^{\left. {}\right. }h_{\rho
]}^{\lambda }-2h_{\mu }^{\lambda }\partial _{\lbrack \rho }h_{\lambda ]\nu
}-h\partial _{\lbrack \mu }h_{\rho ]\nu }\right) .  \label{pfr12}
\end{eqnarray}%
Formulas (\ref{pfr10})--(\ref{pfr12}) enable us to write
\begin{equation}
S_{2}^{\mathrm{h-\psi }}=\int d^{11}x\left( b_{2}^{\mathrm{h-\psi }%
}+b_{1}^{\mathrm{h-\psi }}+b_{0}^{\mathrm{h-\psi }}\right) .  \label{pfr13}
\end{equation}

Finally, we solve (\ref{ecdf2p7}) in its local form, namely (\ref{ecdf2p7l}%
). Taking into account one more time the concrete form of the first-order
deformation, (\ref{descfinfirstorder}), we observe that
\begin{equation}
\left( S_{1},S_{1}\right) ^{\mathrm{int}}=2\left( S_{1}^{\mathrm{h-A}%
},S_{1}^{\mathrm{A-\psi }}\right) +2\left( S_{1}^{\mathrm{h-\psi }},S_{1}^{%
\mathrm{A-\psi }}\right) ,  \label{xint2}
\end{equation}%
where $\Delta ^{\mathrm{int}}$ decomposes as%
\begin{equation}
\Delta ^{\mathrm{int}}=\sum_{I=0}^{2}\Delta _{I}^{\mathrm{int}},\qquad
\mathrm{agh}\left( \Delta _{I}^{\mathrm{int}}\right) =I,\qquad I=\overline{%
0,2},  \label{xint5}
\end{equation}%
with%
\begin{equation}
\Delta _{2}^{\mathrm{int}}=\gamma \left( 2k\tilde{k}C^{\ast \mu \nu }\eta
^{\rho }\bar{\xi}\gamma _{\mu \nu }\psi _{\rho }\right) +\tilde{k}\left( k+%
\bar{k}\right) C^{\ast \mu \nu }\bar{\xi}\gamma _{\mu \alpha }\xi \partial
_{\lbrack \nu }\eta _{\beta ]}\sigma ^{\alpha \beta }+\partial _{\mu }\tau
_{2}^{\mathrm{int~}\mu },  \label{xint6}
\end{equation}%
\begin{eqnarray}
&&\Delta _{1}^{\mathrm{int}}=\delta \left[ 2k\tilde{k}C^{\ast \mu \nu }\eta
^{\rho }\bar{\xi}\gamma _{\mu \nu }\psi _{\rho }-\frac{\mathrm{i}\bar{k}%
\tilde{k}}{36}\xi ^{\ast }\left( \gamma _{\mu \nu \rho \lambda \sigma }\xi
\eta ^{\sigma }+8\gamma _{\mu \nu \rho }\xi \eta _{\lambda }\right) F^{\mu
\nu \rho \lambda }\right]   \notag \\
&&+\gamma \left[ -3k\tilde{k}A^{\ast \mu \nu \rho }\left( h_{\rho }^{\lambda
}\bar{\psi}_{\lambda }\gamma _{\mu \nu }\xi +2\eta ^{\lambda }\bar{\psi}%
_{\mu }\gamma _{\nu \rho }\psi _{\lambda }\right) +\frac{\mathrm{i}\bar{k}%
\tilde{k}}{72}\psi ^{\ast \mu }\gamma _{\nu \rho \lambda \sigma \omega }\xi
h_{\mu }^{\nu }F^{\rho \lambda \sigma \omega }\right.   \notag \\
&&-\frac{\mathrm{i}k\tilde{k}}{9}\psi ^{\ast \lbrack \mu }\gamma ^{\nu \rho
\lambda ]}\xi \left( h_{\mu }^{\sigma }\partial _{\sigma }A_{\nu \rho
\lambda }-\frac{3}{2}A_{\mu \nu \sigma }\partial _{\lbrack \rho }^{\left.
{}\right. }h_{\lambda ]}^{\sigma }\right) -\frac{\mathrm{i}\bar{k}\tilde{k}}{%
36}\psi ^{\ast \mu }\gamma _{\mu \nu \rho \lambda \sigma }\psi ^{\omega
}F^{\nu \rho \lambda \sigma }\eta _{\omega }  \notag \\
&&+\frac{\mathrm{i}k\tilde{k}}{18}\psi ^{\ast \mu }\gamma _{\mu \nu \rho
\lambda \sigma }\xi \left( h_{\omega }^{\nu }\partial ^{\omega }A^{\rho
\lambda \sigma }-\frac{3}{2}A^{\nu \rho \omega }\partial _{\left. {}\right.
}^{[\lambda }h_{\omega }^{\sigma ]}\right) -\frac{\mathrm{i}\bar{k}\tilde{k}%
}{9}\psi ^{\ast \mu }\gamma ^{\nu \rho \lambda }\left( \xi h_{\mu }^{\sigma
}F_{\sigma \nu \rho \lambda }\right.   \notag \\
&&\left. \left. -2\psi ^{\sigma }F_{\mu \nu \rho \lambda }\eta _{\sigma
}\right) \right] +\tilde{k}\left( k+\bar{k}\right) \left\{ 6A^{\ast \mu \nu
\rho }\left[ \eta ^{\lambda }\left( \partial _{\lbrack \mu }\bar{\psi}%
_{\lambda ]}\right) \gamma _{\nu \rho }\xi +\bar{\psi}_{\nu }\gamma _{\rho
\beta }\xi \partial _{\lbrack \mu }\eta _{\alpha ]}\sigma ^{\alpha \beta
}\right. \right.   \notag \\
&&\left. +\frac{1}{4}\bar{\xi}\gamma _{\rho \omega }\xi \partial _{\lbrack
\mu }^{\left. {}\right. }h_{\nu ]}^{\omega }\right] -\frac{\mathrm{i}}{18}%
\psi ^{\ast \lbrack \mu }\gamma ^{\nu \rho \lambda ]}\xi \left( \eta
^{\sigma }\partial _{\sigma }F_{\mu \nu \rho \lambda }+2F_{\mu \nu \rho
\alpha }\partial _{\lbrack \lambda }\eta _{\beta ]}\sigma ^{\alpha \beta
}\right)   \notag \\
&&\left. +\frac{\mathrm{i}}{36}\psi ^{\ast \mu }\gamma _{\mu \nu \rho
\lambda \sigma }\xi \left( \eta ^{\omega }\partial _{\omega }F^{\nu \rho
\lambda \sigma }+2F^{\nu \rho \lambda \omega }\partial _{\lbrack \beta }\eta
_{\omega ]}\sigma ^{\sigma \beta }\right) \right\} +\partial _{\mu }\tau
_{1}^{\mathrm{int~}\mu },  \label{xint7}
\end{eqnarray}%
and%
\begin{eqnarray}
&&\Delta _{0}^{\mathrm{int}}=\delta \left[ -3k\tilde{k}A^{\ast \mu \nu \rho
}\left( h_{\rho }^{\lambda }\bar{\psi}_{\lambda }\gamma _{\mu \nu }\xi
+2\eta ^{\lambda }\bar{\psi}_{\mu }\gamma _{\nu \rho }\psi _{\lambda
}\right) +\frac{\mathrm{i}\bar{k}\tilde{k}}{72}\psi ^{\ast \mu }\gamma _{\nu
\rho \lambda \sigma \omega }\xi h_{\mu }^{\nu }F^{\rho \lambda \sigma \omega
}\right.   \notag \\
&&-\frac{\mathrm{i}k\tilde{k}}{9}\psi ^{\ast \lbrack \mu }\gamma ^{\nu \rho
\lambda ]}\xi \left( h_{\mu }^{\sigma }\partial _{\sigma }A_{\nu \rho
\lambda }-\frac{3}{2}A_{\mu \nu \sigma }\partial _{\lbrack \rho }^{\left.
{}\right. }h_{\lambda ]}^{\sigma }\right) -\frac{\mathrm{i}\bar{k}\tilde{k}}{%
36}\psi ^{\ast \mu }\gamma _{\mu \nu \rho \lambda \sigma }\psi ^{\omega
}F^{\nu \rho \lambda \sigma }\eta _{\omega }  \notag \\
&&+\frac{\mathrm{i}k\tilde{k}}{18}\psi ^{\ast \mu }\gamma _{\mu \nu \rho
\lambda \sigma }\xi \left( h_{\omega }^{\nu }\partial ^{\omega }A^{\rho
\lambda \sigma }-\frac{3}{2}A^{\nu \rho \omega }\partial _{\left. {}\right.
}^{[\lambda }h_{\omega }^{\sigma ]}\right) -\frac{\mathrm{i}\bar{k}\tilde{k}%
}{9}\psi ^{\ast \mu }\gamma ^{\nu \rho \lambda }\left( \xi h_{\mu }^{\sigma
}F_{\sigma \nu \rho \lambda }\right.   \notag \\
&&\left. \left. -2\psi ^{\sigma }F_{\mu \nu \rho \lambda }\eta _{\sigma
}\right) \right] +\gamma \left[ \frac{\bar{k}\tilde{k}}{48}\bar{\psi}^{\mu
}\gamma _{\mu \nu \rho \lambda \sigma \omega }\psi ^{\nu }\left( F^{\rho
\lambda \sigma \omega }h-6A^{\rho \lambda \theta }\partial _{\left.
{}\right. }^{[\sigma }h_{\theta }^{\omega ]}\right. \right.   \notag \\
&&\left. +4h_{\theta }^{\rho }\partial ^{\theta }A^{\lambda \sigma \omega
}\right) -\frac{\bar{k}\tilde{k}}{24}\bar{\psi}^{\mu }\gamma _{\mu \nu \rho
\lambda \sigma \omega }\psi ^{\theta }h_{\theta }^{\nu }F^{\rho \lambda
\sigma \omega }-\frac{k\tilde{k}}{4}\bar{\psi}^{\mu }\gamma ^{\nu \rho }\psi
^{\lambda }\left( hF_{\mu \nu \rho \lambda }-2h_{\lambda }^{\sigma }F_{\mu
\nu \rho \sigma }\right.   \notag \\
&&\left. \left. +2h_{\nu }^{\sigma }\partial _{\sigma }A_{\mu \rho \lambda
}-2h_{\mu }^{\sigma }\partial _{\sigma }A_{\nu \rho \lambda }\right) -\frac{k%
\tilde{k}}{4}\bar{\psi}^{[\mu }\gamma ^{\nu \rho }\psi ^{\lambda ]}A_{\nu
\rho \sigma }\partial _{\lbrack \mu }h_{\lambda ]}^{\sigma }\right]   \notag
\\
&&+\tilde{k}\left( k+\bar{k}\right) \left[ \frac{1}{24}\bar{\psi}^{\mu
}\gamma _{\mu \nu \rho \lambda \sigma \omega }\left( \psi ^{\nu }\eta
_{\theta }\partial ^{\theta }F^{\rho \lambda \sigma \omega }-\xi h_{\theta
}^{\nu }\partial ^{\theta }F^{\rho \lambda \sigma \omega }-2\xi F^{\nu \rho
\lambda \theta }\partial _{\left. {}\right. }^{[\sigma }h_{\theta }^{\omega
]}\right) \right.   \notag \\
&&+\frac{1}{12}\bar{\psi}^{\mu }\gamma _{\mu \nu \rho \lambda \sigma \alpha
}\psi ^{\nu }F^{\rho \lambda \sigma \omega }\partial _{\lbrack \beta }\eta
_{\omega ]}\sigma ^{\alpha \beta }-\frac{1}{2}\bar{\psi}^{\mu }\gamma ^{\nu
\rho }\left( \psi ^{\lambda }F_{\mu \nu \lambda \alpha }\partial _{\lbrack
\rho }\eta _{\beta ]}\sigma ^{\alpha \beta }\right.   \notag \\
&&\left. -\xi F_{\mu \nu \lambda \sigma }\partial _{\left. {}\right.
}^{[\lambda }h_{\rho }^{\sigma ]}\right) -\frac{1}{2}\left( \partial
^{\lbrack \mu }\bar{\psi}^{\nu ]}\right) \gamma ^{\rho \lambda }\left( \psi
^{\sigma }F_{\mu \nu \rho \lambda }\eta _{\sigma }-2\psi ^{\sigma }F_{\mu
\rho \lambda \sigma }\eta _{\nu }\right.   \notag \\
&&\left. \left. -\frac{1}{2}\xi hF_{\mu \nu \rho \lambda }+\xi h_{\mu
}^{\sigma }F_{\sigma \nu \rho \lambda }\right) \right] .  \label{xint8}
\end{eqnarray}%
Reprising the same steps like in the previous cases, we conclude that (\ref%
{ecdf2p7l}) cannot hold unless the parameters $\bar{k}$, $k$, and $\tilde{k}$
satisfy the algebraic equation%
\begin{equation}
\tilde{k}\left( k+\bar{k}\right) =0.  \label{ecc6}
\end{equation}%
Assuming (\ref{ecc6}) holds, we insert (\ref{xint6})--(\ref{xint8}) into (%
\ref{ecdf2p7l}) and identify the nonintegrated density of the second-order
deformation in the interacting sector (describing simultaneous interactions
among graviton, gravitini, and $3$-form) under the form%
\begin{equation}
b_{2}^{\mathrm{int}}=-k\tilde{k}C^{\ast \mu \nu }\eta ^{\rho }\bar{\xi}%
\gamma _{\mu \nu }\psi _{\rho }+\frac{\mathrm{i}\bar{k}\tilde{k}}{72}\xi
^{\ast }\left( \gamma _{\mu \nu \rho \lambda \sigma }\xi \eta ^{\sigma
}+8\gamma _{\mu \nu \rho }\xi \eta _{\lambda }\right) F^{\mu \nu \rho
\lambda },  \label{xint9}
\end{equation}%
\begin{eqnarray}
&&b_{1}^{\mathrm{int}}=3k\tilde{k}A^{\ast \mu \nu \rho }\left( \eta
^{\lambda }\bar{\psi}_{\mu }\gamma _{\nu \rho }\psi _{\lambda }+\frac{1}{2}%
h_{\rho }^{\lambda }\bar{\psi}_{\lambda }\gamma _{\mu \nu }\xi \right)
\notag \\
&&+\frac{\mathrm{i}k\tilde{k}}{18}\psi ^{\ast \lbrack \mu }\gamma ^{\nu \rho
\lambda ]}\xi \left( h_{\mu }^{\sigma }\partial _{\sigma }A_{\nu \rho
\lambda }-\frac{3}{2}A_{\mu \nu \sigma }\partial _{\lbrack \rho }^{\left.
{}\right. }h_{\lambda ]}^{\sigma }\right)   \notag \\
&&-\frac{\mathrm{i}k\tilde{k}}{36}\psi ^{\ast \mu }\gamma _{\mu \nu \rho
\lambda \sigma }\xi \left( h_{\omega }^{\nu }\partial ^{\omega }A^{\rho
\lambda \sigma }-\frac{3}{2}A^{\nu \rho \omega }\partial _{\left. {}\right.
}^{[\lambda }h_{\omega }^{\sigma ]}\right)   \notag \\
&&-\frac{\mathrm{i}\bar{k}\tilde{k}}{144}\psi ^{\ast \mu }\gamma _{\nu \rho
\lambda \sigma \omega }\xi h_{\mu }^{\nu }F^{\rho \lambda \sigma \omega }+%
\frac{\mathrm{i}\bar{k}\tilde{k}}{72}\psi ^{\ast \mu }\gamma _{\mu \nu \rho
\lambda \sigma }\psi ^{\omega }F^{\nu \rho \lambda \sigma }\eta _{\omega }
\notag \\
&&+\frac{\mathrm{i}\bar{k}\tilde{k}}{18}\psi ^{\ast \mu }\gamma ^{\nu \rho
\lambda }\left( \xi h_{\mu }^{\sigma }F_{\sigma \nu \rho \lambda }-2\psi
^{\sigma }F_{\mu \nu \rho \lambda }\eta _{\sigma }\right) ,  \label{xint10}
\end{eqnarray}%
\begin{eqnarray}
&&b_{0}^{\mathrm{int}}=\frac{\bar{k}\tilde{k}}{48}\bar{\psi}^{\mu }\gamma
_{\mu \nu \rho \lambda \sigma \omega }\left( \psi ^{\theta }h_{\theta }^{\nu
}-\frac{1}{2}\psi ^{\nu }h\right) F^{\rho \lambda \sigma \omega }  \notag \\
&&+\frac{k\tilde{k}}{16}\bar{\psi}^{\mu }\gamma _{\mu \nu \rho \lambda
\sigma \omega }\psi ^{\nu }\left( A^{\rho \lambda \theta }\partial _{\left.
{}\right. }^{[\sigma }h_{\theta }^{\omega ]}-\frac{2}{3}h_{\theta }^{\rho
}\partial ^{\theta }A^{\lambda \sigma \omega }\right)   \notag \\
&&+\frac{k\tilde{k}}{8}\bar{\psi}^{\mu }\gamma ^{\nu \rho }\psi ^{\lambda
}\left( hF_{\mu \nu \rho \lambda }-2h_{\lambda }^{\sigma }F_{\mu \nu \rho
\sigma }+2h_{\nu }^{\sigma }\partial _{\sigma }A_{\mu \rho \lambda }\right.
\notag \\
&&\left. -2h_{\mu }^{\sigma }\partial _{\sigma }A_{\nu \rho \lambda }\right)
+\frac{k\tilde{k}}{8}\bar{\psi}^{[\mu }\gamma ^{\nu \rho }\psi ^{\lambda
]}A_{\nu \rho \sigma }\partial _{\lbrack \mu }^{\left. {}\right. }h_{\lambda
]}^{\sigma },  \label{xint11}
\end{eqnarray}%
with the help of which we have that%
\begin{equation}
S_{2}^{\mathrm{int}}=\int d^{11}x\left( b_{2}^{\mathrm{int}}+b_{1}^{%
\mathrm{int}}+b_{0}^{\mathrm{int}}\right) .  \label{xint12}
\end{equation}

In conclusion, we determined all the nontrivial constituents of the
second-order deformation given by (\ref{rf1}).

\section{Redefinition of first- and second-order deformations}

We showed in the previous section that the consistency of the first-order
deformation at order two in the coupling constant implies a simple algebraic
system for the six parameterizing constants, defined by Eqs. (\ref{ecc1}%
), (\ref{ecc2}), (\ref{ecc3}), (\ref{ecc4}), (\ref{ecc5}), and
(\ref{ecc6}).
There are two types of nontrivial solutions, namely%
\begin{equation}
k=-1\ \mathrm{or}\ k=0,\qquad \tilde{k}=\bar{k}=m=0,\qquad \Lambda ,q=%
\mathrm{arbitrary},  \label{solntr1}
\end{equation}%
\begin{equation}
k=-\bar{k}=-1,\qquad \tilde{k}_{1,2}=\pm \frac{\mathrm{i}\sqrt{2}}{8},\qquad
q_{1,2}=-\frac{4\tilde{k}_{1,2}}{\left( 12\right) ^{4}},\qquad m=0=\Lambda .
\label{solntr2}
\end{equation}%
The former type is less interesting from the point of view of interactions
since it maximally allows the graviton to be coupled to the $3$-form (if $%
k=-1$).

For this reason in the sequel we will extensively focus on the latter
solution, (\ref{solntr2}), which forbids both the presence of the
cosmological term for the spin-$2$ field and the appearance of gravitini
`mass' constant. In this case the first-order deformation of the solution to
the master equation is expressed by relation (\ref{descfinfirstorder}),
where:

\begin{enumerate}
\item[(i)] the density of $S_{1}^{\mathrm{h}}$ reads as in formula (47) from
Ref.~\cite{pI};

\item[(ii)] $S_{1}^{\mathrm{\psi }}=0$ (follows from relation (50) given in
Ref.~\cite{pII} where we set $m=0$);

\item[(iii)] the sum $S_{1}^{\mathrm{A}}+S_{1}^{\mathrm{h-A}}$ is furnished
by formula (118) from Ref.~\cite{pI} in which we take $q\rightarrow
-4\tilde{k}_{i}/\left( 12\right) ^{4}$;

\item[(iv)] the density of $S_{1}^{\mathrm{h-\psi }}$ is the sum among the
right-hand sides of formulas (20), (21), and (22) in
Ref.~\cite{pIII} where in addition we put $\bar{k}=1$;

\item[(v)] $S_{1}^{\mathrm{A-\psi }}$ is pictured by relation (110) from
Ref.~\cite{pII} modulo the replacement $\tilde{k}\rightarrow
\tilde{k}_{i}$.
\end{enumerate}

\noindent Consequently, the second-order deformation of the solution to the
master equation is still (\ref{rf1}), up to the following specifications:

\begin{enumerate}
\item $S_{2}^{\mathrm{h}}$ follows from (\ref{sopf}) restricted to $\Lambda
=0$;

\item the density of $S_{2}^{\mathrm{\psi }}$ is equal to the sum among the
right-hand sides of formulas (71), (72), and (74) from
Ref.~\cite{pIII} where we put $\bar{k}=1$;

\item $S_{2}^{\mathrm{A}}=0$, in agreement with (\ref{so3f});

\item $S_{2}^{\mathrm{h-A}}$ comes from relation (117) reported in Ref.~\cite{pI}
modulo the change $q\rightarrow -4\tilde{k}_{i}/\left( 12\right)
^{4}$;

\item $S_{2}^{\mathrm{A-\psi }}$ reads as in (\ref{rf40}) with $k=-\bar{k}%
=-1 $;

\item $S_{2}^{\mathrm{h-\psi }}$ is expressed by (\ref{pfr13}) for $\bar{k}%
=1 $ and $m=0$;

\item $S_{2}^{\mathrm{int}}$ is pictured by (\ref{xint12}), with $k=-\bar{k}%
=-1$ and $\tilde{k}\rightarrow \tilde{k}_{i}$ given by (\ref{solntr2}).
\end{enumerate}

In order to compare the interacting model resulting from our cohomological
approach with the results known from the literature \cite{scherk,wit}%
, it is necessary to redefine the first-order deformation through a trivial,
$s$-exact term, which does not modify either the cohomological class of $%
S_{1}$ or the physical contents of the coupled theory%
\begin{equation}
S_{1}\rightarrow \hat{S}_{1}=S_{1}+sK,  \label{redef}
\end{equation}%
with%
\begin{equation}
K=-\left( \xi ^{\ast }\psi ^{\mu }\eta _{\mu }+\frac{1}{2}\psi ^{\ast \mu
}\psi ^{\nu }h_{\mu \nu }\right) .  \label{k}
\end{equation}%
The above redefinition brings contributions only to the mixed
graviton-gravitini sector, so we can write%
\begin{equation}
\hat{S}_{1}=S_{1}^{\mathrm{h}}+S_{1}^{\mathrm{A}}+S_{1}^{\mathrm{h-A}%
}+S_{1}^{\mathrm{A-\psi }}+\hat{S}_{1}^{\mathrm{h-\psi }},  \label{d1bar}
\end{equation}%
where%
\begin{equation}
S_{1}^{\mathrm{h}}=\int d^{11}x\left[ \mathcal{L}_{1}^{\mathrm{EH}%
}+h^{\ast \mu \rho }\left( h_{\mu \nu }\partial _{\rho }\eta ^{\nu
}-\eta ^{\nu }\partial _{\lbrack \mu }h_{\nu ]\rho }\right)+
\frac{1}{2}\eta ^{\ast
\mu }\eta ^{\nu }\partial _{\left[ \mu \right. }\eta _{\left. \nu \right] }%
\right] ,  \label{foh}
\end{equation}%
\begin{equation}
S_{1}^{\mathrm{A}}=-\frac{4\tilde{k}_{i}}{\left( 12\right) ^{4}}\int
d^{11}x\varepsilon ^{\mu _{1}\ldots \mu _{11}}A_{\mu _{1}\mu _{2}\mu
_{3}}F_{\mu _{4}\ldots \mu _{7}}F_{\mu _{8}\ldots \mu _{11}},
\label{foa}
\end{equation}%
\begin{eqnarray}
&&S_{1}^{\mathrm{h-A}}=\int d^{11}x\left\{ \frac{1}{12}F^{\mu \nu
\rho \lambda }\left[ F_{\mu \nu \rho \sigma }h_{\lambda }^{\sigma
}-3\partial _{\mu }\left( A_{\nu \rho \sigma }h_{\lambda }^{\sigma }\right) -%
\frac{1}{8}F_{\mu \nu \rho \lambda }h\right] \right.   \notag \\
&&+\frac{3}{2}A^{\ast \mu \nu \rho }\left( \frac{2}{3}\eta ^{\lambda
}\partial _{\lambda }A_{\mu \nu \rho }+A_{\mu \nu }^{\;\;\;\lambda }\partial
_{\lbrack \rho }\eta _{\lambda ]}-h_{\rho \lambda }\partial ^{\lambda
}C_{\mu \nu }-C_{\mu \lambda }\partial _{\lbrack \nu }^{\left. {}\right.
}h_{\rho ]}^{\lambda }\right)   \notag \\
&&+C^{\ast \mu \nu }\left[ \left( \partial _{\rho }C_{\mu \nu }\right) \eta
^{\rho }+C_{\mu }^{\;\;\rho }\partial _{\lbrack \nu }\eta _{\rho ]}+h_{\nu
\rho }\partial ^{\rho }C_{\mu }+\frac{1}{2}C^{\rho }\partial _{\lbrack \mu
}h_{\nu ]\rho }\right]   \notag \\
&&\left. +\frac{1}{2}C^{\ast \mu }\left( 2\eta _{\nu }\partial ^{\nu }C_{\mu
}+C^{\nu }\partial _{\lbrack \mu }\eta _{\nu ]}-h_{\mu \nu }\partial ^{\nu
}C\right) +C^{\ast }\left( \partial ^{\mu }C\right) \eta _{\mu }\right\} ,
\label{foha}
\end{eqnarray}%
\begin{eqnarray}
S_{1}^{\mathrm{A-\psi }} &=&-\tilde{k}_{i}\int d^{11}x\left( \frac{1%
}{4}\bar{\psi}_{\mu }\gamma _{\nu \rho }\psi _{\lambda }F^{\mu \nu \rho
\lambda }+\frac{1}{2\cdot 4!}\bar{\psi}^{\alpha }\gamma _{\alpha \beta \mu
\nu \rho \lambda }\psi ^{\beta }F^{\mu \nu \rho \lambda }\right.   \notag \\
&&-\frac{\mathrm{i}}{9}\psi ^{\ast \mu }F_{\mu \nu \rho \lambda }\gamma
^{\nu \rho \lambda }\xi +\frac{\mathrm{i}}{3\cdot 4!}\psi ^{\ast \mu }F^{\nu
\rho \lambda \sigma }\gamma _{\mu \nu \rho \lambda \sigma }\xi   \notag \\
&&\left. +3A^{\ast \mu \nu \rho }\bar{\xi}\gamma _{\mu \nu }\psi _{\rho }-%
\frac{1}{2}C^{\ast \mu \nu }\bar{\xi}\gamma _{\mu \nu }\xi \right) ,
\label{foap}
\end{eqnarray}%
\begin{eqnarray}
&&\hat{S}_{1}^{\mathrm{h-\psi }}=\int d^{11}x\left\{ \frac{\mathrm{i%
}}{4}\left[ \frac{1}{2}\bar{\psi}^{\mu }\left( \gamma ^{\rho }\psi ^{\nu
}-2\sigma ^{\nu \rho }\gamma _{\lambda }\psi ^{\lambda }\right) \partial
_{\lbrack \mu }h_{\nu ]\rho }-h\bar{\psi}_{\mu }\gamma ^{\mu \nu \rho
}\partial _{\nu }\psi _{\rho }\right. \right.   \notag \\
&&\left. +\bar{\psi}_{\mu }\gamma ^{\mu \nu \rho }\left( \partial ^{\lambda
}\psi _{\rho }\right) h_{\nu \lambda }-\frac{1}{2}\bar{\psi}_{\mu }\gamma
^{\mu \nu \rho }\psi ^{\lambda }\partial _{\lbrack \nu }h_{\rho ]\lambda }%
\right] +\frac{1}{4}\left[ \mathrm{i}h^{\ast \mu \nu }\bar{\xi}\gamma _{\mu
}\psi _{\nu }\right.   \notag \\
&&+\frac{1}{2}\psi ^{\ast \mu }\gamma ^{\alpha \beta }\left( \psi _{\mu
}\partial _{\lbrack \alpha }\eta _{\beta ]}-\xi \partial _{\lbrack \alpha
}h_{\beta ]\mu }\right) +4\psi ^{\ast \mu }\left( \partial _{\nu }\psi _{\mu
}\right) \eta ^{\nu }+2\psi ^{\ast \mu }\psi ^{\nu }\partial _{\lbrack \mu
}\eta _{\nu ]}  \notag \\
&&\left. \left. -2\psi ^{\ast \mu }\left( \partial ^{\nu }\xi \right) h_{\mu
\nu }\right] +\xi ^{\ast }\left( \partial _{\mu }\xi \right) \eta ^{\mu }-%
\frac{1}{8}\left( \frac{\mathrm{i}}{2}\eta ^{\ast \mu }\bar{\xi}\gamma _{\mu
}\xi -\xi ^{\ast }\gamma ^{\mu \nu }\xi \partial _{\lbrack \mu }\eta _{\nu
]}\right) \right\} .  \label{fohp}
\end{eqnarray}%
In (\ref{foh}) the notation $\mathcal{L}_{1}^{\mathrm{EH}}$ means the cubic
vertex of the Einstein-Hilbert Lagrangian.

Redefinition (\ref{redef}) induces a modification in the expression of the
second-order deformation. Indeed, let us denote by $\hat{S}_{2}$ the
second-order deformation associated with $\hat{S}_{1}$, namely%
\begin{equation}
\left( \hat{S}_{1},\hat{S}_{1}\right) +2s\hat{S}_{2}=0.  \label{rdf1}
\end{equation}%
Since $S_{2}$ is solution to the equation%
\begin{equation}
\left( S_{1},S_{1}\right) +2sS_{2}=0,  \label{rdf2}
\end{equation}%
then (\ref{rdf1}) and (\ref{rdf2}) provide%
\begin{equation*}
2s\left( \hat{S}_{2}-S_{2}\right) +\left( \hat{S}_{1},\hat{S}_{1}\right)
-\left( S_{1},S_{1}\right) =0,
\end{equation*}%
or, equivalently (due to the bilinearity of the antibracket)%
\begin{equation}
2s\left( \hat{S}_{2}-S_{2}\right) +\left( \hat{S}_{1},\hat{S}%
_{1}-S_{1}\right) +\left( \hat{S}_{1}-S_{1},S_{1}\right) =0.  \label{rdf3}
\end{equation}%
Substituting (\ref{redef}) in (\ref{rdf3}), we infer the equation%
\begin{equation}
2s\left( \hat{S}_{2}-S_{2}\right) +\left( \hat{S}_{1},sK\right) +\left(
sK,S_{1}\right) =0.  \label{rdf4}
\end{equation}%
Recalling the fact that the BRST differential behaves like a derivation with
respect to the antibracket plus the $s$-closeness of both $S_{1}$ and $\hat{S%
}_{1}$, we find that (\ref{rdf4}) becomes%
\begin{equation}
s\left[ \hat{S}_{2}-S_{2}+\left( \frac{\hat{S}_{1}+S_{1}}{2},K\right) \right]
=0,  \label{rdf5}
\end{equation}%
which further produces
\begin{equation}
\hat{S}_{2}=S_{2}-\left( \frac{\hat{S}_{1}+S_{1}}{2},K\right) .  \label{rdf6}
\end{equation}%
Performing the necessary computations with the help of (\ref{k}), we obtain
that
\begin{equation}
\hat{S}_{2}=S_{2}^{\mathrm{h}}+S_{2}^{\mathrm{h-A}}+S_{2}^{\mathrm{A-\psi }}+%
\hat{S}_{2}^{\mathrm{\psi }}+\hat{S}_{2}^{\mathrm{h-\psi }}+\hat{S}_{2}^{%
\mathrm{int}},  \label{rdf7}
\end{equation}%
where%
\begin{eqnarray}
&&S_{2}^{\mathrm{h}}=\int d^{11}x\left\{ \mathcal{L}_{2}^{\mathrm{EH%
}}-\frac{1}{4}h^{\ast \mu \nu }\left[ h_{\mu }^{\lambda }\partial _{\nu
}\left( h_{\rho \lambda }\eta ^{\lambda }\right) \right. \right.   \notag \\
&&\left. +\frac{1}{2}h_{\rho \lambda }\left( \partial ^{\lambda }h_{\mu \nu
}\right) \eta ^{\rho }+\frac{3}{2}\left( \partial _{(\mu }h_{\nu )\lambda
}-\partial _{\lambda }h_{\mu \nu }\right) h_{\rho }^{\lambda }\eta ^{\rho }%
\right]   \notag \\
&&\left. +\frac{1}{8}\eta ^{\ast \mu }\eta ^{\nu }\left( h_{\mu }^{\rho
}\partial _{\lbrack \nu }\eta _{\rho ]}-h_{\nu }^{\rho }\partial _{\lbrack
\mu }\eta _{\rho ]}-\eta ^{\rho }\partial _{\lbrack \nu }h_{\rho ]\mu
}\right) \right\} ,  \label{soh}
\end{eqnarray}%
\begin{eqnarray}
&&S_{2}^{\mathrm{h-A}}=\frac{1}{2}\int d^{11}x\left\{ C^{\ast
}h_{\mu \nu }\eta ^{\mu }\partial ^{\nu }C+C^{\ast \mu }\left[ \frac{3}{4}%
h_{\mu \rho }h_{\nu }^{\rho }\partial ^{\nu }C-\frac{1}{2}C^{\nu }\eta
^{\rho }\partial _{\lbrack \mu }h_{\nu ]\rho }\right. \right.   \notag \\
&&\left. +\frac{1}{4}C^{\nu }\left( h_{\mu }^{\rho }\partial _{\lbrack \nu
}\eta _{\rho ]}-h_{\nu }^{\rho }\partial _{\lbrack \mu }\eta _{\rho
]}\right) -h^{\nu \rho }\eta _{\rho }\partial _{\nu }C_{\mu }\right]
-C^{\ast \mu \nu }\left[ C^{\rho }h_{\nu }^{\lambda }\partial _{\lbrack \mu
}h_{\lambda ]\rho }\right.   \notag \\
&&+\frac{1}{2}C^{\rho }\partial _{\mu }\left( h_{\nu \lambda }h_{\rho
}^{\lambda }\right) -\frac{3}{2}h_{\mu \lambda }h^{\rho \lambda }\partial
_{\rho }C_{\nu }+\eta ^{\rho }\left( \partial _{\lbrack \mu }^{\left.
{}\right. }h_{\rho ]}^{\lambda }-\partial _{\lbrack \xi }h_{\rho ]\mu
}\sigma ^{\xi \lambda }\right) C_{\nu \lambda }  \notag \\
&&\left. -\frac{1}{2}C_{\mu \rho }\left( h^{\rho \lambda }\partial _{\lbrack
\nu }\eta _{\lambda ]}-h_{\nu \lambda }\partial ^{\lbrack \rho }\eta
^{\lambda ]}\right) +\left( \partial _{\rho }C_{\mu \nu }\right) \eta
_{\lambda }h^{\rho \lambda }\right]   \notag \\
&&+\frac{3}{2}A^{\ast \mu \nu \rho }\left[ C_{\rho \xi }\partial _{\mu
}\left( h_{\nu \lambda }h^{\lambda \xi }\right) +\frac{3}{2}h_{\rho \xi
}h^{\lambda \xi }\partial _{\lambda }C_{\mu \nu }+2C_{\mu \lambda }h_{\nu
}^{\xi }\partial _{\lbrack \xi }^{\left. {}\right. }h_{\rho ]}^{\lambda
}\right.   \notag \\
&&-\frac{1}{2}A_{\mu \nu \lambda }\left( h^{\lambda \xi }\partial _{\lbrack
\rho }\eta _{\xi ]}+h_{\rho \xi }\partial ^{\lbrack \lambda }\eta ^{\xi
]}+2\sigma ^{\lambda \pi }\eta ^{\xi }\partial _{\lbrack \rho }h_{\pi ]\xi
}\right) +A_{\mu \nu \lambda }h_{\rho \xi }\partial ^{\lbrack \lambda }\eta
^{\xi ]}  \notag \\
&&\left. -\frac{2}{3}h^{\lambda \xi }\eta _{\xi }\partial _{\lambda }A_{\mu
\nu \rho }\right] +\frac{1}{8}F^{\mu \nu \rho \lambda }F_{\mu \nu \xi \pi }%
\left[ h_{\rho }^{\xi }h_{\lambda }^{\pi }-\frac{1}{3!}\delta _{\rho }^{\xi
}\delta _{\lambda }^{\pi }\left( \frac{1}{4}h^{2}-h^{\alpha \beta }h_{\alpha
\beta }\right) \right.   \notag \\
&&\left. -\frac{1}{3}\delta _{\rho }^{\xi }h_{\lambda \sigma }h^{\pi \sigma }%
\right] +\frac{1}{16}F^{\mu \nu \rho \lambda }\left[ A_{\xi \rho \lambda
}\partial _{\mu }\left( h_{\nu \pi }h^{\lambda \pi }\right) -h_{\mu \pi
}h^{\xi \pi }\left( \partial _{\nu }A_{\xi \rho \lambda }\right. \right.
\notag \\
&&\left. +\frac{4}{3}\partial _{\xi }A_{\nu \rho \lambda }\right) +A_{\mu
\nu \xi }\left( 4h_{\rho }^{\pi }\partial _{\lbrack \pi }^{\left. {}\right.
}h_{\lambda ]}^{\xi }-h\partial _{\lbrack \rho }^{\left. {}\right.
}h_{\lambda ]}^{\xi }\right) -\frac{2}{3}h_{\lambda }^{\xi }h\partial _{\xi
}A_{\mu \nu \rho }  \notag \\
&&\left. -2A_{\mu \xi \pi }\partial _{\nu }\left( h_{\rho }^{\xi }h_{\lambda
}^{\pi }\right) +2h_{\rho }^{\xi }h_{\lambda }^{\pi }\partial _{\xi }A_{\pi
\mu \nu }\right] +q_{i}\varepsilon ^{\mu _{1}\ldots \mu _{11}}\left( \frac{1%
}{2}hA_{\mu _{1}\mu _{2}\mu _{3}}F_{\mu _{4}\ldots \mu _{7}}\right.   \notag
\\
&&\left. -4h_{\mu _{1}}^{\xi }A_{\mu _{2}\mu _{3}\mu _{4}}F_{\mu _{5}\ldots
\mu _{7}\xi }+3h_{\mu _{1}}^{\xi }A_{\xi \mu _{2}\mu _{3}}F_{\mu _{4}\ldots
\mu _{7}}\right) F_{\mu _{8}\ldots \mu _{11}}  \notag \\
&&\left. +\frac{1}{16}\partial _{\xi }\left( h_{[\mu }^{\pi }A_{\nu \rho
]\pi }^{\left. {}\right. }\right) \left[ \partial ^{\rho }\left( h_{\tau
}^{[\xi }A_{\left. {}\right. }^{\mu \nu ]\tau }\right) -\frac{1}{3}\partial
^{\xi }\left( h_{\tau }^{[\mu }A_{\left. {}\right. }^{\nu \rho ]\tau
}\right) \right] \right\} ,  \label{soha}
\end{eqnarray}%
\begin{eqnarray}
&&S_{2}^{\mathrm{A-\psi }}=\frac{\mathrm{i}}{16}\int d^{11}x\left[
C^{\ast }C^{\mu }\bar{\xi}\gamma _{\mu }\xi -C^{\ast \mu }\left( C_{\mu \nu }%
\bar{\xi}\gamma ^{\nu }\xi +C^{\nu }\xi \gamma _{(\mu }\psi _{\nu )}\right)
\right.   \notag \\
&&\left. +C^{\ast \mu \nu }\left( A_{\mu \nu \rho }\bar{\xi}\gamma ^{\rho
}\xi -2C_{\mu }^{\quad \rho }\bar{\xi}\gamma _{(\nu }\psi _{\rho )}\right)
-3A^{\ast \mu \nu \rho }A_{\mu \nu }^{\quad \lambda }\bar{\xi}\gamma _{(\rho
}\psi _{\lambda )}\right] .  \label{soap}
\end{eqnarray}%
According to (\ref{rdf7}) only the last three components from the
second-order deformation change. Thus, the Rarita-Schwinger contribution
passes into
\begin{eqnarray}
&&\hat{S}_{2}^{\mathrm{\psi }}=-\frac{1}{2^{5}}\int d^{11}x\left\{
\mathrm{i}\xi ^{\ast }\left( \gamma ^{\mu \nu }\xi \bar{\xi}\gamma
_{\mu }\psi _{\nu }+\frac{1}{16}\psi ^{\mu }\bar{\xi}\gamma _{\mu
}\xi \right)
\right.   \notag \\
&&-\frac{1}{3^{2}\cdot 2^{4}}\left( 11\cdot 17\psi _{\nu }^{\ast }\gamma
^{\mu }\bar{\psi}^{\ast \nu }-29\psi _{\nu }^{\ast }\gamma ^{\mu \nu \rho }%
\bar{\psi}_{\rho }^{\ast }-62\psi _{\nu }^{\ast }\gamma ^{\nu }\bar{\psi}%
^{\ast \mu }\right) \bar{\xi}\gamma _{\mu }\xi   \notag \\
&&-\frac{1}{3^{2}\cdot 2^{3}}\left( 14\cdot 37\psi _{\mu }^{\ast }\bar{\psi}%
_{\nu }^{\ast }+7\psi ^{\ast \rho }\gamma _{\mu \nu \rho \lambda }\bar{\psi}%
^{\ast \lambda }-68\psi _{\mu }^{\ast }\gamma _{\nu \rho }\bar{\psi}^{\ast
\rho }\right.   \notag \\
&&\left. -29\psi ^{\ast \rho }\gamma _{\mu \nu }\bar{\psi}_{\rho }^{\ast
}\right) \bar{\xi}\gamma ^{\mu \nu }\xi +\frac{1}{3^{3}\cdot 2^{3}}\left(
\psi ^{\ast \varepsilon }\gamma _{\mu \nu \rho \lambda \sigma }\bar{\psi}%
_{\varepsilon }^{\ast }-56\psi _{\mu }^{\ast }\gamma _{\nu \rho \lambda }%
\bar{\psi}_{\sigma }^{\ast }\right.   \notag \\
&&\left. +\psi ^{\ast \varepsilon }\gamma _{\mu \nu \rho \lambda \sigma
\varepsilon \eta }\bar{\psi}^{\ast \eta }+14\psi _{\sigma }^{\ast }\gamma
_{\mu \nu \rho \lambda \varepsilon }\bar{\psi}^{\ast \varepsilon }\right)
\bar{\xi}\gamma ^{\mu \nu \rho \lambda \sigma }\xi +\mathrm{i}\psi ^{\ast
\mu }\gamma ^{\alpha \beta }\left[ \psi _{\mu }\bar{\xi}\gamma _{\alpha
}\psi _{\beta }\right.   \notag \\
&&\left. -\xi \left( \bar{\psi}_{\mu }\gamma _{\alpha }\psi _{\beta }+\frac{1%
}{2}\bar{\psi}_{\alpha }\gamma _{\mu }\psi _{\beta }\right) \right] +\frac{%
\mathrm{i}}{6}\left( \psi _{\lbrack \mu }^{\ast }\gamma _{\nu \rho \lambda
]}^{\left. {}\right. }\xi -\right.   \notag \\
&&\left. -\frac{1}{2}\psi ^{\ast \sigma }\gamma _{\mu \nu \rho \lambda
\sigma }\xi \right) \bar{\psi}^{\mu }\gamma ^{\nu \rho }\psi ^{\lambda }-%
\frac{\mathrm{i}}{16}\psi ^{\ast \mu }\psi ^{\nu }\bar{\xi}\gamma _{(\mu
}\psi _{\nu )}+\bar{\psi}^{\alpha }\gamma ^{\mu }\psi _{\mu }\bar{\psi}%
_{\alpha }\gamma ^{\nu }\psi _{\nu }  \notag \\
&&-\frac{1}{4}\bar{\psi}_{\alpha }\gamma _{\rho }\psi _{\beta }\left( \bar{%
\psi}^{\alpha }\gamma ^{\rho }\psi ^{\beta }+2\bar{\psi}^{\alpha }\gamma
^{\beta }\psi ^{\rho }+\frac{1}{2}\bar{\psi}_{\mu }\gamma ^{\mu \nu \rho
\alpha \beta }\psi _{\nu }\right)   \notag \\
&&\left. -\frac{1}{8}\bar{\psi}_{\mu }\gamma _{\nu \rho }\psi _{\lambda
}\left( \bar{\psi}^{[\mu }\gamma ^{\nu \rho }\psi ^{\lambda ]}+\frac{1}{2}%
\bar{\psi}_{\alpha }\gamma ^{\alpha \beta \mu \nu \rho \lambda }\psi _{\beta
}\right) \right\} ,  \label{sop}
\end{eqnarray}%
the mixed graviton-gravitini piece takes the form
\begin{eqnarray}
&&\hat{S}_{2}^{\mathrm{h-\psi }}=\int d^{11}x\left\{ \frac{1}{16}%
\left[ \xi ^{\ast }\gamma _{\mu \nu }\xi \left( h_{\lambda }^{\mu }\partial
^{\lbrack \nu }\eta ^{\lambda ]}+\eta ^{\lambda }\partial _{\left. {}\right.
}^{[\mu }h_{\lambda }^{\nu ]}\right) \right. \right.   \notag \\
&&\left. -\mathrm{i}\eta ^{\ast \mu }\bar{\xi}\gamma _{(\mu }\psi _{\nu
)}\eta ^{\nu }\right] -\xi ^{\ast }\left( \partial _{\lbrack \mu }\bar{\psi}%
_{\nu ]}\right) \eta ^{\mu }\eta ^{\nu }+\frac{1}{8}\xi ^{\ast }\gamma
^{\alpha \beta }\psi ^{\mu }\eta _{\mu }\partial _{\lbrack \alpha }\eta
_{\beta ]}  \notag \\
&&-\frac{1}{2}\xi ^{\ast }\left( \partial _{\mu }\xi \right) \eta _{\nu
}h^{\mu \nu }+\frac{\mathrm{i}}{8}h^{\ast \mu \nu }h_{\mu }^{\rho }\bar{\xi}%
\gamma _{(\nu }\psi _{\rho )}+\frac{3}{8}\psi ^{\ast \mu }\left( \partial
_{\rho }\xi \right) h_{\mu \nu }h^{\nu \rho }  \notag \\
&&+\frac{1}{8}\psi ^{\ast \lbrack \mu }\psi ^{\nu ]}\left( h_{\mu }^{\rho
}\partial _{\lbrack \nu }\eta _{\rho ]}-\eta ^{\rho }\partial _{\lbrack \mu
}h_{\nu ]\rho }\right) -\frac{1}{2}\psi ^{\ast \mu }\left( \partial _{\rho
}\psi _{\mu }\right) \eta _{\nu }h^{\nu \rho }  \notag \\
&&+\frac{1}{16}\psi ^{\ast \mu }\gamma ^{\alpha \beta }\psi _{\mu }\left(
h_{\alpha }^{\rho }\partial _{\lbrack \beta }\eta _{\rho ]}-\eta ^{\rho
}\partial _{\lbrack \alpha }h_{\beta ]\rho }\right)   \notag \\
&&+\frac{1}{8}\psi ^{\ast \lambda }\gamma ^{\mu \nu }\xi \left( h_{\lambda
}^{\rho }\partial _{\mu }h_{\nu \rho }-h_{\mu }^{\rho }\partial _{\lbrack
\nu }h_{\rho ]\lambda }-\frac{1}{2}h_{\mu }^{\rho }\partial _{\lambda
}h_{\nu \rho }\right)   \notag \\
&&+\frac{\mathrm{i}}{64}\bar{\psi}^{\mu }\gamma _{\mu \nu \rho \lambda
\sigma }\psi ^{\nu }h^{\rho \omega }\partial ^{\lbrack \lambda }h_{\omega
}^{\sigma ]}+\frac{\mathrm{i}}{8}\bar{\psi}_{\mu }\gamma ^{\mu \nu \rho
}\left( \partial _{\nu }\psi _{\rho }\right) \left( h^{\lambda \sigma
}h_{\lambda \sigma }-\frac{h^{2}}{2}\right)   \notag \\
&&-\frac{\mathrm{i}}{16}\bar{\psi}_{\mu }\gamma ^{\mu \nu \rho }\psi
^{\lambda }h\partial _{\lbrack \nu }h_{\rho ]\lambda }+\frac{\mathrm{i}}{8}%
\bar{\psi}_{\mu }\gamma ^{\mu \nu \rho }\left( \partial _{\lambda }\psi
_{\rho }\right) hh_{\nu }^{\lambda }  \notag \\
&&+\frac{\mathrm{i}}{16}\bar{\psi}^{\alpha }\gamma ^{\rho }\psi ^{\beta }%
\left[ h\left( \partial _{\lbrack \alpha }h_{\beta ]\rho }-2\sigma _{\rho
\beta }\partial _{\lbrack \alpha }h_{\lambda ]}^{\quad \lambda }\right)
-\left( h_{\rho }^{\lambda }\partial _{\lbrack \alpha }h_{\beta ]\lambda
}\right. \right.   \notag \\
&&\left. \left. -h_{\lambda \lbrack \alpha }\partial _{\beta ]}h_{\rho
}^{\lambda }+h_{\lambda \lbrack \alpha }\partial ^{\lambda }h_{\beta ]\rho }+%
\frac{1}{2}\left( \partial _{\rho }h_{\lambda \lbrack \alpha }^{\left.
{}\right. }\right) h_{\beta ]}^{\lambda }\right) \right]   \notag \\
&&+\frac{\mathrm{i}}{8}\bar{\psi}^{\mu }\gamma ^{\sigma }\psi _{\sigma }%
\left[ h^{\nu \rho }\partial _{\lbrack \mu }h_{\nu ]\rho }-h_{\rho \lbrack
\mu }\partial _{\nu ]}h^{\rho \nu }+h_{\rho \lbrack \mu }^{\left. {}\right.
}\partial ^{\rho }h_{\nu ]}^{\nu }+\frac{1}{2}\left( \partial ^{\nu }h_{\rho
\lbrack \mu }^{\left. {}\right. }\right) h_{\nu ]}^{\rho }\right]   \notag \\
&&-\frac{\mathrm{i}}{8}\bar{\psi}_{\alpha }\gamma ^{\alpha \beta \gamma }%
\left[ \left( \partial ^{\mu }\psi ^{\nu }\right) h_{\beta \mu }h_{\gamma
\nu }-h_{\beta \lambda }\partial ^{\lambda }\left( h_{\gamma \sigma }\psi
^{\sigma }\right) +\frac{3}{2}\left( \partial _{\mu }\psi _{\gamma }\right)
h_{\beta \rho }h^{\rho \mu }\right.   \notag \\
&&\left. \left. -\frac{1}{2}\psi _{\lambda }h^{\rho \lambda }\partial
_{\beta }h_{\gamma \rho }-\frac{3}{2}\psi _{\sigma }h_{\gamma \lambda
}\partial _{\beta }h^{\lambda \sigma }\right] \right\} ,  \label{rdf8b}
\end{eqnarray}%
and the terms expressing the simultaneous interactions among all the three
types of fields amount to%
\begin{eqnarray}
&&\hat{S}_{2}^{\mathrm{int}}=\tilde{k}_{i}\int d^{11}x\left\{ \frac{%
\mathrm{i}}{72}F^{\mu \nu \rho \lambda }\left( \xi ^{\ast }\gamma _{\mu \nu
\rho \lambda \sigma }\xi +8\sigma _{\mu \sigma }\xi ^{\ast }\gamma _{\mu \nu
\rho }\xi \right) \eta ^{\sigma }\right.   \notag \\
&&-\frac{\mathrm{i}}{18}\psi ^{\ast \lbrack \mu }\gamma ^{\nu \rho \lambda
]}\xi \left[ F_{\mu \nu \rho \sigma }h_{\lambda }^{\sigma }-3\partial _{\mu
}\left( h_{\nu }^{\sigma }A_{\rho \lambda \sigma }\right) \right]   \notag \\
&&+\frac{\mathrm{i}}{36}\psi _{\mu }^{\ast }\gamma ^{\mu \nu \rho \lambda
\sigma }\xi \left[ F_{\nu \rho \lambda \varepsilon }h_{\sigma }^{\varepsilon
}-3\partial _{\nu }\left( A_{\rho \lambda \varepsilon }h_{\sigma
}^{\varepsilon }\right) \right]   \notag \\
&&-\frac{1}{8}hF_{\mu \nu \rho \lambda }\left( \bar{\psi}^{\mu }\gamma ^{\nu
\rho }\psi ^{\lambda }+\frac{1}{12}\bar{\psi}_{\alpha }\gamma ^{\alpha \beta
\mu \nu \rho \lambda }\psi _{\beta }\right)   \notag \\
&&\left. +\frac{1}{12}\left( \bar{\psi}^{[\mu }\gamma ^{\nu \rho }\psi
^{\lambda ]}+\frac{1}{2}\bar{\psi}_{\alpha }\gamma ^{\alpha \beta \mu \nu
\rho \lambda }\psi _{\beta }\right) \left[ F_{\mu \nu \rho \sigma
}h_{\lambda }^{\sigma }-3\partial _{\mu }\left( h_{\nu }^{\sigma }A_{\rho
\lambda \sigma }\right) \right] \right\} .  \label{rdf8a}
\end{eqnarray}%
In deriving formula (\ref{sop}) we used the identity%
\begin{eqnarray}
&&\bar{\psi}_{\alpha }\gamma _{\rho }\psi _{\beta }\bar{\psi}_{\mu }\gamma
^{\mu \nu \rho \alpha \beta }\psi _{\nu }+\frac{1}{2}\bar{\psi}_{\mu }\gamma
_{\nu \rho }\psi _{\lambda }\bar{\psi}_{\alpha }\gamma ^{\alpha \beta \mu
\nu \rho \lambda }\psi _{\beta }  \notag \\
&&+\frac{1}{36}\left[ \frac{1}{2}\left( 4\bar{\psi}_{\mu }\gamma ^{\mu \nu
\rho \lambda \alpha }\psi ^{\sigma }\bar{\psi}_{\alpha }\gamma _{\nu \rho
\lambda \sigma \beta }\psi ^{\beta }+\bar{\psi}^{\mu }\gamma _{\mu \nu \rho
\lambda \sigma }\psi ^{\nu }\bar{\psi}_{\alpha }\gamma ^{\alpha \beta \rho
\lambda \sigma }\psi _{\beta }\right. \right.   \notag \\
&&\left. +\bar{\psi}^{\mu }\gamma ^{\alpha \beta \rho \lambda \sigma }\psi
^{\nu }\bar{\psi}_{\alpha }\gamma _{\mu \nu \rho \lambda \sigma }\psi
_{\beta }\right) +21\left( -4\bar{\psi}_{\mu }\gamma ^{\mu \nu }\psi ^{\rho }%
\bar{\psi}^{\sigma }\gamma _{\sigma \rho }\psi _{\nu }\right.   \notag \\
&&\left. \left. +\bar{\psi}^{\mu }\gamma _{\mu \nu }\psi ^{\nu }\bar{\psi}%
_{\alpha }\gamma ^{\alpha \beta }\psi _{\beta }+\bar{\psi}^{\mu }\gamma
^{\alpha \beta }\psi ^{\nu }\bar{\psi}_{\alpha }\gamma _{\mu \nu }\psi
_{\beta }\right) \right] =0.  \label{abra}
\end{eqnarray}

\section{Analysis of the deformed theory. Uniqueness of $D=11$, $N=1$ SUGRA}

In Ref.~\cite{epjccornea} (Section 5) it has been shown that the
local BRST cohomologies of the Pauli-Fierz model and respectively of
the linearized version of vielbein formulation of spin-two field
theory are isomorphic. Because the local BRST cohomology (in ghost
numbers zero and one) controls the deformation procedure, it results
that this isomorphism allows one to pass in a consistent manner from
the Pauli-Fierz version to the linearized version of the vielbein
formulation and conversely during the deformation procedure.
Nevertheless, the linearized vielbein formulation possesses more
fields (the antisymmetric part of the linearized vielbein) and more
gauge parameters (Lorentz parameters) than the Pauli-Fierz model,
such that the switch from the former version to the latter is
realized via the above mentioned isomorphism by imposing some
partial gauge-fixing conditions,
which come from the more general ones \cite{siegelfields}%
\begin{equation}
\bar{\delta}_{\epsilon }\sigma _{\mu \lbrack a}e_{b]}^{\;\;\mu }=0.
\label{xx15}
\end{equation}%
In the context of this larger partial gauge-fixing, simple computations lead
to the vielbein fields $e_{a}^{\;\;\mu }$, their inverses $e_{\;\;\mu }^{a}$%
, the inverse of their determinant $e$, and the components of the spin
connection $\omega _{\mu ab}$ (up to the second order in the coupling
constant) in terms of the Pauli-Fierz field as
\begin{eqnarray}
e_{a}^{\;\;\mu } &=&\overset{(0)}{e}_{a}^{\;\;\mu }+\lambda \overset{(1)}{e}%
_{a}^{\;\;\mu }+\lambda ^{2}\overset{(2)}{e}_{a}^{\;\;\mu }+\cdots =\delta
_{a}^{\;\;\mu }-\frac{\lambda }{2}h_{a}^{\;\;\mu }+\frac{3\lambda ^{2}}{8}%
h_{a}^{\;\;\rho }h_{\rho }^{\;\;\mu }+\cdots ,  \label{id1a} \\
e_{\;\;\mu }^{a} &=&\overset{(0)}{e}_{\;\;\mu }^{a}+\lambda \overset{(1)}{e}%
_{\;\;\mu }^{a}+\lambda ^{2}\overset{(2)}{e}_{\;\;\mu }^{a}+\cdots =\delta
_{\;\;\mu }^{a}+\frac{\lambda }{2}h_{\;\;\mu }^{a}-\frac{\lambda ^{2}}{8}%
h_{\;\;\rho }^{a}h_{\;\;\mu }^{\rho }+\cdots ,  \label{id1b} \\
e &=&\overset{(0)}{e}+\lambda \overset{(1)}{e}+\lambda ^{2}\overset{(2)}{e}%
+\cdots =1+\frac{\lambda }{2}h+\frac{\lambda ^{2}}{8}\left( h^{2}-2h_{\mu
\nu }h^{\mu \nu }\right) +\cdots ,  \label{id2} \\
\omega _{\mu ab} &=&\lambda \overset{\left( 1\right) }{\omega }_{\mu
ab}+\lambda ^{2}\overset{\left( 2\right) }{\omega }_{\mu ab}+\cdots ,
\label{uv2a}
\end{eqnarray}%
where
\begin{equation}
\overset{\left( 1\right) }{\omega }_{\mu ab}=-\partial _{\lbrack a}h_{b]\mu
},  \label{uv4}
\end{equation}%
\begin{equation}
\overset{\left( 2\right) }{\omega }_{\mu ab}=-\frac{1}{4}\left[
2h_{c[a}\left( \partial _{b]}h_{\;\;\mu }^{c}\right) -2h_{\left[ a\right.
}^{\;\;\;\nu }\partial _{\nu }h_{\left. b\right] \mu }-\left( \partial _{\mu
}h_{[a}^{\;\;\;\nu }\right) h_{b]\nu }\right] .  \label{uv5}
\end{equation}%
Based on these isomorphisms, we can further pass to the analysis of the
deformed theory obtained in the previous sections.

The component of antighost number equal to zero present in $\hat{S}_{1}$ is
precisely the interacting Lagrangian at order one in the coupling constant
\begin{eqnarray}
&&\mathcal{L}_{1}=\mathcal{L}_{1}^{\mathrm{EH}}+\mathcal{L}_{1}^{\mathrm{h-A}%
}+\left[ -\frac{\mathrm{i}}{4}h\bar{\psi}_{\mu }\gamma ^{\mu \nu \rho
}\partial _{\nu }\psi _{\rho }\right] +\left[ -\frac{\mathrm{i}}{4}\bar{\psi}%
^{\sigma }\gamma ^{\mu \nu \rho }\left( \partial _{\nu }\psi _{\rho }\right)
h_{\mu \sigma }\right]   \notag \\
&&+\left[ \frac{\mathrm{i}}{4}\bar{\psi}^{\sigma }\gamma ^{\mu \nu \rho
}\left( \partial _{\nu }\psi _{\rho }\right) h_{\mu \sigma }+\frac{\mathrm{i}%
}{4}\bar{\psi}_{\mu }\gamma ^{\mu \nu \rho }\left( \partial ^{\lambda }\psi
_{\rho }\right) h_{\nu \lambda }+\frac{\mathrm{i}}{4}\bar{\psi}_{\mu }\gamma
^{\mu \nu \rho }\left( \partial _{\nu }\psi ^{\lambda }\right) h_{\rho
\lambda }\right]   \notag \\
&&+\left[ \frac{\mathrm{i}}{8}\left( \bar{\psi}^{\mu }\gamma ^{\lambda }\psi
^{\nu }-2\sigma ^{\nu \lambda }\bar{\psi}^{\mu }\gamma ^{\rho }\psi _{\rho
}\right) \partial _{\lbrack \mu }h_{\nu ]\lambda }\right]   \notag \\
&&+\left[ -\frac{\mathrm{i}}{8}\left( 2\bar{\psi}_{\mu }\gamma ^{\mu \nu
\rho }\left( \partial _{\nu }\psi ^{\lambda }\right) h_{\rho \lambda }+\bar{%
\psi}_{\rho }\gamma ^{\rho \mu \nu }\psi ^{\lambda }\partial _{\lbrack \mu
}h_{\nu ]\lambda }\right) \right]   \notag \\
&&\left[ -\frac{\tilde{k}_{i}}{48}F_{\mu \nu \rho \lambda }\left( \bar{\psi}%
_{\alpha }\gamma ^{\alpha \beta \mu \nu \rho \lambda }\psi _{\beta }+2\bar{%
\psi}^{[\mu }\gamma ^{\nu \rho }\psi ^{\lambda ]}\right) \right]   \notag \\
&\equiv &\mathcal{L}_{1}^{\mathrm{EH}}+\mathcal{L}_{1}^{\mathrm{h-A}}+\left[
-\frac{\mathrm{i}}{2}\overset{(1)}{e}\overset{(0)}{\bar{\psi}}_{\mu }\overset%
{(0)}{\Gamma }^{\mu \nu \rho }\overset{(0)}{D}_{\nu }\left( \frac{\Omega +%
\hat{\Omega}}{2}\right) \overset{(0)}{\psi }_{\rho }\right]   \notag \\
&&+\left[ -\frac{\mathrm{i}}{2}\overset{(0)}{e}\overset{(1)}{\bar{\psi}}%
_{\mu }\overset{(0)}{\Gamma }^{\mu \nu \rho }\overset{(0)}{D}_{\nu }\left(
\frac{\Omega +\hat{\Omega}}{2}\right) \overset{(0)}{\psi }_{\rho }\right]
\notag \\
&&+\left[ -\frac{\mathrm{i}}{2}\overset{(0)}{e}\overset{(0)}{\bar{\psi}}%
_{\mu }\overset{(1)}{\Gamma }^{\mu \nu \rho }\overset{(0)}{D}_{\nu }\left(
\frac{\Omega +\hat{\Omega}}{2}\right) \overset{(0)}{\psi }_{\rho }\right]
\notag \\
&&+\left[ -\frac{\mathrm{i}}{2}\overset{(0)}{e}\overset{(0)}{\bar{\psi}}%
_{\mu }\overset{(0)}{\Gamma }^{\mu \nu \rho }\overset{(1)}{D}_{\nu }\left(
\frac{\Omega +\hat{\Omega}}{2}\right) \overset{(0)}{\psi }_{\rho }\right]
\notag \\
&&+\left[ -\frac{\mathrm{i}}{2}\overset{(0)}{e}\overset{(0)}{\bar{\psi}}%
_{\mu }\overset{(0)}{\Gamma }^{\mu \nu \rho }\overset{(0)}{D}_{\nu }\left(
\frac{\Omega +\hat{\Omega}}{2}\right) \overset{(1)}{\psi }_{\rho }\right]
\notag \\
&&+\left[ -\frac{\tilde{k}_{i}}{48}\overset{(0)}{e}\overset{\left( 0\right) }%
{\bar{F}}_{\mu \nu \rho \lambda }\left( \overset{(0)}{\bar{\psi}}_{\alpha }%
\overset{(0)}{\Gamma }^{\alpha \beta \mu \nu \rho \lambda }\overset{(0)}{%
\psi }_{\beta }+2\overset{(0)}{\bar{\psi}}^{[\mu }\overset{(0)}{\Gamma }%
^{\nu \rho }\overset{(0)}{\psi }^{\lambda ]}\right) \right] ,  \label{id1}
\end{eqnarray}%
where $\mathcal{L}_{1}^{\mathrm{h-A}}$ and $\overset{\left( 0\right) }{\bar{F%
}}_{\mu \nu \rho \lambda }$ are respectively listed in formulas (124) and
(126) from Ref.~\cite{pI} (with $q\rightarrow q_{i}$ and $q_{i}$ as in (\ref%
{solntr2})). In the above we also made the notations%
\begin{equation}
\overset{(0)}{D}_{\mu }\left( \frac{\Omega +\hat{\Omega}}{2}\right)
=\partial _{\mu },  \label{dc0}
\end{equation}%
\begin{equation}
\overset{(1)}{D}_{\mu }\left( \frac{\Omega +\hat{\Omega}}{2}\right) =\frac{1%
}{16}\left( \overset{\left( 1\right) }{\Omega }_{\mu ab}+\overset{\left(
1\right) }{\hat{\Omega}}_{\mu ab}\right) \gamma ^{ab},  \label{dc1}
\end{equation}%
where $\overset{\left( n\right) }{\Omega }_{\mu ab}$ and $\overset{\left(
n\right) }{\hat{\Omega}}_{\mu ab}$ ($n\geq 1$) are the net contributions of
the quantities
\begin{eqnarray}
\Omega _{\mu ab} &=&\omega _{\mu ab}+K_{\mu ab}\equiv \omega _{\mu ab}+\frac{%
\mathrm{i}\lambda ^{2}}{16}e_{\;\;\mu }^{m}\left[ \bar{\psi}^{n}\gamma
_{mnpab}\psi ^{p}+2\left( \bar{\psi}_{a}\gamma _{m}\psi _{b}+\bar{\psi}%
_{m}\gamma _{\lbrack a}\psi _{b]}\right) \right] ,  \label{conn1} \\
\hat{\Omega}_{\mu ab} &=&\Omega _{\mu ab}-\frac{\mathrm{i}\lambda ^{2}}{16}%
e_{\;\;\mu }^{m}\bar{\psi}^{n}\gamma _{mnpab}\psi ^{p}\equiv \omega _{\mu
ab}+\frac{\mathrm{i}\lambda ^{2}}{8}e_{\;\;\mu }^{m}\left( \bar{\psi}%
_{a}\gamma _{m}\psi _{b}+\bar{\psi}_{m}\gamma _{\lbrack a}\psi _{b]}\right)
\label{conn2}
\end{eqnarray}%
to order $n$ of perturbation theory, with $\omega _{\mu ab}$ given in (\ref%
{uv2a}). Notations $\overset{(n)}{\Gamma }^{\alpha _{1}\cdots \alpha _{k}}$ (%
$k\leq 6$) signify the net contributions of the matrices%
\begin{equation}
\Gamma ^{\alpha _{1}\cdots \alpha _{k}}=e_{a_{1}}^{\;\;\alpha _{1}}\cdots
e_{a_{k}}^{\;\;\alpha _{k}}\gamma ^{a_{1}\cdots a_{k}},  \label{gamacurb}
\end{equation}%
again to order $n$ of perturbation theory and $\overset{(0)}{\psi }_{\mu
}=\psi _{m}$ means the zero-order approximation of the curved spin-vector%
\begin{equation}
\psi _{\mu }=e_{\;\;\mu }^{m}\psi _{m}.  \label{spincurb}
\end{equation}%
Along the same line, the piece of antighost number equal to zero from the
second-order deformation offers us the interacting Lagrangian at order two
in the coupling constant $\mathcal{L}_{2}$
\begin{equation}
\mathcal{L}_{2}=\mathcal{L}_{2}^{\mathrm{EH}}+\mathcal{L}_{2}^{\mathrm{h-A}%
}+T_{0}+T_{1}+T_{2}+T_{3}+T_{4}+T_{5}+T_{6},  \label{ide2}
\end{equation}%
where%
\begin{eqnarray}
T_{0} &=&\left[ \frac{1}{2^{5}}\bar{\psi}_{\mu }\gamma ^{\rho }\psi _{\rho }%
\bar{\psi}^{\mu }\gamma ^{\lambda }\psi _{\lambda }-\frac{1}{2^{9}}\left(
\bar{\psi}^{\mu }\gamma _{\mu \nu \rho \lambda \alpha \beta }\psi ^{\nu
}+2\left( \bar{\psi}_{\alpha }\gamma _{\rho }\psi _{\beta }+\bar{\psi}_{\rho
}\gamma _{\lbrack \alpha }\psi _{\beta ]}\right) \right) \times \right.
\notag \\
&&\left. \times \left( -\bar{\psi}_{\lambda }\gamma ^{\rho \lambda \sigma
\alpha \beta }\psi _{\sigma }+2\left( \bar{\psi}_{\alpha }\gamma _{\rho
}\psi _{\beta }+\bar{\psi}_{\rho }\gamma _{\lbrack \alpha }\psi _{\beta
]}\right) \right) \right]   \notag \\
&\equiv &\left[ -\frac{1}{2}\overset{(0)}{e}_{a}^{\;\;\mu }\overset{(0)}{e}%
_{b}^{\;\;\nu }\left( \overset{(2)}{K}_{\mu }^{\;\;ac}\overset{(2)}{K}_{\nu
\;\;\;c}^{\;\;b}-\overset{(2)}{K}_{\nu }^{\;\;ac}\overset{(2)}{K}_{\mu
\;\;\;c}^{\;\;b}\right) \right] ,  \label{t0}
\end{eqnarray}%
\begin{eqnarray}
T_{1} &=&\left[ -\frac{\mathrm{i}}{16}\bar{\psi}_{\mu }\gamma ^{\mu \nu \rho
}\left( \partial _{\nu }\psi _{\rho }\right) \left( h^{2}-2h_{\alpha \beta
}h^{\alpha \beta }\right) \right] +\left[ -\frac{\mathrm{i}}{8}\bar{\psi}%
^{\sigma }\gamma ^{\mu \nu \rho }\left( \partial _{\nu }\psi _{\rho }\right)
hh_{\mu \sigma }\right]   \notag \\
&&+\left[ \frac{\mathrm{i}}{8}h\left( \bar{\psi}^{\sigma }\gamma ^{\mu \nu
\rho }\left( \partial _{\nu }\psi _{\rho }\right) h_{\mu \sigma }+\bar{\psi}%
_{\mu }\gamma ^{\mu \nu \rho }\left( \partial ^{\lambda }\psi _{\rho
}\right) h_{\nu \lambda }+\bar{\psi}_{\mu }\gamma ^{\mu \nu \rho }\left(
\partial _{\nu }\psi ^{\lambda }\right) h_{\rho \lambda }\right) \right]
\notag \\
&&+\left[ \frac{\mathrm{i}}{16}\left( \bar{\psi}^{\mu }\gamma ^{\lambda
}\psi ^{\nu }-2\sigma ^{\nu \lambda }\bar{\psi}^{\mu }\gamma ^{\rho }\psi
_{\rho }\right) h\partial _{\lbrack \mu }h_{\nu ]\lambda }\right]   \notag \\
&&+\left[ -\frac{\mathrm{i}}{16}h\left( 2\bar{\psi}_{\mu }\gamma ^{\mu \nu
\rho }\left( \partial _{\nu }\psi ^{\lambda }\right) h_{\rho \lambda }+\bar{%
\psi}_{\rho }\gamma ^{\rho \mu \nu }\psi ^{\lambda }\partial _{\lbrack \mu
}h_{\nu ]\lambda }\right) \right]   \notag \\
&\equiv &\left[ -\frac{\mathrm{i}}{2}\overset{(2)}{e}\overset{(0)}{\bar{\psi}%
}_{\mu }\overset{(0)}{\Gamma }^{\mu \nu \rho }\overset{(0)}{D}_{\nu }\left(
\frac{\Omega +\hat{\Omega}}{2}\right) \overset{(0)}{\psi }_{\rho }\right]
\notag \\
&&+\left[ -\frac{\mathrm{i}}{2}\overset{(1)}{e}\overset{(1)}{\bar{\psi}}%
_{\mu }\overset{(0)}{\Gamma }^{\mu \nu \rho }\overset{(0)}{D}_{\nu }\left(
\frac{\Omega +\hat{\Omega}}{2}\right) \overset{(0)}{\psi }_{\rho }\right]
\notag \\
&&+\left[ -\frac{\mathrm{i}}{2}\overset{(1)}{e}\overset{(0)}{\bar{\psi}}%
_{\mu }\overset{(1)}{\Gamma }^{\mu \nu \rho }\overset{(0)}{D}_{\nu }\left(
\frac{\Omega +\hat{\Omega}}{2}\right) \overset{(0)}{\psi }_{\rho }\right]
\notag \\
&&+\left[ -\frac{\mathrm{i}}{2}\overset{(1)}{e}\overset{(0)}{\bar{\psi}}%
_{\mu }\overset{(0)}{\Gamma }^{\mu \nu \rho }\overset{(1)}{D}_{\nu }\left(
\frac{\Omega +\hat{\Omega}}{2}\right) \overset{(0)}{\psi }_{\rho }\right]
\notag \\
&&+\left[ -\frac{\mathrm{i}}{2}\overset{(1)}{e}\overset{(0)}{\bar{\psi}}%
_{\mu }\overset{(0)}{\Gamma }^{\mu \nu \rho }\overset{(0)}{D}_{\nu }\left(
\frac{\Omega +\hat{\Omega}}{2}\right) \overset{(1)}{\psi }_{\rho }\right] ,
\label{t1}
\end{eqnarray}%
\begin{eqnarray}
T_{2} &=&\left[ \frac{\mathrm{i}}{16}\bar{\psi}^{\sigma }\gamma ^{\mu \nu
\rho }\left( \partial _{\nu }\psi _{\rho }\right) h_{\sigma }^{\lambda
}h_{\mu \lambda }\right]   \notag \\
&&+\left[ \frac{\mathrm{i}}{8}\bar{\psi}_{\mu }\gamma ^{\alpha \beta \gamma
}\left( \partial _{\nu }\psi _{\rho }\right) h_{\sigma }^{\mu }\left(
h_{\alpha }^{\sigma }\delta _{\beta }^{\nu }\delta _{\gamma }^{\rho }+\delta
_{\alpha }^{\sigma }h_{\beta }^{\nu }\delta _{\gamma }^{\rho }+\delta
_{\alpha }^{\sigma }\delta _{\beta }^{\nu }h_{\gamma }^{\rho }\right) \right]
\notag \\
&&+\left[ -\frac{\mathrm{i}}{16}\bar{\psi}_{\mu }\left( \delta _{\lambda
}^{\alpha }\delta _{\nu }^{\beta }\gamma _{\rho }+\delta _{\nu }^{\alpha
}\delta _{\rho }^{\beta }\gamma _{\lambda }+\delta _{\rho }^{\alpha }\delta
_{\lambda }^{\beta }\gamma _{\nu }\right) \psi _{\rho }h_{\lambda }^{\mu
}\partial _{\lbrack \alpha }^{\left. {}\right. }h_{\beta ]}^{\nu }\right.
\notag \\
&&\left. +\frac{\mathrm{i}}{16}\bar{\psi}_{\mu }\sigma ^{\alpha \lbrack
\lambda }\gamma ^{\nu \rho ]\beta }\psi _{\rho }h_{\lambda }^{\mu }\partial
_{\lbrack \alpha }h_{\beta ]\nu }\right] +\left[ -\frac{\mathrm{i}}{8}h_{\mu
\sigma }\bar{\psi}^{\sigma }\gamma ^{\mu \nu \rho }\partial _{\nu }\left(
h_{\rho \lambda }\psi ^{\lambda }\right) \right]   \notag \\
&\equiv &\left[ -\frac{\mathrm{i}}{2}\overset{(0)}{e}\overset{(2)}{\bar{\psi}%
}_{\mu }\overset{(0)}{\Gamma }^{\mu \nu \rho }\overset{(0)}{D}_{\nu }\left(
\frac{\Omega +\hat{\Omega}}{2}\right) \overset{(0)}{\psi }_{\rho }\right]
\notag \\
&&+\left[ -\frac{\mathrm{i}}{2}\overset{(0)}{e}\overset{(1)}{\bar{\psi}}%
_{\mu }\overset{(1)}{\Gamma }^{\mu \nu \rho }\overset{(0)}{D}_{\nu }\left(
\frac{\Omega +\hat{\Omega}}{2}\right) \overset{(0)}{\psi }_{\rho }\right]
\notag \\
&&+\left[ -\frac{\mathrm{i}}{2}\overset{(0)}{e}\overset{(1)}{\bar{\psi}}%
_{\mu }\overset{(0)}{\Gamma }^{\mu \nu \rho }\overset{(1)}{D}_{\nu }\left(
\frac{\Omega +\hat{\Omega}}{2}\right) \overset{(0)}{\psi }_{\rho }\right]
\notag \\
&&+\left[ -\frac{\mathrm{i}}{2}\overset{(0)}{e}\overset{(1)}{\bar{\psi}}%
_{\mu }\overset{(0)}{\Gamma }^{\mu \nu \rho }\overset{(0)}{D}_{\nu }\left(
\frac{\Omega +\hat{\Omega}}{2}\right) \overset{(1)}{\psi }_{\rho }\right] ,
\label{t2}
\end{eqnarray}%
\begin{eqnarray}
T_{3} &=&\left[ -\frac{\mathrm{i}}{16}\bar{\psi}_{\mu }\gamma ^{\alpha \beta
\gamma }\left( \partial _{\nu }\psi _{\rho }\right) \left( 3h^{\mu \lambda
}h_{\lambda \alpha }\delta _{\beta }^{\nu }\delta _{\gamma }^{\rho }+3\delta
_{\alpha }^{\mu }h^{\nu \lambda }h_{\lambda \beta }\delta _{\gamma }^{\rho
}\right. \right.   \notag \\
&&\left. \left. +3\delta _{\alpha }^{\mu }\delta _{\beta }^{\nu }h^{\rho
\lambda }h_{\lambda \gamma }+2h_{\alpha }^{\mu }h_{\beta }^{\nu }\delta
_{\gamma }^{\rho }+2h_{\alpha }^{\mu }\delta _{\beta }^{\nu }h_{\gamma
}^{\rho }+2\delta _{\alpha }^{\mu }h_{\beta }^{\nu }h_{\gamma }^{\rho
}\right) \right]   \notag \\
&&+\left[ \frac{\mathrm{i}}{16}\bar{\psi}_{\mu }\left( \delta _{\alpha
}^{\lambda }\delta _{\nu }^{\sigma }\gamma _{\rho }+\delta _{\nu }^{\lambda
}\delta _{\rho }^{\sigma }\gamma _{\alpha }+\delta _{\rho }^{\lambda }\delta
_{\alpha }^{\sigma }\gamma _{\nu }\right) \psi ^{\rho }h^{\mu \alpha
}\partial _{\lbrack \lambda }^{\left. {}\right. }h_{\sigma ]}^{\nu }\right.
\notag \\
&&+\frac{\mathrm{i}}{16}\bar{\psi}^{\alpha }\left( \delta _{\alpha
}^{\lambda }\delta _{\nu }^{\sigma }\gamma _{\rho }+\delta _{\nu }^{\lambda
}\delta _{\rho }^{\sigma }\gamma _{\alpha }+\delta _{\rho }^{\lambda }\delta
_{\alpha }^{\sigma }\gamma _{\nu }\right) \psi ^{\rho }h^{\mu \nu }\partial
_{\lbrack \lambda }h_{\sigma ]\mu }  \notag \\
&&+\frac{\mathrm{i}}{16}\bar{\psi}^{\alpha }\left( \delta _{\alpha
}^{\lambda }\delta _{\nu }^{\sigma }\gamma _{\rho }+\delta _{\nu }^{\lambda
}\delta _{\rho }^{\sigma }\gamma _{\alpha }+\delta _{\rho }^{\lambda }\delta
_{\alpha }^{\sigma }\gamma _{\nu }\right) \psi _{\mu }h^{\mu \rho }\partial
_{\lbrack \lambda }^{\left. {}\right. }h_{\sigma ]}^{\nu }  \notag \\
&&\left. +\frac{\mathrm{i}}{32}\bar{\psi}_{\mu }\gamma ^{\mu \nu \rho \alpha
\beta }\psi _{\nu }h_{\rho }^{\lambda }\partial _{\lbrack \alpha }h_{\beta
]\lambda }\right]   \notag \\
&&+\left[ \frac{\mathrm{i}}{8}\bar{\psi}_{\mu }\gamma ^{\alpha \beta \gamma
}\left( h_{\alpha }^{\mu }\delta _{\beta }^{\nu }\delta _{\gamma }^{\rho
}+\delta _{\alpha }^{\mu }h_{\beta }^{\nu }\delta _{\gamma }^{\rho }+\delta
_{\alpha }^{\mu }\delta _{\beta }^{\nu }h_{\gamma }^{\rho }\right) \partial
_{\nu }\left( h_{\rho \lambda }\psi ^{\lambda }\right) \right]   \notag \\
&\equiv &\left[ -\frac{\mathrm{i}}{2}\overset{(0)}{e}\overset{(0)}{\bar{\psi}%
}_{\mu }\overset{(2)}{\Gamma }^{\mu \nu \rho }\overset{(0)}{D}_{\nu }\left(
\frac{\Omega +\hat{\Omega}}{2}\right) \overset{(0)}{\psi }_{\rho }\right]
\notag \\
&&+\left[ -\frac{\mathrm{i}}{2}\overset{(0)}{e}\overset{(0)}{\bar{\psi}}%
_{\mu }\overset{(1)}{\Gamma }^{\mu \nu \rho }\overset{(1)}{D}_{\nu }\left(
\frac{\Omega +\hat{\Omega}}{2}\right) \overset{(0)}{\psi }_{\rho }\right]
\notag \\
&&+\left[ -\frac{\mathrm{i}}{2}\overset{(0)}{e}\overset{(0)}{\bar{\psi}}%
_{\mu }\overset{(1)}{\Gamma }^{\mu \nu \rho }\overset{(0)}{D}_{\nu }\left(
\frac{\Omega +\hat{\Omega}}{2}\right) \overset{(1)}{\psi }_{\rho }\right] ,
\label{t3}
\end{eqnarray}%
\begin{eqnarray}
T_{4} &=&\left[ \frac{\mathrm{i}}{32}\left( \bar{\psi}^{\mu }\gamma ^{\rho
}\psi ^{\nu }-2\bar{\psi}^{\mu }\gamma ^{\sigma }\psi _{\sigma }\sigma ^{\nu
\rho }-\frac{1}{2}\bar{\psi}_{\alpha }\gamma ^{a\beta \mu \nu \rho }\psi
_{\beta }\right) \times \right.   \notag \\
&&\times \left( 2h_{\lambda \lbrack \mu }\partial _{\nu ]}h_{\rho }^{\lambda
}-2h_{\lambda \lbrack \mu }\partial ^{\lambda }h_{\nu ]\rho }-\left(
\partial _{\rho }h_{[\mu }^{\lambda }\right) h_{\nu ]\lambda }^{\left.
{}\right. }-\frac{\mathrm{i}}{4}\left( \frac{1}{2}\bar{\psi}^{\lambda
}\gamma _{\mu \nu \rho \lambda \sigma }\psi ^{\sigma }\right. \right.
\notag \\
&&\left. \left. \left. +2\left( \bar{\psi}_{\mu }\gamma _{\rho }\psi _{\nu }+%
\bar{\psi}_{\rho }\gamma _{\lbrack \mu }\psi _{\nu ]}\right) \right) \right) %
\right]   \notag \\
&&\left[ -\frac{\mathrm{i}}{16}\bar{\psi}^{\mu }\left( \delta _{\mu
}^{\alpha }\delta _{\nu }^{\beta }\gamma _{\rho }+\delta _{\nu }^{\alpha
}\delta _{\rho }^{\beta }\gamma _{\mu }+\delta _{\rho }^{\alpha }\delta
_{\mu }^{\beta }\gamma _{\nu }\right) \psi _{\lambda }h^{\rho \lambda
}\partial _{\lbrack \alpha }^{\left. {}\right. }h_{\beta ]}^{\nu }\right.
\notag \\
&&\left. +\frac{\mathrm{i}}{16}\bar{\psi}_{\mu }\sigma ^{\alpha \lbrack \mu
}\gamma ^{\nu \rho ]\beta }\psi ^{\lambda }h_{\rho \lambda }\partial
_{\lbrack \alpha }h_{\beta ]\nu }\right] +\left[ \frac{\mathrm{i}}{16}\bar{%
\psi}_{\mu }\gamma ^{\mu \nu \rho }\partial _{\nu }\left( h_{\rho \lambda
}h^{\lambda \sigma }\psi _{\sigma }\right) \right]   \notag \\
&\equiv &\left[ -\frac{\mathrm{i}}{2}\overset{(0)}{e}\overset{(0)}{\bar{\psi}%
}_{\mu }\overset{(0)}{\Gamma }^{\mu \nu \rho }\overset{(2)}{D}_{\nu }\left(
\frac{\Omega +\hat{\Omega}}{2}\right) \overset{(0)}{\psi }_{\rho }\right]
\notag \\
&&+\left[ -\frac{\mathrm{i}}{2}\overset{(0)}{e}\overset{(0)}{\bar{\psi}}%
_{\mu }\overset{(0)}{\Gamma }^{\mu \nu \rho }\overset{(1)}{D}_{\nu }\left(
\frac{\Omega +\hat{\Omega}}{2}\right) \overset{(1)}{\psi }_{\rho }\right]
\notag \\
&&+\left[ -\frac{\mathrm{i}}{2}\overset{(0)}{e}\overset{(0)}{\bar{\psi}}%
_{\mu }\overset{(0)}{\Gamma }^{\mu \nu \rho }\overset{(0)}{D}_{\nu }\left(
\frac{\Omega +\hat{\Omega}}{2}\right) \overset{(2)}{\psi }_{\rho }\right] ,
\label{t4}
\end{eqnarray}%
\begin{eqnarray}
T_{5} &=&\frac{1}{3\cdot 2^{10}}\bar{\psi}_{[\mu }\gamma _{\nu \rho }\psi
_{\lambda ]}\left( \bar{\psi}_{\alpha }\gamma ^{\alpha \beta \mu \nu \rho
\lambda }\psi _{\beta }+2\bar{\psi}^{[\mu }\gamma ^{\nu \rho }\psi ^{\lambda
]}\right)   \notag \\
&\equiv &-\frac{\tilde{k}_{i}^{2}}{3\cdot 2^{5}}\overset{(0)}{e}\overset{(0)}%
{\bar{\psi}}_{[\mu }\overset{(0)}{\Gamma }_{\nu \rho }\overset{(0)}{\psi }%
_{\lambda ]}\left( \overset{(0)}{\bar{\psi}}_{\alpha }\overset{(0)}{\Gamma }%
^{\alpha \beta \mu \nu \rho \lambda }\overset{(0)}{\psi }_{\beta }+2\overset{%
(0)}{\bar{\psi}}^{[\mu }\overset{(0)}{\Gamma }^{\nu \rho }\overset{(0)}{\psi
}^{\lambda ]}\right) ,  \label{t5}
\end{eqnarray}%
\begin{eqnarray}
T_{6} &=&\left[ -\frac{\tilde{k}_{i}}{3\cdot 2^{5}}hF_{\mu \nu \rho \lambda
}\left( \bar{\psi}_{\alpha }\gamma ^{\alpha \beta \mu \nu \rho \lambda }\psi
_{\beta }+2\bar{\psi}^{[\mu }\gamma ^{\nu \rho }\psi ^{\lambda ]}\right) %
\right]   \notag \\
&&+\left[ \frac{\tilde{k}_{i}}{3\cdot 2^{3}}F_{\mu \nu \rho \sigma
}h_{\lambda }^{\sigma }\left( \bar{\psi}_{\alpha }\gamma ^{\alpha \beta \mu
\nu \rho \lambda }\psi _{\beta }+2\bar{\psi}^{[\mu }\gamma ^{\nu \rho }\psi
^{\lambda ]}\right) \right]   \notag \\
&&+\left[ -\frac{\tilde{k}_{i}}{2^{3}}\partial _{\mu }\left( h_{\nu
}^{\sigma }A_{\rho \lambda \sigma }\right) \left( \bar{\psi}_{\alpha }\gamma
^{\alpha \beta \mu \nu \rho \lambda }\psi _{\beta }+2\bar{\psi}^{[\mu
}\gamma ^{\nu \rho }\psi ^{\lambda ]}\right) \right]   \notag \\
&\equiv &\left[ -\frac{\tilde{k}_{i}}{48}\overset{(1)}{e}\overset{\left(
0\right) }{\bar{F}}_{\mu \nu \rho \lambda }\left( \overset{(0)}{\bar{\psi}}%
_{\alpha }\overset{(0)}{\Gamma }^{\alpha \beta \mu \nu \rho \lambda }\overset%
{(0)}{\psi }_{\beta }+2\overset{(0)}{\bar{\psi}}^{[\mu }\overset{(0)}{\Gamma
}^{\nu \rho }\overset{(0)}{\psi }^{\lambda ]}\right) \right]   \notag \\
&&+\left[ -\frac{\tilde{k}_{i}}{48}\overset{(0)}{e}\overset{\left( 0\right) }%
{\bar{F}}_{\mu \nu \rho \lambda }\left( \overset{(1)}{\bar{\psi}}_{\alpha }%
\overset{(0)}{\Gamma }^{\alpha \beta \mu \nu \rho \lambda }\overset{(0)}{%
\psi }_{\beta }+\overset{(0)}{\bar{\psi}}_{\alpha }\overset{(1)}{\Gamma }%
^{\alpha \beta \mu \nu \rho \lambda }\overset{(0)}{\psi }_{\beta }\right.
\right.   \notag \\
&&+\overset{(0)}{\bar{\psi}}_{\alpha }\overset{(0)}{\Gamma }^{\alpha \beta
\mu \nu \rho \lambda }\overset{(1)}{\psi }_{\beta }+2\overset{(1)}{\bar{\psi}%
}^{[\mu }\overset{(0)}{\Gamma }^{\nu \rho }\overset{(0)}{\psi }^{\lambda ]}+2%
\overset{(0)}{\bar{\psi}}^{[\mu }\overset{(1)}{\Gamma }^{\nu \rho }\overset{%
(0)}{\psi }^{\lambda ]}  \notag \\
&&\left. \left. +2\overset{(0)}{\bar{\psi}}^{[\mu }\overset{(0)}{\Gamma }%
^{\nu \rho }\overset{(1)}{\psi }^{\lambda ]}\right) \right]   \notag \\
&&+\left[ -\frac{\tilde{k}_{i}}{48}\overset{(0)}{e}\overset{\left( 1\right) }%
{\bar{F}}_{\mu \nu \rho \lambda }\left( \overset{(0)}{\bar{\psi}}_{\alpha }%
\overset{(0)}{\Gamma }^{\alpha \beta \mu \nu \rho \lambda }\overset{(0)}{%
\psi }_{\beta }+2\overset{(0)}{\bar{\psi}}^{[\mu }\overset{(0)}{\Gamma }%
^{\nu \rho }\overset{(0)}{\psi }^{\lambda ]}\right) \right] ,  \label{t6}
\end{eqnarray}%
with $\mathcal{L}_{2}^{\mathrm{h-A}}$ given in formula (128) from
Ref.~\cite{pI}
(with $q\rightarrow q_{i}$ and $q_{i}$ as in (\ref{solntr2})), and%
\begin{equation}
\overset{(2)}{D}_{\mu }\left( \frac{\Omega +\hat{\Omega}}{2}\right) =\frac{1%
}{16}\left( \overset{\left( 2\right) }{\Omega }_{\mu ab}+\overset{\left(
2\right) }{\hat{\Omega}}_{\mu ab}\right) \gamma ^{ab}.  \label{uv6}
\end{equation}%
We observe from (\ref{t0}) that only now, in the presence of
\emph{all} fields, the quartic gravitini vertex is permitted, by
contrast to the results from Refs.~\cite{pII} and \cite{pIII}, where
it has been shown that gravitini allow no self-interactions in
$D=11$ if separately coupled to a graviton or respectively to a
three-form gauge field.

Relying on (\ref{id1}) and (\ref{ide2}), we observe that the first orders of
the interacting Lagrangian, $\mathcal{L}_{0}+\lambda \mathcal{L}_{1}+\lambda
^{2}\mathcal{L}_{2}+\cdots $, come from the expansion of the following
Lagrangian (expressed in terms of the `curved' spin-vector $\psi _{\mu }$
and the field strength of the `curved' $3$-form $\bar{A}_{\mu \nu \rho }$)%
\begin{eqnarray}
&&\mathcal{L}=\frac{2}{\lambda ^{2}}eR\left( \Omega \left( e\right) \right) -%
\frac{\mathrm{i}e}{2}\bar{\psi}_{\mu }\Gamma ^{\mu \nu \rho }D_{\nu }\left(
\frac{\Omega +\hat{\Omega}}{2}\right) \psi _{\rho }-\frac{e}{48}\bar{F}_{\mu
\nu \rho \lambda }\bar{F}^{\mu \nu \rho \lambda }  \notag \\
&&-\frac{\lambda \tilde{k}_{i}}{96}e\left( \bar{F}_{\mu \nu \rho \lambda }+%
\hat{F}_{\mu \nu \rho \lambda }\right) \left( \bar{\psi}_{\alpha }\Gamma
^{\alpha \beta \mu \nu \rho \lambda }\psi _{\beta }+2\bar{\psi}^{[\mu
}\Gamma ^{\nu \rho }\psi ^{\lambda ]}\right)  \notag \\
&&-\frac{4\lambda \tilde{k}_{i}}{\left( 12\right) ^{4}}\varepsilon ^{\mu
_{1}\mu _{2}\cdots \mu _{11}}\bar{A}_{\mu _{1}\mu _{2}\mu _{3}}\bar{F}_{\mu
_{4}\cdots \mu _{7}}\bar{F}_{\mu _{8}\cdots \mu _{11}}.  \label{lagfull}
\end{eqnarray}%
The notation $D_{\mu }\left( \frac{\Omega +\hat{\Omega}}{2}\right) \psi
_{\rho }$ denotes the full covariant derivatives of $\psi _{\rho }$%
\begin{equation}
D_{\mu }\left( \frac{\Omega +\hat{\Omega}}{2}\right) \psi _{\rho }=\partial
_{\mu }\psi _{\rho }+\frac{1}{8}\left( \frac{\Omega _{\mu ab}+\hat{\Omega}%
_{\mu ab}}{2}\right) \gamma ^{ab}\psi _{\rho }  \label{xx20}
\end{equation}%
and
\begin{equation}
\hat{F}_{\mu \nu \rho \lambda }=\bar{F}_{\mu \nu \rho \lambda }+\lambda
\tilde{k}_{i}\bar{\psi}_{[\mu }\Gamma _{\nu \rho }\psi _{\lambda ]}.
\label{superfield}
\end{equation}%
The field strength $\bar{F}_{\mu \nu \rho \lambda }$ reads as in
relation (130) from Ref.~\cite{pI} and the Levi-Civita symbol
$\varepsilon ^{\mu _{1}\mu _{2}\cdots \mu _{11}}$ is defined via
formula (132) from the same reference.

At the first sight it seems that we obtained two different interacting
theories, respectively corresponding to the two different values of $\tilde{k%
}$ and $q$ from (\ref{solntr2}), $\tilde{k}_{1}=\frac{\mathrm{i}\sqrt{2}}{8}$%
, $q_{1}=-\frac{\mathrm{i}\sqrt{2}}{2\cdot \left( 12\right) ^{4}}$ and $%
\tilde{k}_{2}=-\frac{\mathrm{i}\sqrt{2}}{8}$, $q_{2}=\frac{\mathrm{i}\sqrt{2}%
}{2\cdot \left( 12\right) ^{4}}$ respectively. Nevertheless, this is not the
case since the two models are correlated through the transformation $\bar{A}%
_{\mu \nu \rho }\longrightarrow -\bar{A}_{\mu \nu \rho }$, so (\ref{lagfull}%
) is the $D=11$, $N=1$ SUGRA Lagrangian for both choices (see also Refs.~\cite%
{invarianceD=11} and \cite{nieuwen}).

The pieces linear in the antifields from the deformed solution to the master
equation give us the deformed gauge transformations for the original fields
(the indexes $\mu $, $\nu $, $\alpha $, $\beta $, $\gamma $ are flat) as
\begin{eqnarray}
&&\bar{\delta}_{\epsilon ,\varepsilon }h_{\mu \nu }=\partial _{(\mu
}\epsilon _{\nu )}+\lambda \left[ \frac{1}{2}h_{\rho (\mu }\partial _{\nu
)}\epsilon ^{\rho }-\frac{1}{2}\epsilon ^{\rho }\partial _{(\mu }h_{\nu
)\rho }+\epsilon ^{\rho }\partial _{\rho }h_{\mu \nu }+\frac{\mathrm{i}}{8}%
\bar{\varepsilon}\gamma _{(\mu }\psi _{\nu )}\right]  \notag \\
&&+\lambda ^{2}\left[ \frac{3}{8}\epsilon ^{\rho }h_{\rho }^{\lambda
}\partial _{(\mu }h_{\nu )\lambda }-\frac{1}{2}\epsilon ^{\rho }h_{\rho
}^{\lambda }\partial _{\lambda }h_{\mu \nu }-\frac{1}{8}h_{\rho (\mu
}\partial _{\nu )}\left( h^{\rho \lambda }\epsilon _{\lambda }\right) \right.
\notag \\
&&\left. +\frac{\mathrm{i}}{16}h_{(\mu }^{\rho }\bar{\varepsilon}\gamma
_{\nu )}^{\left. {}\right. }\psi _{\rho }+\frac{\mathrm{i}}{16}\bar{%
\varepsilon}\gamma _{\rho }\psi _{(\mu }^{\left. {}\right. }h_{\nu )}^{\rho }%
\right]  \notag \\
&&+\cdots  \notag \\
&=&\overset{(0)}{\bar{\delta}}_{\epsilon ,\varepsilon }h_{\mu \nu }+\lambda
\overset{(1)}{\bar{\delta}}_{\epsilon ,\varepsilon }h_{\mu \nu }+\lambda ^{2}%
\overset{(2)}{\bar{\delta}}_{\epsilon ,\varepsilon }h_{\mu \nu }+\cdots ,
\label{trgraviton}
\end{eqnarray}%
\begin{eqnarray}
&&\bar{\delta}_{\epsilon ,\varepsilon }\psi _{\mu }=\partial _{\mu
}\varepsilon +\lambda \left[ -\frac{1}{2}h_{\mu }^{\nu }\partial _{\nu
}\varepsilon +\left( \partial _{\alpha }\psi _{\mu }\right) \epsilon
^{\alpha }+\frac{1}{2}\psi ^{\nu }\partial _{\lbrack \mu }\epsilon _{\nu
]}\right.  \notag \\
&&+\frac{1}{8}\gamma ^{\alpha \beta }\psi _{\mu }\partial _{\lbrack \alpha
}\epsilon _{\beta ]}-\frac{1}{8}\gamma ^{\alpha \beta }\varepsilon \partial
_{\lbrack \alpha }h_{\beta ]\mu }+\frac{\mathrm{i}\tilde{k}_{i}}{9}\left(
\gamma ^{\nu \rho \lambda }\varepsilon F_{\mu \nu \rho \lambda }\right.
\notag \\
&&\left. \left. -\frac{1}{8}\gamma _{\mu \nu \rho \lambda \sigma
}\varepsilon F^{\nu \rho \lambda \sigma }\right) \right] +\lambda ^{2}\left[
\frac{3}{8}h_{\mu \nu }h^{\nu \rho }\partial _{\rho }\varepsilon -\frac{1}{32%
}\gamma ^{\alpha \beta }\varepsilon \left( -2h_{\mu }^{\rho }\partial
_{\lbrack \alpha }h_{\beta ]\rho }\right. \right.  \notag \\
&&\left. +2h_{\rho \lbrack \alpha }\left( \partial _{\beta ]}h_{\mu }^{\rho
}\right) -2h_{\rho \lbrack \alpha }\left( \partial ^{\rho }h_{\beta ]\mu
}\right) -\left( \partial _{\mu }h_{\rho \lbrack \alpha }^{\left. {}\right.
}\right) h_{\beta ]}^{\rho }\right)  \notag \\
&&+\frac{1}{16}\gamma ^{\rho \lambda }\psi _{\mu }\left( h_{\rho }^{\sigma
}\partial _{\lbrack \lambda }\epsilon _{\sigma ]}-\epsilon ^{\sigma
}\partial _{\lbrack \rho }h_{\lambda ]\sigma }\right) +\frac{1}{8}\psi
^{\rho }\left( h_{\mu }^{\lambda }\partial _{\lbrack \rho }\epsilon
_{\lambda ]}-h_{\rho }^{\lambda }\partial _{\lbrack \mu }\epsilon _{\lambda
]}\right)  \notag \\
&&-\frac{1}{4}\psi ^{\nu }\epsilon ^{\rho }\partial _{\lbrack \mu }h_{\nu
]\rho }-\frac{1}{2}\left( \partial _{\alpha }\psi _{\mu }\right) \epsilon
_{\beta }h^{\alpha \beta }-\frac{\mathrm{i}}{16}\psi ^{\nu }\bar{\varepsilon}%
\gamma _{(\mu }\psi _{\nu )}  \notag \\
&&-\frac{\mathrm{i}}{64}\gamma ^{\alpha \beta }\psi _{\mu }\bar{\varepsilon}%
\gamma _{\lbrack \alpha }\psi _{\beta ]}+\frac{\mathrm{i}}{64}\gamma
^{\alpha \beta }\varepsilon \left( \bar{\psi}_{\alpha }\gamma _{\mu }\psi
_{\beta }+\bar{\psi}_{\mu }\gamma _{\lbrack \alpha }\psi _{\beta ]}\right)
\notag \\
&&-\frac{\mathrm{i}}{9\cdot 2^{5}}\gamma ^{\nu \rho \lambda }\varepsilon
\bar{\psi}_{[\mu }\gamma _{\nu \rho }\psi _{\lambda ]}+\frac{\mathrm{i}%
\tilde{k}_{i}}{18}\gamma ^{\nu \rho \lambda }\varepsilon \left( h_{[\mu
}^{\sigma }F_{\nu \rho \lambda ]\sigma }^{\left. {}\right. }+\partial
_{\lbrack \mu }\left( h_{\nu }^{\sigma }A_{\rho \lambda ]\sigma }\right)
\right)  \notag \\
&&\left. +\frac{\mathrm{i}}{9\cdot 2^{8}}\gamma _{\mu \nu \rho \lambda
\sigma }\varepsilon \bar{\psi}^{[\nu }\gamma ^{\rho \lambda }\psi ^{\sigma
]}-\frac{\mathrm{i}\tilde{k}_{i}}{144}\gamma _{\mu \nu \rho \lambda \sigma
}\varepsilon \left( h_{\varepsilon }^{[\nu }F_{\left. {}\right. }^{\rho
\lambda \sigma ]\varepsilon }+\partial ^{\lbrack \nu }\left( h_{\varepsilon
}^{\rho }A^{\lambda \sigma ]\varepsilon }\right) \right) \right]  \notag \\
&&+\cdots  \notag \\
&=&\overset{(0)}{\bar{\delta}}_{\epsilon ,\varepsilon }\psi _{\mu }+\lambda
\overset{(1)}{\bar{\delta}}_{\epsilon ,\varepsilon }\psi _{\mu }+\lambda ^{2}%
\overset{(2)}{\bar{\delta}}_{\epsilon ,\varepsilon }\psi _{\mu }+\cdots ,
\label{trgravitino}
\end{eqnarray}%
\begin{eqnarray}
&&\bar{\delta}_{\epsilon ,\varepsilon }A_{\alpha \beta \gamma }=\partial
_{\lbrack \alpha }\varepsilon _{\beta \gamma ]}+\lambda \left[ \epsilon
^{\delta }\partial _{\delta }A_{\alpha \beta \gamma }+\frac{1}{2}%
A_{\;\;[\alpha \beta }^{\delta }\delta _{\gamma ]}^{\sigma }\partial
_{\lbrack \sigma }\epsilon _{\delta ]}\right.  \notag \\
&&\left. -\frac{1}{2}\left( \partial ^{\delta }\varepsilon _{\lbrack \alpha
\beta }\right) h_{\gamma ]\delta }+\frac{1}{2}\varepsilon _{\;\;[\alpha
}^{\delta }\partial _{\beta }^{\left. {}\right. }h_{\gamma ]\delta }^{\left.
{}\right. }-\tilde{k}_{i}\bar{\xi}\gamma _{\lbrack \alpha \beta }\psi
_{\gamma ]}\right]  \notag \\
&&+\lambda ^{2}\left[ -\frac{1}{8}\varepsilon _{\;\;[\alpha }^{\delta
}\left( \partial _{\beta }^{\left. {}\right. }h_{\gamma ]}^{\sigma }\right)
h_{\delta \sigma }+\frac{3}{8}\varepsilon _{\;\;[\alpha }^{\delta }h_{\beta
}^{\sigma }\partial _{\gamma ]}^{\left. {}\right. }h_{\delta \sigma }\right.
\notag \\
&&+\frac{3}{8}\left( \partial ^{\delta }\varepsilon _{\lbrack \alpha \beta
}\right) h_{\gamma ]}^{\sigma }h_{\delta \sigma }-\frac{1}{4}\left( \partial
_{\delta }h_{[\alpha }^{\sigma }\right) h_{\beta }^{\delta }\varepsilon
_{\gamma ]\sigma }  \notag \\
&&-\frac{1}{8}A_{\;\;[\alpha \beta }^{\delta }\delta _{\gamma ]}^{\omega
}\left( \partial _{\lbrack \omega }\epsilon _{\sigma ]}\right) h_{\delta
}^{\sigma }+\frac{1}{8}A_{\;\;[\alpha \beta }^{\delta }h_{\gamma ]}^{\sigma
}\partial _{\lbrack \delta }\epsilon _{\sigma ]}  \notag \\
&&-\frac{1}{4}A_{\;\;[\alpha \beta }^{\delta }\delta _{\gamma ]}^{\omega
}\partial \left( _{\lbrack \omega }h_{\delta ]}^{\sigma }\right) \epsilon
_{\sigma }-\frac{1}{2}\left( \partial ^{\delta }A_{\alpha \beta \gamma
}\right) h_{\delta }^{\sigma }\epsilon _{\sigma }-\frac{\mathrm{i}}{16}%
A_{\delta \lbrack \alpha \beta }\bar{\varepsilon}\gamma _{\gamma ]}\psi
^{\delta }  \notag \\
&&\left. -\frac{\mathrm{i}}{16}\bar{\varepsilon}\gamma ^{\delta }\psi
_{\lbrack \alpha }A_{\beta \gamma ]\delta }\right] +\cdots  \notag \\
&=&\overset{\left( 0\right) }{\bar{\delta}}_{\epsilon ,\varepsilon
}A_{\alpha \beta \gamma }+\lambda \overset{\left( 1\right) }{\bar{\delta}}%
_{\epsilon ,\varepsilon }A_{\alpha \beta \gamma }+\lambda ^{2}\overset{%
\left( 2\right) }{\bar{\delta}}_{\epsilon ,\varepsilon }A_{\alpha \beta
\gamma }+\cdots .  \label{tr3f}
\end{eqnarray}%
If we introduce the notation%
\begin{equation}
g_{\mu \nu }=\sigma _{\mu \nu }+\lambda h_{\mu \nu },  \label{metrica}
\end{equation}%
then (\ref{trgraviton}) imply some gauge transformations for the metric
tensor of the form%
\begin{equation}
\frac{1}{\lambda }\bar{\delta}_{\epsilon ,\varepsilon }g_{\mu \nu }=\bar{%
\epsilon}_{(\mu ;\nu )}+\frac{\mathrm{i}\lambda }{8}\bar{\varepsilon}\Gamma
_{(\mu }\psi _{\nu )},  \label{trge}
\end{equation}%
where
\begin{equation}
\bar{\epsilon}_{\mu ;\nu }=\partial _{\mu }\bar{\epsilon}_{\nu }-\mathit{%
\Gamma }_{\mu \nu }^{\rho }\bar{\epsilon}_{\rho }.  \label{dercovpar}
\end{equation}%
Here, $\mathit{\Gamma }_{\mu \nu }^{\rho }$ are precisely the (affine)
connection coefficients associated with the metric (\ref{metrica})%
\begin{equation}
\mathit{\Gamma }_{\mu \nu }^{\rho }=g^{\rho \lambda }\mathit{\Gamma }%
_{\lambda \mu \nu },  \label{connmet2}
\end{equation}%
where $g^{\rho \lambda }$ are the elements of the inverse of (\ref{metrica}%
), and%
\begin{equation}
\mathit{\Gamma }_{\lambda \mu \nu }=\frac{1}{2}\left( \partial _{\mu
}g_{\nu \lambda }+\partial _{\nu }g_{\mu \lambda }-\partial
_{\lambda }g_{\mu \nu }\right)  \label{connmet1}
\end{equation}%
stand for the standard Christoffel symbols of the first kind. In (\ref{trge}%
) quantities $\bar{\epsilon}_{\mu }$ are the `curved' gauge parameters of
the spin-$2$ field%
\begin{eqnarray}
\bar{\epsilon}_{\mu } &=&e_{\;\;\mu }^{a}\epsilon _{a}=\overset{(0)}{\bar{%
\epsilon}}_{\mu }+\lambda \overset{(1)}{\bar{\epsilon}}_{\mu }+\lambda ^{2}%
\overset{(2)}{\bar{\epsilon}}_{\mu }+\cdots  \notag \\
&=&\left( \delta _{\;\;\mu }^{a}+\frac{\lambda }{2}h_{\;\;\mu }^{a}-\frac{%
\lambda ^{2}}{8}h_{\;\;\rho }^{a}h_{\;\;\mu }^{\rho }+\cdots \right)
\epsilon _{a}.  \label{parcurb}
\end{eqnarray}%
Using expansions (\ref{id1a})--(\ref{id1b}) and transformations (\ref%
{trgraviton}), one can show perturbatively that the gauge transformations of
the vielbein fields and of their inverses read as
\begin{equation}
\frac{1}{\lambda }\bar{\delta}_{\epsilon ,\varepsilon }e_{a}^{\;\;\mu }=\bar{%
\epsilon}^{\rho }\partial _{\rho }e_{a}^{\;\;\mu }-e_{a}^{\;\;\rho }\partial
_{\rho }\bar{\epsilon}^{\mu }+\epsilon _{a}^{\;\;b}e_{b}^{\;\;\mu }-\frac{%
\mathrm{i}\lambda }{8}\bar{\varepsilon}\Gamma ^{\mu }\psi _{a}  \label{id6}
\end{equation}%
and%
\begin{equation}
\frac{1}{\lambda }\bar{\delta}_{\epsilon ,\varepsilon }e_{\;\;\mu }^{a}=\bar{%
\epsilon}^{\rho }\partial _{\rho }e_{\;\;\mu }^{a}+e_{\;\;\rho }^{a}\partial
_{\mu }\bar{\epsilon}^{\rho }+\epsilon _{\;\;b}^{a}e_{\;\;\mu }^{b}+\frac{%
\mathrm{i}\lambda }{8}\bar{\varepsilon}\gamma ^{a}\psi _{\mu }  \label{id6a}
\end{equation}%
respectively. Indeed, the translation and rotation gauge parameters allow
the perturbative developments
\begin{equation}
\bar{\epsilon}^{\mu }=\overset{(0)}{\bar{\epsilon}}^{\mu }+\lambda \overset{%
(1)}{\bar{\epsilon}}^{\mu }+\cdots =\left( \delta _{a}^{\;\;\mu }-\frac{%
\lambda }{2}h_{a}^{\;\;\mu }+\cdots \right) \epsilon ^{a}\equiv
e_{a}^{\;\;\mu }\epsilon ^{a}  \label{uv15}
\end{equation}%
and%
\begin{eqnarray}
\epsilon _{ab} &=&\overset{(0)}{\epsilon }_{ab}+\lambda \overset{(1)}{%
\epsilon }_{ab}+\cdots =\frac{1}{2}\partial _{\lbrack a}\epsilon _{b]}
\notag \\
&&-\frac{\lambda }{4}\left( \epsilon ^{c}\partial _{\lbrack a}h_{b]c}-\frac{1%
}{2}h_{[a}^{c}\partial _{b]}\epsilon _{c}-\frac{1}{2}\left( \partial
_{c}\epsilon _{\lbrack a}\right) h_{b]}^{c}+\frac{\mathrm{i}}{4}\bar{%
\varepsilon}\gamma _{\lbrack a}\psi _{b]}\right) +\cdots  \label{uv12}
\end{eqnarray}%
respectively. Using now (\ref{id1b}) combined with (\ref{trgraviton}), it
follows that
\begin{equation}
\overset{(0)}{\bar{\delta}}_{\epsilon ,\varepsilon }e_{\;\;\mu }^{a}=\overset%
{(0)}{\bar{\delta}}_{\epsilon ,\varepsilon }\delta _{\;\;\mu }^{a}=0,
\label{tre0}
\end{equation}%
\begin{eqnarray}
&&\overset{(1)}{\bar{\delta}}_{\epsilon ,\varepsilon }e_{\;\;\mu }^{a}=\frac{%
1}{2}\overset{(0)}{\bar{\delta}}_{\epsilon ,\varepsilon }h_{\;\;\mu }^{a}=%
\frac{1}{2}\left( \partial ^{a}\epsilon ^{b}+\partial ^{b}\epsilon
^{a}\right) \sigma _{b\mu }  \notag \\
&=&0+\partial _{\mu }\epsilon ^{a}+\frac{1}{2}\partial ^{\lbrack a}\epsilon
^{b]}\sigma _{b\mu }  \notag \\
&=&\overset{(0)}{\bar{\epsilon}}^{\rho }\partial _{\rho }\overset{(0)}{e}%
_{\;\;\mu }^{a}+\overset{(0)}{e}_{\;\;\rho }^{a}\partial _{\mu }\overset{(0)}%
{\bar{\epsilon}}^{\rho }+\overset{(0)}{\epsilon }_{\;\;b}^{a}\overset{(0)}{e}%
_{\;\;\mu }^{b},  \label{tre1}
\end{eqnarray}%
\begin{eqnarray}
&&\overset{(2)}{\bar{\delta}}_{\epsilon ,\varepsilon }e_{\;\;\mu }^{a}=\frac{%
1}{2}\overset{(1)}{\bar{\delta}}_{\epsilon ,\varepsilon }h_{\;\;\mu }^{a}-%
\frac{1}{8}\left( \overset{(0)}{\bar{\delta}}_{\epsilon ,\varepsilon
}h_{\;\;\rho }^{a}\right) h_{\;\;\mu }^{\rho }-\frac{1}{8}h_{\;\;\rho }^{a}%
\overset{(0)}{\bar{\delta}}_{\epsilon ,\varepsilon }h_{\;\;\mu }^{\rho }
\notag \\
&=&\frac{1}{4}\left[ h_{m\mu }\partial ^{a}\epsilon ^{m}+h_{m}^{a}\partial
_{\mu }\epsilon ^{m}-\epsilon ^{m}\partial _{\mu }h_{m}^{a}-\epsilon
^{m}\partial ^{a}h_{m\mu }\right.  \notag \\
&&\left. +2\epsilon ^{m}\partial _{m}h_{\;\;\mu }^{a}+\frac{\mathrm{i}}{4}%
\bar{\varepsilon}\gamma ^{(a}\psi ^{b)}\sigma _{b\mu }\right]  \notag \\
&&-\frac{1}{8}\left( \partial ^{a}\epsilon _{m}+\partial _{m}\epsilon
^{a}\right) h_{\;\;\mu }^{m}-\frac{1}{8}h_{m}^{a}\left( \partial
^{m}\epsilon ^{n}+\partial ^{n}\epsilon ^{m}\right) \sigma _{n\mu }  \notag
\\
&=&\frac{1}{2}\epsilon ^{m}\partial _{m}h_{\;\;\mu }^{a}-\frac{1}{2}\partial
_{\mu }\left( h_{m}^{a}\epsilon ^{m}\right) +\frac{1}{2}h_{m}^{a}\partial
_{\mu }\epsilon ^{m}+\frac{1}{4}\partial ^{\lbrack a}\epsilon ^{b]}h_{m\mu
}+\left( -\frac{1}{4}\epsilon _{m}\partial ^{\lbrack a}h^{b]m}\right.  \notag
\\
&&\left. +\frac{1}{8}h^{m[a}\partial ^{b]}\epsilon _{m}+\frac{1}{8}\left(
\partial _{m}\epsilon ^{\lbrack a}\right) h^{b]m}-\frac{\mathrm{i}}{16}\bar{%
\varepsilon}\gamma ^{\lbrack a}\psi ^{b]}\right) \sigma _{b\mu }+\frac{%
\mathrm{i}}{8}\bar{\varepsilon}\gamma ^{a}\psi ^{b}\sigma _{b\mu }  \notag \\
&=&\overset{(0)}{\bar{\epsilon}}^{\rho }\partial _{\rho }\overset{(1)}{e}%
_{\;\;\mu }^{a}+\overset{(0)}{e}_{\;\;\rho }^{a}\partial _{\mu }\overset{(1)}%
{\bar{\epsilon}}^{\rho }+\overset{(1)}{e}_{\;\;\rho }^{a}\partial _{\mu }%
\overset{(0)}{\bar{\epsilon}}^{\rho }+\overset{(0)}{\epsilon }_{\;\;b}^{a}%
\overset{(1)}{e}_{\;\;\mu }^{b}  \notag \\
&&+\overset{(1)}{\epsilon }_{\;\;b}^{a}\overset{(0)}{e}_{\;\;\mu }^{b}+\frac{%
\mathrm{i}}{8}\bar{\varepsilon}\gamma ^{a}\overset{(0)}{\psi }_{\mu },
\label{tre2}
\end{eqnarray}%
and thus (\ref{tre0})--(\ref{tre2}) are nothing but the first three orders
of the gauge transformations (\ref{id6a}).

As we specified before, all the original fields bear flat indices, so in (%
\ref{trgraviton})--(\ref{tr3f}) $A_{\alpha \beta \gamma }$ means $A_{abc}$
and $\psi _{\mu }$ is $\psi _{m}$. The first three orders of the gauge
transformations for the gravitini, (\ref{trgravitino}), can be put under the
form%
\begin{equation}
\overset{\left( 0\right) }{\bar{\delta}}_{\epsilon ,\varepsilon }\psi _{m}=%
\overset{(0)}{e}_{m}^{\;\;\mu }\overset{(0)}{D}_{\mu }\left( \hat{\Omega}%
\right) \varepsilon ,  \label{uw0}
\end{equation}%
\begin{eqnarray}
&&\overset{\left( 1\right) }{\bar{\delta}}_{\epsilon ,\varepsilon }\psi _{m}=%
\overset{(1)}{e}_{m}^{\;\;\mu }\overset{(0)}{D}_{\mu }\left( \hat{\Omega}%
\right) \varepsilon +\overset{(0)}{e}_{m}^{\;\;\mu }\overset{(1)}{D}_{\mu
}\left( \hat{\Omega}\right) \varepsilon +\left( \partial _{\mu }\psi
_{m}\right) \overset{(0)}{\bar{\epsilon}}^{\mu }+\overset{(0)}{\epsilon }%
_{mn}\psi ^{n}  \notag \\
&&+\frac{1}{4}\gamma ^{ab}\psi _{m}\overset{(0)}{\epsilon }_{ab}+\frac{%
\mathrm{i}\tilde{k}_{i}}{9}\overset{(0)}{e}_{m}^{\;\;\mu }\overset{(0)}{%
\Gamma }^{\nu \rho \lambda }\varepsilon \overset{(0)}{F}_{\mu \nu \rho
\lambda }-\frac{\mathrm{i}\tilde{k}_{i}}{72}\overset{(0)}{e}_{m}^{\;\;\mu }%
\overset{(0)}{\Gamma }_{\mu \nu \rho \lambda \sigma }\varepsilon \overset{(0)%
}{F}^{\nu \rho \lambda \sigma },  \label{uw1}
\end{eqnarray}%
\begin{eqnarray}
&&\overset{\left( 2\right) }{\bar{\delta}}_{\epsilon ,\varepsilon }\psi _{m}=%
\overset{(2)}{e}_{m}^{\;\;\mu }\overset{(0)}{D}_{\mu }\left( \hat{\Omega}%
\right) \varepsilon +\overset{(1)}{e}_{m}^{\;\;\mu }\overset{(1)}{D}_{\mu
}\left( \hat{\Omega}\right) \varepsilon +\overset{(0)}{e}_{m}^{\;\;\mu }%
\overset{(2)}{D}_{\mu }\left( \hat{\Omega}\right) \varepsilon +\left(
\partial _{\mu }\psi _{m}\right) \overset{(1)}{\bar{\epsilon}}^{\mu }  \notag
\\
&&+\overset{(1)}{\epsilon }_{mn}\psi ^{n}+\frac{1}{4}\gamma ^{ab}\psi _{m}%
\overset{(1)}{\epsilon }_{ab}+\frac{\mathrm{i}}{8}\left( \bar{\varepsilon}%
\gamma ^{a}\psi _{m}\right) \psi _{a}+\frac{\mathrm{i}\tilde{k}_{i}}{9}%
\overset{(1)}{e}_{m}^{\;\;\mu }\overset{(0)}{\Gamma }^{\nu \rho \lambda
}\varepsilon \overset{(0)}{F}_{\mu \nu \rho \lambda }  \notag \\
&&+\frac{\mathrm{i}\tilde{k}_{i}}{9}\overset{(0)}{e}_{m}^{\;\;\mu }\overset{%
(1)}{\Gamma }^{\nu \rho \lambda }\varepsilon \overset{(0)}{F}_{\mu \nu \rho
\lambda }+\frac{\mathrm{i}\tilde{k}_{i}}{9}\overset{(0)}{e}_{m}^{\;\;\mu }%
\overset{(0)}{\Gamma }^{\nu \rho \lambda }\varepsilon \overset{(1)}{F}_{\mu
\nu \rho \lambda }-\frac{\mathrm{i}\tilde{k}_{i}}{72}\overset{(1)}{e}%
_{m}^{\;\;\mu }\overset{(0)}{\Gamma }_{\mu \nu \rho \lambda \sigma
}\varepsilon \overset{(0)}{F}^{\nu \rho \lambda \sigma }  \notag \\
&&-\frac{\mathrm{i}\tilde{k}_{i}}{72}\overset{(0)}{e}_{m}^{\;\;\mu }\overset{%
(1)}{\Gamma }_{\mu \nu \rho \lambda \sigma }\varepsilon \overset{(0)}{F}%
^{\nu \rho \lambda \sigma }-\frac{\mathrm{i}\tilde{k}_{i}}{72}\overset{(0)}{e%
}_{m}^{\;\;\mu }\overset{(0)}{\Gamma }_{\mu \nu \rho \lambda \sigma
}\varepsilon \overset{(1)}{F}^{\nu \rho \lambda \sigma }  \notag \\
&&-\frac{\mathrm{i}}{9\cdot 2^{5}}\gamma ^{abc}\varepsilon \left( \bar{\psi}%
_{[m}\gamma _{ab}\psi _{c]}\right) +\frac{\mathrm{i}}{9\cdot 2^{8}}\gamma
_{mabcd}\varepsilon \left( \bar{\psi}^{[a}\gamma ^{bc}\psi ^{d]}\right) ,
\label{uw2}
\end{eqnarray}%
where
\begin{equation}
D_{\mu }\left( \hat{\Omega}\right) \varepsilon =\partial _{\mu }\varepsilon +%
\frac{1}{8}\hat{\Omega}_{\mu ab}\gamma ^{ab}\varepsilon .
\label{derivcovarmaj}
\end{equation}%
Consequently, we can state that formulas (\ref{uw0})--(\ref{uw2}) originate
in the perturbative expansion of the expression%
\begin{eqnarray}
&&\bar{\delta}_{\epsilon ,\varepsilon }\psi _{m}=e_{m}^{\;\;\mu }D_{\mu
}\left( \hat{\Omega}\right) \varepsilon +\lambda \left[ \left( \partial
_{\mu }\psi _{m}\right) \bar{\epsilon}^{\mu }+\epsilon _{mn}\psi ^{n}+\frac{1%
}{4}\gamma ^{ab}\psi _{m}\epsilon _{ab}\right.  \notag \\
&&\left. +\frac{\mathrm{i}\tilde{k}_{i}}{9}e_{m}^{\;\;\mu }\Gamma ^{\nu \rho
\lambda }\varepsilon \hat{F}_{\mu \nu \rho \lambda }-\frac{\mathrm{i}\tilde{k%
}_{i}}{72}e_{m}^{\;\;\mu }\Gamma _{\mu \nu \rho \lambda \sigma }\varepsilon
\hat{F}^{\nu \rho \lambda \sigma }+\frac{\mathrm{i}\lambda }{8}\left( \bar{%
\varepsilon}\gamma ^{a}\psi _{m}\right) \psi _{a}\right] .
\label{trfullmajflat}
\end{eqnarray}%
Taking into account (\ref{id6}), from (\ref{trfullmajflat}) we deduce the
form of the gauge transformations for `curved' gravitini, $\psi _{\mu }=\psi
_{m}e_{\;\;\mu }^{m}$, as%
\begin{eqnarray}
&&\bar{\delta}_{\epsilon ,\varepsilon }\psi _{\mu }=D_{\mu }\left( \hat{%
\Omega}\right) \varepsilon +\lambda \left[ \left( \partial _{\rho }\psi
_{\mu }\right) \bar{\epsilon}^{\rho }+\psi _{\rho }\partial _{\mu }\bar{%
\epsilon}^{\rho }+\frac{1}{4}\gamma ^{ab}\psi _{\mu }\epsilon _{ab}\right.
\notag \\
&&\left. +\frac{\mathrm{i}\tilde{k}_{i}}{9}\Gamma ^{\nu \rho \lambda
}\varepsilon \hat{F}_{\mu \nu \rho \lambda }-\frac{\mathrm{i}\tilde{k}_{i}}{%
72}\Gamma _{\mu \nu \rho \lambda \sigma }\varepsilon \hat{F}^{\nu \rho
\lambda \sigma }\right] .  \label{trfullmajcurb}
\end{eqnarray}%
We reprise the same procedure with respect to the $3$-form. The first three
orders of (\ref{tr3f}) (with $\alpha \beta \gamma \rightarrow abc$) can be
organized as%
\begin{equation}
\overset{\left( 0\right) }{\bar{\delta}}_{\epsilon ,\varepsilon
}A_{abc}=\left( \overset{(0)}{D}_{\mu }\left( \omega \right) \varepsilon
_{\lbrack ab}^{\left. {}\right. }\right) \overset{(0)}{e}_{c]}^{\;\;\mu },
\label{ap03f}
\end{equation}%
\begin{eqnarray}
&&\overset{\left( 1\right) }{\bar{\delta}}_{\epsilon ,\varepsilon
}A_{abc}=\left( \overset{(1)}{D}_{\mu }\left( \omega \right) \varepsilon
_{\lbrack ab}^{\left. {}\right. }\right) \overset{(0)}{e}_{c]}^{\;\;\mu
}+\left( \overset{(0)}{D}_{\mu }\left( \omega \right) \varepsilon _{\lbrack
ab}^{\left. {}\right. }\right) \overset{(1)}{e}_{c]}^{\;\;\mu }  \notag \\
&&+\left( \partial _{\mu }A_{abc}\right) \overset{(0)}{\bar{\epsilon}}^{\mu
}+A_{\;\;[ab}^{m}\overset{(0)}{\epsilon }_{c]m}-\tilde{k}_{i}\bar{\varepsilon%
}\gamma _{\lbrack ab}\psi _{c]},  \label{ap13f}
\end{eqnarray}%
\begin{eqnarray}
&&\overset{\left( 2\right) }{\bar{\delta}}_{\epsilon ,\varepsilon
}A_{abc}=\left( \overset{(2)}{D}_{\mu }\left( \omega \right) \varepsilon
_{\lbrack ab}^{\left. {}\right. }\right) \overset{(0)}{e}_{c]}^{\;\;\mu
}+\left( \overset{(1)}{D}_{\mu }\left( \omega \right) \varepsilon _{\lbrack
ab}^{\left. {}\right. }\right) \overset{(1)}{e}_{c]}^{\;\;\mu }  \notag \\
&&+\left( \overset{(0)}{D}_{\mu }\left( \omega \right) \varepsilon _{\lbrack
ab}^{\left. {}\right. }\right) \overset{(2)}{e}_{c]}^{\;\;\mu }+\left(
\partial _{\mu }A_{abc}\right) \overset{(1)}{\bar{\epsilon}}^{\mu
}+A_{\;\;[ab}^{m}\overset{(1)}{\epsilon }_{c]m}  \notag \\
&&-\frac{\mathrm{i}}{8}\left( \bar{\varepsilon}\gamma ^{m}\psi _{\lbrack
a}\right) A_{bc]m},  \label{ap23f}
\end{eqnarray}%
where we denoted by $\overset{(0)}{D}_{\mu }$, $\overset{(1)}{D}_{\mu }$,
and $\overset{(2)}{D}_{\mu }$ the net contributions of orders zero, one, and
two respectively of the covariant derivative%
\begin{equation}
D_{\mu }\left( \omega \right) \varepsilon _{ab}=\partial _{\mu }\varepsilon
_{ab}+\frac{1}{2}\left( \varepsilon _{a}^{\;\;m}\omega _{\mu bm}-\varepsilon
_{b}^{\;\;m}\omega _{\mu am}\right) .  \label{dercov3f}
\end{equation}%
Therefore, relations (\ref{ap03f})--(\ref{ap23f}) are nothing but the first
three orders of the general formula
\begin{eqnarray}
\bar{\delta}_{\epsilon ,\varepsilon }A_{abc} &=&\left( D_{\mu }\left( \omega
\right) \varepsilon _{\lbrack ab}^{\left. {}\right. }\right) e_{c]}^{\;\;\mu
}+\lambda \left[ \left( \partial _{\mu }A_{abc}\right) \bar{\epsilon}^{\mu
}+A_{\;\;[ab}^{m}\epsilon _{c]m}\right.  \notag \\
&&\left. -\tilde{k}_{i}\bar{\varepsilon}\gamma _{\lbrack ab}\psi _{c]}-\frac{%
\mathrm{i}\lambda }{8}\left( \bar{\varepsilon}\gamma ^{m}\psi _{\lbrack
a}\right) A_{bc]m}\right] .  \label{tr3fflat}
\end{eqnarray}%
Due to (\ref{id6}) and (\ref{tr3fflat}), we obtain the gauge transformations
of the `curved' $3$-form, $\bar{A}_{\mu \nu \rho }$, are given by
\begin{equation}
\bar{\delta}_{\epsilon ,\varepsilon }\bar{A}_{\mu \nu \rho }=\partial
_{\lbrack \mu }\bar{\varepsilon}_{\nu \rho ]}+\lambda \left[ \bar{\epsilon}%
^{\lambda }\partial _{\lambda }\bar{A}_{\mu \nu \rho }+A_{\lambda \lbrack
\mu \nu }\left( \partial _{\rho ]}\bar{\epsilon}^{\lambda }\right) -\tilde{k}%
_{i}\bar{\varepsilon}\Gamma _{\lbrack \mu \nu }\psi _{\rho ]}\right] ,
\label{tr3fcurb}
\end{equation}%
where
\begin{equation*}
\bar{A}_{\mu \nu \rho }=e_{\;\;\mu }^{a}e_{\;\;\nu }^{b}e_{\;\;\rho
}^{c}A_{abc},\qquad \bar{\varepsilon}_{\mu \nu }=e_{\;\;\mu }^{a}e_{\;\;\nu
}^{b}\varepsilon _{ab}.
\end{equation*}

So far, we proved that the only consistent interactions in $D=11$ for a spin-%
$2$ field, a massless $3$-form, and a massless (Rarita-Schwinger) spinor
vector complying with our working hypotheses are nothing but the first
orders of the Lagrangian formulation of $D=11$, $N=1$ SUGRA (action (\ref%
{lagfull}) and gauge transformations (\ref{id6}),
(\ref{trfullmajcurb}), and (\ref{tr3fcurb})). The uniqueness of
$D=11$, $N=1$ SUGRA to all orders in the coupling constant can be
shown using exactly the same procedure like in Section 6 of
Ref.~\cite{pI}. Thus, it can be proved that the complete deformed
solution of the master equation for a spin-$2$ field, a massless
$3$-form, and a massless Rarita-Schwinger spinor, consistent at all
orders in the
coupling constant%
\begin{equation}
\hat{S}=S_{0}+\lambda \hat{S}_{1}+\lambda ^{2}\hat{S}_{2}+\cdots ,
\label{soldef}
\end{equation}%
coincides at each order with the solution of the master equation for $D=11$,
$N=1$ SUGRA modulo a redefinition of the coupling constant of the type%
\begin{equation}
\lambda \longrightarrow \lambda \left( 1+k_{2}\lambda ^{2}+k_{3}\lambda
^{3}+\cdots \right) ,  \label{modpardef}
\end{equation}%
where $\left( k_{m}\right) _{m\geq 2}$ are some arbitrary, real constants.

\section{Conclusion}

To conclude with, in this paper we have completed the cohomological BRST
approach to the consistent interactions in eleven spacetime dimensions that
can be added to a free theory describing a massless spin-$2$ field, a
massless (Rarita-Schwinger) spin-$3/2$ field, and an Abelian $3$-form gauge
field. The couplings are obtained under the hypotheses of smoothness in the
coupling constant, locality, Lorentz covariance, Poincar\'{e} invariance,
and the derivative order assumption (the maximum derivative order of the
interacting Lagrangian density is equal to two, with the precaution that
each interacting field equation contains at most one spacetime derivative
acting on gravitini). Our main result is that if we decompose the metric
like $g_{\mu \nu }=\sigma _{\mu \nu }+\lambda h_{\mu \nu }$, then we can
couple the $3$-form and the gravitini to $h_{\mu \nu }$ in the space of
formal series with the maximum derivative order equal to two in $h_{\mu \nu
} $ such that the resulting interactions agree with the well-known $D=11$, $%
N=1 $ SUGRA couplings in the vielbein formulation. Only now, in the presence
of \emph{all} fields, the cosmological term and the gravitini `mass'
constant are forbidden and the quartic gravitini vertex is unfolded.
Although at a first sight it seems that two different theories emerge
(corresponding to the two different values of $\tilde{k}$ from (\ref{solntr2}%
)), in fact each of them describes $D=11$, $N=1$ SUGRA since they can be
obtained one from the other by the simple $3$-form redefinition $%
A_{abc}\rightarrow -A_{abc}$. Our approach is thus a systematic,
cohomological proof of the uniqueness of $D=11$, $N=1$ SUGRA.

\section*{Acknowledgments}

The authors wish to thank Constantin Bizdadea and Odile Saliu for useful
discussions and comments. This work is partially supported by the European
Commission FP6 program MRTN-CT-2004-005104 and by the grant AT24/2005 with
the Romanian National Council for Academic Scientific Research
(C.N.C.S.I.S.) and the Romanian Ministry of Education and Research (M.E.C.).

\end{document}